\documentclass[twoside, 11pt, journal, onecolumn]{IEEEtran}

\usepackage{epsfig}
\usepackage{graphicx}
\usepackage{amsmath}
\usepackage{amssymb}
\usepackage{float}
\usepackage{url}
\newtheorem{definition}{Definition}

\newtheorem{proposition}{Proposition}
\newtheorem{corollary}{Corollary}
\newtheorem{theorem}{Theorem}

\newtheorem{remark}{Remark}
\newtheorem{example}{Example}

\newcommand{\dsum}{\displaystyle\sum}

\newcommand{\naturals}{\ensuremath{\mathbb{N}}}
\newcommand{\reals}{\ensuremath{\mathbb{R}}}

\newcommand{\pr}{\ensuremath{\mathbb{P}}}
\newcommand{\expectation}{\ensuremath{\mathbb{E}}}

\begin{document}
\markboth{Un-published manuscript. Last
updated: September 4, 2012.}{I. SASON: An Information-Theoretic Perspective of the Poisson Approximation via the Chen-Stein Method}

\title{ \huge{An Information-Theoretic Perspective of the Poisson Approximation via the
Chen-Stein Method}}

\author{\IEEEauthorblockN{\vspace*{0.2cm} Igal Sason\\
Department of Electrical Engineering\\
Technion - Israel Institute of Technology\\
Haifa 32000, Israel\\
\hspace*{-0.4cm} E-mail: sason@ee.technion.ac.il}}

\maketitle

\vspace*{-0.5cm}
\begin{abstract}
The first part of this work considers the entropy of the sum of
(possibly dependent and non-identically distributed) Bernoulli
random variables. Upper bounds on the error that follows from an
approximation of this entropy by the entropy of a Poisson
random variable with the same mean are derived via the Chen-Stein
method. The second part of this work derives new lower bounds on the
total variation distance and relative entropy between the distribution
of the sum of independent Bernoulli random variables and the
Poisson distribution. The starting point of the derivation of the new
bounds in the second part of this work is an introduction of a new
lower bound on the total variation distance, whose derivation generalizes
and refines the analysis by Barbour and Hall (1984), based on the Chen-Stein method
for the Poisson approximation. A new lower bound on the relative entropy
between these two distributions is introduced, and this lower bound is
compared to a previously reported upper bound on the relative entropy
by Kontoyiannis et al. (2005). The derivation of the new lower bound
on the relative entropy follows from the new lower bound on the total
variation distance, combined with a distribution-dependent
refinement of Pinsker's inequality by Ordentlich and Weinberger (2005).
Upper and lower bounds on the Bhattacharyya parameter, Chernoff
information and Hellinger distance between the distribution of
the sum of independent Bernoulli random variables and the Poisson
distribution with the same mean are derived as well via some
relations between these quantities with the total variation distance
and the relative entropy. The analysis in this work combines elements of
information theory with the Chen-Stein method for the Poisson approximation.
The resulting bounds are easy to compute, and their applicability
is exemplified.
\end{abstract}



\section{Introduction}
\label{section: introduction}
Convergence to the Poisson distribution, for the number of occurrences of possibly
dependent events, naturally arises in various applications. Following the work of
Poisson, there has been considerable interest in how well the Poisson distribution
approximates the binomial distribution. This approximation was treated by a limit
theorem in \cite[Chapter~8]{Feller_book1950}, and later some non-asymptotic
results have considered the accuracy of this approximation. Among these old and interesting
results, Le Cam's inequality \cite{Le Cam_1960} provides
an upper bound on the total variation distance between the distribution of
the sum $S_n = \sum_{i=1}^n X_i$ of $n$ independent Bernoulli random variables $\{X_i\}_{i=1}^n$,
where $X_i \sim \text{Bern}(p_i)$, and a Poisson distribution $\text{Po}(\lambda)$ with
mean $\lambda = \sum_{i=1}^n p_i$. This inequality states that

\vspace*{-0.2cm}
$$ d_{\text{TV}}\bigl(P_{S_n}, \text{Po}(\lambda)\bigr) \triangleq \frac{1}{2} \,
\sum_{k=0}^{\infty} \, \Bigl| \pr(S_n = k) - \frac{e^{-\lambda} \lambda^k}{k!} \Bigr|
\leq \sum_{i=1}^n p_i^2$$
so if, e.g., $X_i \sim \text{Bern}\bigl(\frac{\lambda}{n}\bigr)$ for every $i \in \{1, \ldots, n\}$
(referring to the case that $S_n$ is binomially distributed) then this upper bound is equal to
$\frac{\lambda^2}{n}$, thus decaying to zero as
$n$ tends to infinity. This upper bound was later improved, e.g., by Barbour and Hall
(see \cite[Theorem~1]{BarbourH_1984}), replacing the above upper bound by
$\left(\frac{1-e^{-\lambda}}{\lambda}\right) \sum_{i=1}^n p_i^2$ and therefore improving
it by a factor of $\frac{1}{\lambda}$ when $\lambda$ is large. This improved upper bound
was also proved by Barbour and Hall to be essentially tight (see \cite[Theorem~2]{BarbourH_1984})
with the following lower bound on the total variation distance:

\vspace*{-0.3cm}
$$d_{\text{TV}}\bigl(P_{S_n}, \text{Po}(\lambda)\bigr) \geq \frac{1}{32} \,
\min\Bigl\{1, \frac{1}{\lambda}\Bigr\} \, \sum_{i=1}^n p_i^2$$
so the upper and lower bounds on the total variation distance differ by a factor
of at most 32, irrespectively of the value of $\lambda$ (it is noted that
in \cite[Remark~3.2.2]{BarbourHJ_book_1992}, the factor
$\frac{1}{32}$ in the lower bound was claimed to be improvable to $\frac{1}{14}$
with no explicit proof). The Poisson approximation and later also the
compound Poisson approximation have been extensively treated in the literature (see,
e.g., the reference list in \cite{BarbourHJ_book_1992} and this paper).

Among modern methods, the Chen-Stein method forms a powerful probabilistic tool that is used
to calculate error bounds when the Poisson approximation serves to assess the distribution of
a sum of (possibly dependent) Bernoulli random variables \cite{Chen_1975}. This method is
based on the simple property of the Poisson distribution where $Z \sim \text{Po}(\lambda)$
with $\lambda \in (0, \infty)$ if and only if
$ \lambda \, \expectation[f(Z+1)] - \expectation[Z \, f(Z)]= 0 $
for all bounded functions $f$ that are defined on $\naturals_0 \triangleq \{0, 1, \ldots \}$.
This method provides a rigorous analytical treatment, via error bounds, to the
case where $W$ has approximately the Poisson distribution $\text{Po}(\lambda)$ where it can be
expected that $ \lambda \, \expectation[f(W+1)] - \expectation[W \, f(W)] \approx 0 $
for an arbitrary bounded function $f$ that is defined on $\naturals_0$. The interested
reader is referred to several comprehensive surveys on the
Chen-Stein method in \cite{ArratiaGG_Tutorial90},
\cite{BarbourHJ_book_1992}, \cite[Chapter~2]{BarbourC_book_2005},
\cite{Probability_Surveys_2005}, \cite[Chapter~2]{RossP_book07} and \cite{Ross_Tutorial11}.

%

During the last decade, information-theoretic methods were exploited to establish
convergence to Poisson and compound Poisson limits in suitable paradigms. An
information-theoretic study of the convergence rate of the binomial-to-Poisson
distribution, in terms of the relative entropy between the binomial and Poisson
distributions, was provided in \cite{HarremoesR_2004}, and maximum entropy results
for the binomial, Poisson and compound Poisson distributions were studied in
\cite{Harremoes_2001}, \cite{Johnson_2007}, \cite{KarlinR_81}, \cite{Shepp_Olkin_1981},
\cite{Yu_IT08}, \cite{Yu_IT09_paper1} and \cite{Yu_IT09_paper2}. The law of small
numbers refers to the phenomenon that, for random variables $\{X_i\}_{i=1}^n$ defined on
$\naturals_0$, the sum $\sum_{i=1}^n X_i$ is approximately Poisson distributed with mean
$\lambda = \sum_{i=1}^n p_i$ if (qualitatively) the following conditions hold:
$\pr(X_i = 0)$ is close to~1, $\pr(X_i = 1)$ is uniformly small,
$\pr(X_i > 1)$ is negligible as compared to $\pr(X_i=1)$, and
$\{X_i\}_{i=1}^n$ are weakly dependent (see \cite{Freedman_1974},
\cite{Serfling_1975} and \cite{Serfling_1978}).
An information-theoretic study of the law of small numbers was provided in
\cite{KontoyiannisHJ_2005} via the derivation of upper bounds on the relative
entropy between the distribution of the sum of possibly dependent Bernoulli
random variables and the Poisson distribution with the same mean. An extension
of the law of small numbers to a thinning limit theorem for convolutions of
discrete distributions that are defined on $\naturals_0$ was introduced in
\cite{Thinning_IT2010}, followed by an analysis of the convergence rate and
some non-asymptotic results. Further work in this direction was studied in
\cite{JohnsonY_IT10}, and the work in  \cite{BarbourJKM_EJP_2010}
provides an information-theoretic study for the problem of compound Poisson
approximation, which parallels the earlier study for the Poisson approximation
in \cite{KontoyiannisHJ_2005}. A recent follow-up to the works in \cite{BarbourJKM_EJP_2010}
and \cite{KontoyiannisHJ_2005} is provided in \cite{Ley_Swan_2011_paper1} and
\cite{Ley_Swan_2011_paper2}, considering connections between Stein characterizations
and Fisher information functionals.
Nice surveys on the line of work on information-theoretic aspects of the Poisson
approximation are introduced in \cite[Chapter~7]{Johnson_2007} and \cite{Kontoyiannis_slides2006}.
Furthermore, \cite[Chapter~2]{DasGupta_2008} surveys some commonly-used metrics between
probability measures with some pointers to the Poisson approximation.

This paper provides an information-theoretic study of the Poisson approximation
via the Chen-Stein method.
The novelty of this paper is considered to be in the following aspects:
\begin{itemize}
\item Consider the entropy of a sum of (possibly dependent and
non-identically distributed) Bernoulli random variables. Upper bounds
on the error that follows from an approximation of this entropy by
the entropy of a Poisson random variable with the same mean are derived
via the Chen-Stein method (see
Theorem~\ref{theorem: upper bound on the Poisson approximation of the entropy}
and its related results in Section~\ref{section: Error bounds on the
entropy of the sum of Bernoulli random variables}).
The use of these new bounds is exemplified
for some interesting applications of the Chen-Stein method in
\cite{ArratiaGG_AOP} and \cite{ArratiaGG_Tutorial90}.
\item Improved lower bounds on the relative entropy between the distribution of a sum
of independent Bernoulli random variables and the Poisson distribution with the same
mean are derived (see Theorem~\ref{theorem: improved lower bound on the relative entropy}
in Section~\ref{section: improved lower bounds on the total variation distance etc.}).
These new bounds are obtained by combining a derivation of some sharpened
lower bounds on the total variation distance (see Theorem~\ref{theorem: improved
lower bound on the total variation distance} and some related results in
Section~\ref{section: improved lower bounds on the total variation distance etc.})
that improve the original lower bound in \cite[Theorem~2]{BarbourH_1984}, and a
probability-dependent refinement of Pinsker's inequality \cite{OrdentlichW_IT2005}.
The new lower bounds are compared with existing upper bounds.
\item New upper and lower bounds on the Chernoff information and Bhattacharyya
parameter are also derived in
Section~\ref{section: improved lower bounds on the total variation distance etc.}
via the introduction of new bounds on the Hellinger distance and relative entropy.
The use of the new lower bounds on the relative entropy and Chernoff information
is exemplified in the context of binary hypothesis testing. The impact of the
improvements of these new bounds is studied as well.
\end{itemize}

To the best of our knowledge, among the publications of the {\em IEEE Trans. on
Information Theory}, the Chen-Stein method for Poisson approximation was used so
far only in two occasions. In \cite{Wyner_IT97}, this probabilistic method was used
by A. J. Wyner to analyze the redundancy and the distribution of the phrase lengths
in one of the versions of the Lempel-Ziv data compression algorithm. In the second
occasion, this method was applied in \cite{FranceschettiM_IT2006} in the context of
random networks. In \cite{FranceschettiM_IT2006}, the authors relied on existing
upper bounds on the total variation distance, applying them to analyze
the asymptotic distribution of the number of isolated nodes in a random grid network
where nodes are always active. The first part of this paper relies (as well) on some
existing upper bounds on the total variation distance, with the purpose of obtaining
error bounds on the Poisson approximation of the entropy for a sum of (possibly dependent)
Bernoulli random variables, or more generally for a sum of non-negative,
integer-valued and bounded random variables (this work relies
on stronger versions of the upper bounds in \cite[Theorems~2.2 and~2.4]{FranceschettiM_IT2006}).

The paper is structured as follows:
Section~\ref{section: Error bounds on the entropy of the sum of Bernoulli random variables}
forms the first part of this work where the entropy of the sum of
Bernoulli random variables is considered.
Section~\ref{section: improved lower bounds on the total variation distance etc.}
provides the second part of this work where new lower bounds on the
total variation distance and relative entropy between the distribution
of the sum of independent Bernoulli random variables and the
Poisson distribution are derived. The derivation of the new
and improved lower bounds on the total variation distance relies
on the Chen-Stein method for the Poisson approximation, and it
generalizes and tightens the analysis that was used to derive the
original lower bound on the total variation distance in
\cite{BarbourH_1984}. The derivation of the new lower bound
on the relative entropy follows from the new lower bounds on the
total variation distance, combined with a distribution-dependent
refinement of Pinsker's inequality in \cite{OrdentlichW_IT2005}.
The new lower bound on the relative entropy is compared to a
previously reported upper bound on the relative entropy from
\cite{KontoyiannisHJ_2005}.
Upper and lower bounds on the Bhattacharyya parameter, Chernoff
information and the Hellinger, local and Kolmogorov-Smirnov distances
between the distribution of the sum of independent Bernoulli random
variables and the Poisson distribution with the same mean are also derived in
Section~\ref{section: improved lower bounds on the total variation distance etc.}
via some relations between these quantities with the total variation distance
and the relative entropy. The analysis in this work combines elements of
information theory with the Chen-Stein method for Poisson approximation.
The use of these new bounds is exemplified in the two parts of this work,
partially relying on some interesting applications of the Chen-Stein method
for the Poisson approximation that were introduced in \cite{ArratiaGG_AOP}
and \cite{ArratiaGG_Tutorial90}. The bounds that are derived
in this work are easy to compute, and their applicability is exemplified.
Throughout the paper, the logarithms are expressed on the natural base (on base $e$).

\section{Error Bounds on the Entropy of the Sum of Bernoulli
Random Variables}
\label{section: Error bounds on the entropy of the sum of Bernoulli
random variables}

This section considers the entropy of a sum
of (possibly dependent and non-identically distributed)
Bernoulli random variables.
Section~\ref{subsection: First part of the review of
some known results} provides a review of some reported results
on the Poisson approximation, whose derivation relies on the Chen-Stein method,
that are relevant to the analysis in this section. The
original results of this section are introduced from
Section~\ref{subsection: A New Result on the Entropy of Discrete Random Variables}
which provides an upper bound on the entropy difference between
two discrete random variables in terms of their total variation
distance. This bound is later in this section in the context of
the Poisson approximation.
Section~\ref{subsection: New error bounds on the entropy}
introduces some explicit upper bounds on the error that
follows from the approximation of the entropy of a sum
of Bernoulli random variables by the entropy of a Poisson
random variable with the same mean. Some applications of
the new bounds are exemplified in Section~\ref{subsection:
Examples for the use of the mew error bounds on the entropy},
and these bounds are proved in
Section~\ref{subsection: Proofs of the new bounds in the
first part of this paper}. Finally, a generalization of these
bounds is introduced in Section~\ref{subsection: generalization
of the bounds on the entropy for the sum of integer-valued random variables}
to address the case of the Poisson approximation for the entropy of a
sum of non-negative, integer-valued and bounded random variables.

\subsection{Review of Some Essential Results for the Analysis in
Section~\ref{section: Error bounds on the entropy of the sum of Bernoulli
random variables}}
\label{subsection: First part of the review of some known results}

Throughout the paper, we use the term `distribution' to refer to the discrete
probability mass function of an integer-valued random variable. In the following,
we review briefly some known results that are used for the analysis later in this
section.
\begin{definition}
Let $P$ and $Q$ be two probability measures defined on a set $\mathcal{X}$.
Then, the total variation distance between $P$ and $Q$ is defined by
\begin{equation}
d_{\text{TV}}(P, Q)  \triangleq \sup_{\text{Borel} \,
A \subseteq \mathcal{X}} |P(A) - Q(A)|
\label{eq: total variation distance}
\end{equation}
where the supermum is taken w.r.t. all the Borel subsets $A$ of
$\mathcal{X}$.
If $\mathcal{X}$ is a countable set then \eqref{eq: total variation distance}
is simplified to
\begin{equation}
d_{\text{TV}}(P, Q) = \frac{1}{2} \sum_{x \in \mathcal{X}} |P(x) - Q(x)| =
\frac{||P-Q||_1}{2}
\label{eq: the L1 distance is twice the total variation distance}
\end{equation}
so the total variation distance is equal to one-half of the $L_1$-distance
between the two probability distributions.
\label{definition: total variation distance}
\end{definition}

The following theorem combines \cite[Theorems~1 and 2]{BarbourH_1984},
and its proof relies on the Chen-Stein method:

\begin{theorem}
Let $W = \sum_{i=1}^n X_i$ be a sum of $n$ independent Bernoulli random
variables with $\expectation(X_i) = p_i$ for $i \in \{1, \ldots, n\}$,
and $\expectation(W) = \lambda$. Then, the total variation distance
between the probability distribution of $W$ and the Poisson
distribution with mean $\lambda$ satisfies
\begin{equation}
\frac{1}{32} \, \Bigl(1 \wedge \frac{1}{\lambda}\Bigr) \, \sum_{i=1}^n p_i^2
\leq d_{\text{TV}}(P_W, \text{Po}(\lambda)) \leq
\left(\frac{1-e^{-\lambda}}{\lambda}\right) \, \sum_{i=1}^n p_i^2
\label{eq: bounds on the total variation distance - Barbour and Hall 1984}
\end{equation}
where $a \wedge b \triangleq \min\{a,b\}$ for every $a, b \in \reals$.
\label{theorem: bounds on the total variation distance - Barbour and Hall 1984}
\end{theorem}

\begin{remark}
The ratio between the upper and lower bounds in
Theorem~\ref{theorem: bounds on the total variation distance - Barbour and Hall 1984}
is not larger than~32, irrespectively of the values of $\{p_i\}$. This shows that,
for independent Bernoulli random variables, these bounds are essentially tight.
The upper bound in
\eqref{eq: bounds on the total variation distance - Barbour and Hall 1984} improves
Le Cam's inequality (see \cite{Le Cam_1960}, \cite{Steele_1994})) which states that
$ d_{\text{TV}}(P_W, \text{Po}(\lambda)) \leq \sum_{i=1}^n p_i^2 $
so the improvement, for large values of $\lambda$, is approximately by the factor
$\frac{1}{\lambda}$.
\end{remark}

Theorem~\ref{theorem: bounds on the total variation distance - Barbour and Hall 1984}
provides a non-asymptotic result for the Poisson approximation of sums of independent
binary random variables via the use of the Chen-Stein method. In general, this method
enables to analyze the Poisson approximation for sums of dependent random variables. To
this end, the following notation was used in \cite{ArratiaGG_AOP} and
\cite{ArratiaGG_Tutorial90}:

Let $I$ be a countable index set, and for $\alpha \in I$, let $X_{\alpha}$ be a Bernoulli
random variable with
\begin{equation}
p_{\alpha} \triangleq \pr(X_{\alpha}=1) = 1-\pr(X_{\alpha}=0) > 0.
\label{eq: probabilities of the Bernoulli random variables}
\end{equation}
Let
\begin{equation}
W \triangleq \sum_{\alpha \in I} X_{\alpha}, \quad \lambda \triangleq \expectation(W)
= \sum_{\alpha \in I} p_{\alpha}
\label{eq: Bernoulli sums and their mean}
\end{equation}
where it is assumed that $\lambda \in (0, \infty)$.
For every $\alpha \in I$, let $B_{\alpha}$ be a subset of $I$ that is chosen such that
$\alpha \in B_{\alpha}$. This subset is interpreted in \cite{ArratiaGG_AOP}
as the neighborhood of dependence for $\alpha$ in the sense that $X_{\alpha}$ is independent
or weakly dependent of all of the $X_{\beta}$ for $\beta \notin B_{\alpha}$. Furthermore,
the following coefficients were defined in \cite[Section~2]{ArratiaGG_AOP}:
\begin{eqnarray}
&& b_1 \triangleq \sum_{\alpha \in I} \sum_{\beta \in B_{\alpha}} p_{\alpha} p_{\beta}
\label{eq: b1} \\[0.1cm]
&& b_2 \triangleq \sum_{\alpha \in I} \sum_{\beta \in B_{\alpha} \setminus \{\alpha\}}
p_{\alpha, \beta},
\quad p_{\alpha, \beta} \triangleq \expectation(X_{\alpha} X_{\beta}) \label{eq: b2} \\[0.1cm]
&& b_3 \triangleq \sum_{\alpha \in I} s_{\alpha}, \quad \quad
s_{\alpha} \triangleq \expectation \bigl| \expectation(X_{\alpha} - p_{\alpha} \, | \,
\sigma(\{X_{\beta}\})_{\beta \in I \setminus B_{\alpha}}) \bigr|  \label{eq: b3}
\end{eqnarray}
where $\sigma(\cdot)$ in the conditioning of \eqref{eq: b3} denotes the $\sigma$-algebra
that is generated by the random variables inside the parenthesis.
In the following, we cite \cite[Theorem~1]{ArratiaGG_AOP} which essentially implies
that when $b_1, b_2$ and $b_3$ are all small, then the total number $W$ of events is
approximately Poisson distributed.
\begin{theorem}
Let $W = \sum_{\alpha \in I} X_{\alpha}$ be a sum of (possibly dependent and
non-identically distributed) Bernoulli random variables $\{X_{\alpha}\}_{\alpha \in I}$.
Then, with the notation in
\eqref{eq: probabilities of the Bernoulli random variables}--\eqref{eq: b3},
the following upper bound on the total variation distance holds:
\vspace*{-0.1cm}
\begin{equation}
d_{\text{TV}}(P_W, \text{Po}(\lambda)) \leq (b_1 + b_2) \left( \frac{1-e^{-\lambda}}{\lambda}\right)
+ b_3 \Bigl(1 \wedge \frac{1.4}{\sqrt{\lambda}}\Bigr).
\label{eq: upper bound on the total variation distance by Arratia et al.}
\end{equation}
\label{theorem: upper bound on the total variation distance by Arratia et al.}
\end{theorem}

\begin{remark}
A comparison of the right-hand side of
\eqref{eq: upper bound on the total variation distance by Arratia et al.}
with the bound in \cite[Theorem~1]{ArratiaGG_AOP} shows a difference in
a factor of 2 between the two upper bounds. This follows from a
difference in a factor of~2 between the two definitions of the total variation
distance in \cite[Section~2]{ArratiaGG_AOP} and
Definition~\ref{definition: total variation distance} here. It is noted,
however, that Definition~\ref{definition: total variation distance} in
this work is consistent, e.g., with \cite{BarbourH_1984} and \cite{BarbourHJ_book_1992}.
\end{remark}

\begin{remark}
Theorem~\ref{theorem: upper bound on the total variation distance by Arratia et al.}
forms a generalization of the upper bound in
Theorem~\ref{theorem: bounds on the total variation distance - Barbour and Hall 1984}
by choosing $B_{\alpha} = \{\alpha\}$ for $\alpha \in I \triangleq \{1, \ldots, n\}$
(note that, due to the independence assumption of the Bernoulli random variables in
Theorem~\ref{theorem: bounds on the total variation distance - Barbour and Hall 1984},
the neighborhood of dependence of $\alpha$ is $\alpha$ itself). In this setting,
under the independence assumption,
$$ b_1 = \sum_{i=1}^n p_i^2, \quad b_2 = b_3 = 0$$
which therefore gives, from
\eqref{eq: upper bound on the total variation distance by Arratia et al.}, the upper
bound on the right-hand side of
\eqref{eq: bounds on the total variation distance - Barbour and Hall 1984}.
\label{remark: Generalization of Theorem 1 of Barbour and Hall}
\end{remark}


\vspace*{0.1cm}
Before proceeding to this analysis, the following maximum entropy result of the Poisson
distribution is introduced.

\begin{theorem}
The Poisson distribution $\text{Po}(\lambda)$ has the maximal entropy among all probability
distributions with mean $\lambda$ that can be obtained as sums of independent Bernoulli RVs:
\begin{eqnarray}
&& \hspace*{-0.5cm} H(\text{Po}(\lambda)) =
\sup_{S \in B_{\infty}(\lambda)} H(S) \nonumber \\
&& \hspace*{-0.5cm} B_{\infty}(\lambda) \triangleq
\bigcup_{n \in \naturals} B_n(\lambda) \nonumber \\
&& \hspace*{-0.5cm} B_n(\lambda) \triangleq \left\{S: \,
S = \sum_{i=1}^n X_i, \; X_i \sim \text{Bern}(p_i)\;
\text{independent}, \; \sum_{i=1}^n p_i = \lambda \right\}.
\label{eq: maximum entropy result for the Poisson distribution}
\end{eqnarray}
Furthermore, since the supremum of the entropy over the set $B_n(\lambda)$ is monotonic
increasing in $n$, then
$$ H(\text{Po}(\lambda)) = \lim_{n \rightarrow \infty} \sup_{S \in B_n(\lambda)} H(S).$$
For $n \in \naturals$, the maximum entropy distribution in the class $B_n(\lambda)$
is the Binomial distribution of the sum of $n$ i.i.d. Bernoulli random variables  $\text{Ber}\Bigl(\frac{\lambda}{n}\Bigr)$, so
\begin{equation*}
H(\text{Po}(\lambda)) = \lim_{n \rightarrow \infty}
H\Bigl(\text{Binomial}\Bigl(n, \frac{\lambda}{n}\Bigr)\Bigr).
\end{equation*}
\label{theorem: maximum entropy result for the Poisson distribution}
\end{theorem}

\begin{remark}
Theorem~\ref{theorem: maximum entropy result for the Poisson distribution}
partially appears in \cite[Proposition~2.1]{KarlinR_81}
(see \cite[Eq.~(2.20)]{KarlinR_81}). This theorem
follows directly from \cite[Theorems~7 and 8]{Harremoes_2001}.
\end{remark}

\begin{remark}
The maximum entropy result for the Poisson distribution in
Theorem~\ref{theorem: maximum entropy result for the Poisson distribution}
was strengthened in \cite{Johnson_2007} by showing that the supermum on
the right-hand side of \eqref{eq: maximum entropy result for the Poisson distribution}
can be extended to the larger set of ultra-log-concave probability mass functions
(that includes the binomial distribution). This
result for the Poisson distribution was generalized in \cite{JohnsonKM_ISIT2009} and
\cite{JohnsonKM_2012} to maximum entropy results for discrete compound Poisson distributions.
\end{remark}

\vspace*{0.1cm}
{\em Calculation of the entropy of a Poisson random variable}:
In the next sub-section, we consider the approximation of the entropy of a sum
of Bernoulli random variables by the entropy of a Poisson random variable with
the same mean. To this end, it is required to evaluate the entropy of
$Z \sim \text{Po}(\lambda)$. It is straightforward to verify that
\begin{equation}
H(Z) = \lambda \log\left(\frac{e}{\lambda}\right) + \sum_{k=1}^{\infty}
\frac{\lambda^k e^{-\lambda} \log k!}{k!}
\label{eq: entropy of Poisson distribution}
\end{equation}
so the entropy of the Poisson distribution (in nats) is given in terms of an infinite
series that has no closed-form expression. Sequences of simple upper and lower bounds on
this entropy, which are asymptotically tight, were derived in \cite{AdellLY_IT2010}.
In particular, from \cite[Theorem~2]{AdellLY_IT2010},
\begin{equation}
-\frac{31}{24 \lambda^2} - \frac{33}{20 \lambda^3} - \frac{1}{20\lambda^4}
\leq H(Z) - \frac{1}{2} \, \log(2\pi e \lambda) + \frac{1}{12 \lambda}
\leq \frac{5}{24 \lambda^2} + \frac{1}{60 \lambda^3}
\label{eq: bounds on the entropy of a Poisson RV with large mean}
\end{equation}
which gives tight bounds on the entropy of $Z \sim \text{Po}(\lambda)$ for
large values of $\lambda$. For $\lambda \geq 20$, the entropy of $Z$ is
approximated by the average of its upper and lower bounds in
\eqref{eq: bounds on the entropy of a Poisson RV with large mean}, asserting
that the relative error of this approximation is less than $0.1\%$ (and it
decreases like $\frac{1}{\lambda^2}$ while increasing the value of $\lambda$). For
$\lambda \in (0, 20)$, a truncation of the infinite
series on the right-hand side of \eqref{eq: entropy of Poisson distribution} after
its first $\lceil 10 \lambda \rceil$ terms gives an accurate approximation.

\subsection{A New Bound on the Entropy Difference of Two Discrete Random Variables}
\label{subsection: A New Result on the Entropy of Discrete Random Variables}
The following theorem provides a new upper bound on the entropy difference between
two discrete random variables in terms of their total variation distance. This
theorem relies on the bound of Ho and Yeung in \cite[Theorem~6]{entropy_difference_and_variational_distance_IT2010}
 that forms an improvement over the previously reported bound in
\cite[Theorem~17.3.3]{Cover_Thomas} or \cite[Lemma~2.7]{Csiszar_Korner_book}.
The following new bound is later used in this section in the context of the
Poisson approximation.
\begin{theorem}
Let $\mathcal{A} = \{a_1, a_2, \ldots\}$ be a countable infinite set. Let
$X$ and $Y$ be two discrete random variables where $X$ takes values from
a finite set $\mathcal{X} = \{a_1, \ldots, a_m\}$, for some $m \in \naturals$,
and $Y$ takes values from the entire set $\mathcal{A}$. Assume that
\begin{equation}
d_{\text{TV}}(X, Y) \leq \eta
\label{eq: eta as a bound on the total variation distance}
\end{equation}
for some $\eta \in [0,1)$, and let
\begin{equation}
M \triangleq \max \left\{ m+1, \frac{1}{1-\eta} \right\}.
\label{eq: definition of the parameter M}
\end{equation}
Furthermore, let $\mu > 0$ be set such that
\begin{equation}
-\sum_{i=M}^{\infty} P_Y(a_i) \, \log P_Y(a_i) \leq \mu
\label{eq: condition on the tail of the entropy}
\end{equation}
then
\begin{equation}
| H(X) - H(Y) | \leq \eta \, \log(M-1) + h(\eta) + \mu
\label{eq: new upper bound on the entropy difference of two discrete RVs}
\end{equation}
where $h$ denote the binary entropy function.
\label{theorem: L1 bound on the entropy}
\end{theorem}

\vspace*{0.2cm}
\begin{proof}
Let $\widetilde{Y}$ be a random variable that is defined to be
equal to $Y$ if $Y \in \{a_1, \ldots, a_{M-1}\}$, and it is
set to be equal to $a_M$ if $Y=a_i$ for some $i \geq M$. Hence, the probability
mass function of $\widetilde{Y}$ is related to that of $Y$ as follows
\begin{equation}
P_{\widetilde{Y}}(a_i) = \left\{ \begin{array}{ll}
                          P_Y(a_i)  & \mbox{if $i \in \{1, \ldots, M-1\}$} \\[0.1cm]
                          \sum_{j=M}^{\infty} P_Y(a_j)  & \mbox{if $i=M$.}
                          \end{array}
                          \right.
\label{eq: probability mass function of Y tilde}
\end{equation}
Since $P_X(a_i) = 0$ for every $i > m$ and $M \geq m+1$, then it follows from
\eqref{eq: probability mass function of Y tilde} that
\begin{eqnarray}
&& d_{\text{TV}}(X, \widetilde{Y}) \nonumber \\
&& = \frac{1}{2} \sum_{i=1}^m |P_X(a_i) - P_{\widetilde{Y}}(a_i)|
+ \frac{1}{2} \sum_{i=m+1}^{M-1} P_{\widetilde{Y}}(a_i)
+ \frac{1}{2} \, P_{\widetilde{Y}}(a_M) \nonumber \\
&& = \frac{1}{2} \sum_{i=1}^m |P_X(a_i) - P_Y(a_i)| +
\frac{1}{2} \sum_{i=m+1}^{\infty} P_Y(a_i) \nonumber \\
&& = d_{\text{TV}}(X, Y).
\label{eq: the two total variation distances are equal}
\end{eqnarray}
Hence, $X$ and $\widetilde{Y}$ are two discrete random variables that take
values from the set $\{a_1, \ldots, a_M\}$ (note that it includes the set
$\mathcal{X}$) and $d_{\text{TV}}(X, \widetilde{Y}) \leq \eta$ (see
\eqref{eq: eta as a bound on the total variation distance} and
\eqref{eq: the two total variation distances are equal}). The bound in \cite[Theorem~6]{entropy_difference_and_variational_distance_IT2010}
therefore implies that if $\eta \leq 1 - \frac{1}{M}$ (which is indeed the case,
due to the way $M$ is defined in \eqref{eq: definition of the parameter M}),
then
\begin{equation}
|H(X) - H(\widetilde{Y})| \leq \eta \, \log(M-1) + h(\eta).
\label{eq: bound on the entropy difference in terms of the total variation distance
by Ho and Yeung}
\end{equation}
Since $\widetilde{Y}$ is a deterministic function of $Y$ then $H(Y) = H(Y, \widetilde{Y}) \geq H(\widetilde{Y})$, and therefore \eqref{eq: condition on the tail of the entropy} and
\eqref{eq: probability mass function of Y tilde} imply that
\begin{eqnarray}
&& | H(\widetilde{Y}) - H(Y) | \nonumber \\
&& = H(Y) - H(\widetilde{Y}) \nonumber \\[0.1cm]
&& = - \sum_{i=M}^{\infty} P_Y (a_i) \, \log P_Y(a_i) +
\left( \sum_{i=M}^{\infty} P_Y(a_i) \right)
\log \left( \sum_{i=M}^{\infty} P_Y(a_i) \right) \nonumber \\
&& \leq - \sum_{i=M}^{\infty} P_Y (a_i) \, \log P_Y(a_i) \nonumber \\
&& \leq \mu.
\label{eq: bound on the entropy difference of Y and Y tilde}
\end{eqnarray}
Finally, the bound in
\eqref{eq: new upper bound on the entropy difference of two discrete RVs}
follows from
\eqref{eq: bound on the entropy difference in terms of the total variation distance
by Ho and Yeung}, \eqref{eq: bound on the entropy difference of Y and Y tilde}
and the triangle inequality.
\end{proof}

\subsection{New Error Bounds on the Entropy of Sums of Bernoulli Random Variables}
\label{subsection: New error bounds on the entropy}
The new bounds on the entropy of sums of Bernoulli random variables are introduced
in the following. Their use is exemplified in
Section~\ref{subsection: Examples for the use of the mew error bounds on the entropy},
and their proofs appear in
Section~\ref{subsection: Proofs of the new bounds in the first part of this paper}.

\vspace*{0.1cm}
\begin{theorem}
Let $I$ be an arbitrary finite index set with $|I| \triangleq n$. Under the
assumptions of
Theorem~\ref{theorem: upper bound on the total variation distance by Arratia et al.}
and the notation used in
Eqs.~\eqref{eq: probabilities of the Bernoulli random variables}--\eqref{eq: b3}, let
\begin{eqnarray}
&& \eta \triangleq (b_1 + b_2) \left(\frac{1-e^{-\lambda}}{\lambda}\right)
+ b_3 \Bigl(1 \wedge \frac{1.4}{\sqrt{\lambda}}\Bigr)
\label{eq: function eta in the upper bound on the Poisson approximation of the entropy}
\\[0.2cm]
&& M \triangleq \max \left\{n+2, \frac{1}{1-\eta} \right\}
\label{eq: function M in the upper bound on the Poisson approximation of the entropy}
\\[0.2cm]
&& \mu \triangleq \left[ \Bigl(\lambda \log \Bigl(\frac{e}{\lambda}\Bigr)\Bigr)_+ \,
+ \lambda^2 + \frac{6 \log(2\pi) + 1}{12} \right]
\, \exp \left\{-\left[\lambda + (M-2) \, \log\left(\frac{M-2}{\lambda e} \right)
\right] \right\}
\label{eq: function mu in the upper bound on the Poisson approximation of the entropy}
\end{eqnarray}
where, in \eqref{eq: function mu in the upper bound on the Poisson approximation of the entropy},
$(x)_+ \triangleq \max\{x, 0\}$ for every $x \in \reals$.
Let $Z \sim \text{Po}(\lambda)$ be a Poisson random variable with mean $\lambda$. If
$\eta < 1$, then the difference between the entropies of $Z$ and $W$ satisfies the
following inequality:
\begin{equation}
|H(Z) - H(W)| \leq \eta \, \log(M-1) + h(\eta) + \mu.
\label{eq: upper bound on the Poisson approximation of the entropy}
\end{equation}
\label{theorem: upper bound on the Poisson approximation of the entropy}
\end{theorem}

\vspace*{0.1cm}
The following corollary refers to the entropy of a sum of independent Bernoulli
random variables:
\begin{corollary}
Consider the setting in Theorem~\ref{theorem: upper bound on the Poisson
approximation of the entropy}, and assume that the Bernoulli random variables
$\{X_{\alpha}\}_{\alpha \in I}$ are also independent. Then, the following
inequality holds:
\begin{equation}
0 \leq H(Z) - H(W) \leq \eta \, \log(M-1) + h(\eta) + \mu
\label{eq: an error bound on the Poisson approximation of the entropy
for independent Bernoulli RVs}
\end{equation}
where $\eta$ in
\eqref{eq: function eta in the upper bound on the Poisson approximation of the entropy}
is specialized to
\begin{equation}
\eta \triangleq \left( \frac{1-e^{-\lambda}}{\lambda} \right) \, \sum_{\alpha \in I}
p_{\alpha}^2.
\label{eq: upper bound on the total variation distance for independent summands}
\end{equation}
\label{corollary: upper bound on the Poisson approximation of the entropy
for independent RVs}
\end{corollary}

\vspace*{0.1cm}
The following bound forms a possible improvement of the result in
Corollary~\ref{corollary: upper bound on the Poisson approximation
of the entropy for independent RVs}.

\begin{proposition}
Assume that the conditions in Corollary~\ref{corollary: upper bound on the
Poisson approximation of the entropy for independent RVs} are satisfied.
Then, inequality \eqref{eq: an error bound on the Poisson approximation of
the entropy for independent Bernoulli RVs} holds with the new parameter
\begin{equation}
\eta \triangleq \theta \, \min
\left\{ 1 - e^{-\lambda}, \; \frac{3}{4e (1-\sqrt{\theta})^{3/2}}
\right\}
\label{eq: improved eta}
\end{equation}
where
\begin{eqnarray}
&& \lambda \triangleq \sum_{\alpha \in I} p_{\alpha} \label{eq: lambda} \\
&& \theta \triangleq \frac{1}{\lambda} \sum_{\alpha \in I} p_{\alpha}^2.
\label{eq: theta}
\end{eqnarray}
\label{proposition: a possibly improved error bound on the Poisson approximation
of the entropy for independent Bernoulli RVs}
\end{proposition}

\begin{remark}
From \eqref{eq: lambda} and \eqref{eq: theta}, it follows that
$0 \leq \theta \leq \max_{\alpha \in I} p_{\alpha} \triangleq p_{\max}$.
The condition that $\eta < 1$ is mild since it is a meaningful
upper bound on the total variation distance (which is bounded by 1).
\end{remark}

\begin{remark}
Proposition~\ref{proposition: a possibly improved error bound on
the Poisson approximation of the entropy for independent Bernoulli RVs}
improves the bound in Corollary~\ref{corollary: upper bound on the
Poisson approximation of the entropy for independent RVs}
only if $\theta$ is below a certain value that depends
on $\lambda$. The maximal improvement that is
obtained by Proposition~\ref{proposition: a possibly improved
error bound on the Poisson approximation of the entropy for
independent Bernoulli RVs}, as compared to
Corollary~\ref{corollary: upper bound on the Poisson approximation
of the entropy for independent RVs}, is in the case where
$\theta \rightarrow 0$ and $\lambda \rightarrow \infty$,
and the corresponding improvement in the value of $\eta$
is by a factor of $\frac{3}{4e} \approx 0.276$.
\end{remark}

\subsection{Applications of the New Error Bounds on the Entropy}
\label{subsection: Examples for the use of the mew error bounds on the entropy}

In the following, the use of
Theorem~\ref{theorem: upper bound on the Poisson approximation of the entropy}
is exemplified for the estimation of the entropy of sums
of (possibly dependent) Bernoulli random variables.
It starts with a simple example where the summands are independent binary random
variables, and some interesting examples from \cite[Section~3]{ArratiaGG_AOP}
and \cite[Section~4]{ArratiaGG_Tutorial90} are considered next. These examples
are related to sums of dependent Bernoulli random variables, where the use of
Theorem~\ref{theorem: upper bound on the Poisson approximation of the entropy}
is exemplified for the calculation of error bounds on the entropy via
the Chen-Stein method.


\vspace*{0.1cm}
\begin{example}[sums of independent Bernoulli random variables] Let
$W = \sum_{i=1}^n X_i$ be a sum of $n$ independent Bernoulli random
variables where $X_i \sim \text{Bern}(p_i)$ for $i=1, \ldots, n$.
The calculation of the entropy of $W$ involves the numerical computation
of the probabilities
$$\bigl(P_W(0), P_W(1), \ldots, P_W(n)\bigr) = (1-p_1, p_1) \ast
(1-p_2, p_2) \ast \ldots (1-p_n, p_n)$$
whose computational complexity is high for very large values of $n$,
especially if the probabilities $p_1, \ldots, p_n$ are not the same.
The bounds in
Corollary~\ref{corollary: upper bound on the Poisson approximation
of the entropy for independent RVs}
and Proposition~\ref{proposition: a possibly improved error bound on
the Poisson approximation of the entropy for independent Bernoulli RVs}
provide rigorous upper bounds on the accuracy of the Poisson
approximation for $H(W)$. Lets exemplify this in the case where
$$p_i = 2ai, \quad \forall \, i \in \{1, \ldots, n\},
\; a = 10^{-10}, \; n = 10^8$$ then
$$ \lambda = \sum_{i=1}^n p_i = an(n+1) = 1,000,000.01 \approx 10^6$$
and from \eqref{eq: theta}
$$ \theta = \frac{1}{\lambda} \sum_{i=1}^n p_i^2 =
\frac{2a(2n+1)}{3} = 0.0133.$$
The entropy of the Poisson random variable $Z \sim \text{Po}(\lambda)$
is evaluated via the bounds in
\eqref{eq: bounds on the entropy of a Poisson RV with large mean}
(since $\lambda \gg 1$, these bounds are tight), and they imply that
$H(Z) = 8.327 \, \text{nats}$.
From Corollary~\ref{corollary: upper bound on the Poisson approximation
of the entropy for independent RVs}
(see Eq.~\eqref{eq: an error bound on the Poisson approximation of the
entropy for independent Bernoulli RVs} where $I = \{1, \ldots, n\}$),
it follows that $0 \leq H(Z) - H(W) \leq 0.316 \, \text{nats}$, and
Proposition~\ref{proposition: a possibly improved error bound on the
Poisson approximation of the entropy for independent Bernoulli RVs}
improves it to $0 \leq H(Z) - H(W) \leq 0.110 \, \text{nats}$. Hence,
$H(W) \approx 8.272 \, \text{nats}$ with a relative error of at most
$0.7\%.$
\end{example}

\vspace*{0.1cm}
\begin{example}[random graphs]
This problem, which appears in \cite[Example~1]{ArratiaGG_AOP}, is described
as follows: On the cube $\{0,1\}^n$, assume that each of the
$n 2^{n-1}$ edges is assigned a random direction by tossing a fair
coin. Let $k \in \{0, 1, \ldots, n\}$ be fixed, and denote by
$W \triangleq W(k,n)$ the random variable that is equal to the
number of vertices at which exactly $k$ edges point outward (so
$k=0$ corresponds to the event where all $n$ edges, from a certain
vertex, point inward). Let $I$ be the set of all $2^n$ vertices,
and $X_{\alpha}$ be the indicator that vertex $\alpha \in I$ has
exactly $k$ of its edges directed outward. Then
$W = \sum_{\alpha \in I} X_{\alpha}$ with $$X_{\alpha} \sim
\text{Bern}(p), \quad p = 2^{-n} {{n}\choose{k}},
\quad \forall \alpha \in I.$$
This implies that $\lambda = {{n}\choose{k}}$ (since $|I|=2^n$).
Clearly, the neighborhood of dependence of a vertex $\alpha \in I$,
denoted by $B_{\alpha}$, is the set of vertices that are directly
connected to $\alpha$ (including $\alpha$ itself since
Theorem~\ref{theorem: upper bound on the total variation distance
by Arratia et al.} requires that $\alpha \in B_{\alpha}$). It is
noted, however, that $B_{\alpha}$ in \cite[Example~1]{ArratiaGG_AOP}
was given by $B_{\alpha} = \{\beta: \, |\beta - \alpha| = 1\}$ so
it excluded the vertex $\alpha$. From \eqref{eq: b1}, this difference
implies that $b_1$ in their example should be modified to
\begin{eqnarray}
&& b_1 = |I| \, |B_{\alpha}| \, \left(2^{-n}
{{n}\choose{k}}\right)^2 \nonumber \\[0.2cm]
&& \hspace*{0.4cm} = 2^{-n} (n+1) {{n}\choose{k}}^2
\end{eqnarray}
so $b_1$ is larger than its value in \cite[p.~14]{ArratiaGG_AOP}
by a factor of $1+\frac{1}{n}$ which has a negligible effect if
$n \gg 1$. As is noted in \cite[p.~14]{ArratiaGG_AOP}, if $\alpha$
and $\beta$ are two vertices that are connected by an edge,
then a conditioning on the direction of this edge gives that
$$p_{\alpha, \beta} \triangleq \expectation(X_\alpha X_\beta) =
2^{2-2n} \, {{n-1}\choose{k}} \, {{n-1}\choose{k-1}}, \quad
\forall \, \alpha \in I, \; \; \beta \in B_\alpha \setminus
\{\alpha\}$$ and therefore, from \eqref{eq: b2},
$$ b_2 = n \, 2^{2-n} \, {{n-1}\choose{k}} \, {{n-1}\choose{k-1}}.$$
Finally, as is noted in \cite[Example~1]{ArratiaGG_AOP}, $b_3=0$
(this is because the conditional expectation of $X_{\alpha}$
given $(X_\beta)_{\beta \in I \setminus B_{\alpha}}$ is, similarly
to the un-conditional expectation, equal to $p_{\alpha}$; i.e.,
the directions of the edges outside the neighborhood of dependence
of $\alpha$ are irrelevant to the directions of the edges connecting
the vertex $\alpha$).

In the following,
Theorem~\ref{theorem: upper bound on the Poisson approximation of the entropy}
is applied to get a rigorous error bound on the Poisson approximation of the
entropy $H(W)$.
Table~\ref{table: a random graph problem} presents numerical results for
the approximated value of $H(W)$, and the maximal relative error that is
associated with this approximation. Note that, by symmetry,
the cases with $W(k,n)$ and $W(n-k,n)$ are equivalent, so
$H\bigl(W(k,n)\bigr) = H\bigl(W(n-k,n)\bigr).$

\begin{table*}[here!]
\caption{Numerical results for the Poisson approximations of the entropy $H(W)$
($W = W(k,n)$) by the entropy $H(Z)$ where $Z \sim \text{Po}(\lambda)$, jointly with
the associated error bounds of these approximations. These error bounds are calculated from
Theorem~\ref{theorem: upper bound on the Poisson approximation of the entropy} for
the random graph problem in Example~\ref{example: a random graph problem}.}
\label{table: a random graph problem}
\centering
\renewcommand{\arraystretch}{1.5}
\begin{tabular}{|c|c|c|c|c|} \hline
$n$ & $k$ (or $n-k$)  &  $ \lambda = {{n}\choose{k}} $  & Approximation of $H(W)$ & Maximal relative error \\  \hline
30 & 27 & $4.060 \cdot 10^3$ & 5.573 \text{nats} & 0.1\% \\
30 & 26 & $2.741 \cdot 10^4$ & 6.528 \text{nats} & 0.5\% \\
30 & 25 & $1.425 \cdot 10^5$ & 7.353 \text{nats} & 2.3\% \\ \hline
50 & 48 & $1.225 \cdot 10^3$ & 4.974 \text{nats} & $7.6 \cdot 10^{-10}$\\
50 & 46 & $2.303 \cdot 10^5$ & 7.593 \text{nats} & $9.5 \cdot 10^{-8}$ \\
50 & 44 & $1.589 \cdot 10^7$ & 9.710 \text{nats} & $5.2 \cdot 10^{-6}$ \\
50 & 42 & $5.369 \cdot 10^8$ & 11.470 \text{nats} & $1.5 \cdot 10^{-4}$ \\
50 & 40 & $1.027 \cdot 10^{10}$ & 12.945 \text{nats} & $2.5 \cdot 10^{-3}$ \\ \hline
100 & 95 & $7.529 \cdot 10^7$ & 10.487 \text{nats} & $7.9 \cdot 10^{-20}$ \\
100 & 90 & $1.731 \cdot 10^{13}$ & 16.660 \text{nats} & $1.2 \cdot 10^{-14}$ \\
100 & 85 & $2.533 \cdot 10^{17}$ & 21.456 \text{nats} & $1.3 \cdot 10^{-10}$ \\
100 & 80 & $5.360 \cdot 10^{20}$ & 25.284 \text{nats} & $2.4 \cdot 10^{-7}$ \\
100 & 75 & $2.425 \cdot 10^{23}$ & 28.342 \text{nats} & $9.6 \cdot 10^{-5}$ \\
100 & 70 & $2.937 \cdot 10^{25}$ & 30.740 \text{nats} & $1.1\%$ \\ \hline
\end{tabular}
\end{table*}
\label{example: a random graph problem}
\end{example}

\vspace*{0.1cm}
\begin{example}[maxima of dependent Gaussian random variables]
Consider a finite sequence of possibly dependent Gaussian random variables. The Chen-Stein
method was used in \cite[Section~4.4]{ArratiaGG_Tutorial90} and \cite{Holst_Janson_AOP89} to
derive explicit upper bounds on the total variation distance between the distribution of the
number of times $(W)$ where this sequence exceeds a given level and the Poisson distribution with
the same mean.
The following example relies on the analysis in \cite[Section~4.4]{ArratiaGG_Tutorial90}, and
it aims to provide a rigorous estimate of the entropy of the random variable that counts the
number of times that the sequence of Gaussian random variables exceeds a given level. This
estimation is done as an application of
Theorem~\ref{theorem: upper bound on the Poisson approximation of the entropy}. In order
to sharpen the error bound on the entropy, we derive a tightened upper bound on the coefficient $b_2$ in \eqref{eq: b2} for the studied example; this bound on $b_2$ improves the upper
bound in \cite[Eq.~(21)]{ArratiaGG_Tutorial90}, and it therefore also improves the error
bound on the entropy of $W$. Note that the random variable $W$ can be expressed as a sum
of dependent Bernoulli random variables where each of these binary random variables is
an indicator function that the corresponding Gaussian random variable in the sequence exceeds
the fixed level. The probability that a Gaussian random variable with zero mean and a unit
variance exceeds a certain high level is small, and the law of small numbers indicates that the Poisson approximation for $W$ is good if the required level of crossings is high.

By referring to the setting in  \cite[Section~4.4]{ArratiaGG_Tutorial90}, let $\{Z_i\}$ be
a sequence of independent and standard Gaussian random variables (having a zero mean and a unit
variance). Consider a 1-dependent moving average of Gaussian random variables $\{Y_i\}$ that
are defined, for some $\theta \in \reals$, by
\begin{equation}
Y_i \triangleq \frac{Z_i + \theta Z_{i+1}}{\sqrt{1+\theta^2}}, \quad \forall \, i \geq 1.
\label{eq: 1-dependent moving average sequence of Gaussian random variables}
\end{equation}
This implies that $\expectation(Y_i) = 0$, $\expectation(Y_i^2) = 1$, and the lag-1
auto-correlation is equal to
\begin{equation}
\rho \triangleq \expectation(Y_i \, Y_{i+1}) = \frac{\theta}{1+\theta^2} \, .
\label{eq: the correlation rho between the considered jointly Gaussian RVs}
\end{equation}
Let $t > 0$ be a fixed level, $n \in \naturals$, and $W$ be the number of elements in
the sequence $\{Y_1, \ldots, Y_n\}$ that exceed the level $t$. Then, $W = \sum_{i=1}^n X_i$
is the sum of dependent Bernoulli random variables where
$X_i \triangleq 1_{\{Y_i > t\}}$ for $i \in \{1, \ldots, n\}$ (note that $W$ is a sum of
independent Bernoulli random variables only if $\theta = 0$). The expected value of $W$ is
\begin{equation}
\expectation(W) = n \pr(Y_1 > t) = n \, \bigl(1-\Phi(t)\bigr) \triangleq \lambda_n(t)
\label{eq: the expected value of W for the considered sequence of Gaussian RVs}
\end{equation}
where
\begin{equation}
\Phi(t) \triangleq \frac{1}{\sqrt{2\pi}} \, \int_{-\infty}^t e^{-\frac{x^2}{2}}
\, \mathrm{d}t \, , \quad \forall \, t \in \reals
\label{eq: Gaussian cumulative distribution function}
\end{equation}
is the Gaussian cumulative distribution function. Considering the sequence
of Bernoulli random variables $\{X_\alpha\}_{\alpha \in I}$ where $I = \{1, \ldots, n\}$
then, it follows from \eqref{eq: probabilities of the Bernoulli random variables} that
\begin{equation}
p_{\alpha} = \pr(Y_{\alpha} > t) = 1-\Phi(t), \quad \forall \, \alpha \in I.
\label{eq: probability that the Bernoulli random variable is 1 in the problem on
the Gaussian sequence}
\end{equation}
The neighborhood of dependence of an arbitrary $\alpha \in I$ is
$$B_{\alpha} \triangleq \{\alpha - 1, \, \alpha, \, \alpha+1\} \cap I$$
since $Y_{\alpha}$ only depends in $Y_{\alpha-1}, Y_{\alpha}, Y_{\alpha+1}$.
From \eqref{eq: b1},
\eqref{eq: the expected value of W for the considered sequence of Gaussian RVs} and
\eqref{eq: probability that the Bernoulli random variable is 1 in the problem on
the Gaussian sequence}, and also because
$|B_{\alpha}| \leq 3$ for every $\alpha \in I$, then the following upper bound
on $b_1$ (see \eqref{eq: b1}) holds (see \cite[Eq.~(21)]{ArratiaGG_Tutorial90})
\begin{equation}
b_1 \leq |I| \max_{\alpha \in I} \{|B_{\alpha}| \, p_{\alpha}^2\} =
\frac{3 \lambda_n^2(t)}{n}
\label{eq: upper bound on b1 for the problem of the sequence of Gaussian RVs}
\end{equation}
In the following, a tightened upper bound on $b_2$ (as is defined in \eqref{eq: b2})
is derived, which improves the bound in \cite[Eq.~(21)]{ArratiaGG_Tutorial90}.
Since, by definition $X_{\alpha} = 1_{\{Y_{\alpha} > t\}}$,
$X_{\beta} = 1_{\{Y_{\beta} > t\}}$, and (from \eqref{eq: b2})
$p_{\alpha, \beta} \triangleq \expectation(X_{\alpha} X_{\beta})$, then
\begin{equation}
p_{\alpha, \beta} = \pr\bigl(\min\{Y_{\alpha}, Y_{\beta}\} > t \bigr), \quad
\forall \, \alpha \in I, \; \beta \in B_{\alpha} \setminus \{\alpha\}.
\end{equation}
Note that for every $\alpha \in I$ and $\beta \in B_{\alpha} \setminus \{\alpha\}$,
necessarily $\beta = \alpha \pm 1$ so $Y_{\alpha}$ and $Y_{\beta}$ are jointly standard
Gaussian random variables with the correlation $\rho$
in \eqref{eq: the correlation rho between the considered jointly Gaussian RVs} (it therefore
follows that $\rho \in \bigl[-\frac{1}{2}, \frac{1}{2}\bigr]$, achieving these two extreme
values at $\theta = \pm 1$). From \cite[Eq.~(23) in Lemma~1]{ArratiaGG_Tutorial90},
it follows that
\begin{equation}
p_{\alpha, \beta} < \sqrt{\frac{2(1+\rho)}{\pi(1-\rho)}}
\cdot \bigl[\varphi(u) - u \, (1-\Phi(u))\bigr]
\label{eq: upper bound on the that the minimum of two jointly Gaussian random variables
is above an arbitrary level t}
\end{equation}
where
\begin{eqnarray}
&& u \triangleq t \sqrt{\frac{2}{1+\rho}} \; , \label{eq: u} \\
&& \varphi(u) \triangleq \frac{1}{\sqrt{2\pi}} \; \exp\bigl(-\frac{u^2}{2}\bigr)
= \Phi'(u). \label{eq: Gaussian function}
\end{eqnarray}
Finally, since $|I|=n$ and $|B_{\alpha}| \leq 3$ for every $\alpha \in I$, then
\eqref{eq: b2} and \eqref{eq: upper bound on the that the minimum of two jointly
Gaussian random variables is above an arbitrary level t} lead to the following
upper bound:
\begin{eqnarray}
&& b_2 \leq |I| \, \max_{\alpha \in I, \; \beta \in B_{\alpha} \setminus \{\alpha\}}
\bigl\{ \bigl(|B_{\alpha}|-1 \bigr) p_{\alpha, \beta} \bigr\} \nonumber \\
&& \hspace*{0.4cm} \leq 2n \, \sqrt{\frac{2(1+\rho)}{\pi(1-\rho)}}
\cdot \bigl[\varphi(u) - u \, (1-\Phi(u))\bigr]
\label{eq: upper bound on b2 for the problem of the sequence of Gaussian RVs}
\end{eqnarray}
where $\Phi$, $\rho$, $u$ and $\varphi$ are introduced, respectively, in
Eqs.~\eqref{eq: the correlation rho between the considered jointly Gaussian RVs},
\eqref{eq: Gaussian cumulative distribution function}, \eqref{eq: u}
and \eqref{eq: Gaussian function}. This improves the upper bound on $b_2$
in \cite[Eq.~(21)]{ArratiaGG_Tutorial90} where the reason for this improvement
is related to the weakening of an inequality in the transition from
\cite[Eq.~(23)]{ArratiaGG_Tutorial90} to \cite[Eq.~(24)]{ArratiaGG_Tutorial90}.
As is noted in \cite[Eq.~(21)]{ArratiaGG_Tutorial90}, since $Y_{\alpha}$ is
independent of $(Y_{\beta})_{\beta \notin B_{\alpha}}$, then it follows from
\eqref{eq: b3} that $b_3=0$.

Having upper bounds on $b_1$ and $b_2$ (see
\eqref{eq: upper bound on b1 for the problem of the sequence of Gaussian RVs}
and \eqref{eq: upper bound on b2 for the problem of the sequence of Gaussian RVs})
and the exact value of $b_3$, we are ready to use
Theorem~\ref{theorem: upper bound on the Poisson approximation of the entropy}
to get error bounds for the approximation of $H(W)$
by the entropy of a Poisson random variable $Z$ with the same mean (i.e.,
$Z \sim \text{Po}(\lambda_n(t))$ where $\lambda_n(t)$ is introduced in
\eqref{eq: the expected value of W for the considered sequence of Gaussian RVs}).
Table~\ref{table: the Gaussian problem in this paper} presents numerical results
for the Poisson approximation of the entropy, and the associated error bounds.
It also shows the improvement in the error bound due to the
tightening of the upper bound on $b_2$ in
\eqref{eq: upper bound on b2 for the problem of the sequence of Gaussian RVs}
(as compared to its original bound in \cite[Eq.~(21)]{ArratiaGG_Tutorial90}).

\begin{table*}[here!]
\caption{Numerical results for the Poisson approximations of the entropy $H(W)$ in
Example~\ref{example: the entropy of the number of times that a sequence of Gaussian
random variables exceeds a given level}. It is approximated by
$H(Z)$ where $Z \sim \text{Po}(\lambda_n(t))$ in
\eqref{eq: the expected value of W for the considered sequence of Gaussian RVs},
and the associated error bounds are computed from
Theorem~\ref{theorem: upper bound on the Poisson approximation of the entropy}.
The influence of the tightened bound
in \eqref{eq: upper bound on b2 for the problem of the sequence of Gaussian RVs}
is examined by a comparison with the loosened
upper bound on $b_2$ in \cite[Eq.~(21)]{ArratiaGG_Tutorial90}.}
\label{table: the Gaussian problem in this paper}
\centering
\renewcommand{\arraystretch}{1.5}
\begin{tabular}{|c|c|c|c|c|c|} \hline
$n$ & $\theta$ & $t$ (a fixed
&  $\expectation(W) = \lambda_n(t)$
& Poisson Approximation & Maximal relative error with  \\
& (Eq.~\eqref{eq: 1-dependent moving average sequence of Gaussian random variables})
& level)
& (Eq.~\eqref{eq: the expected value of W for the considered sequence of Gaussian RVs})
& of $H(W)$ & tightened and loosened bounds \\  \hline
$10^4$ & $+1$ & 5 & $2.87 \cdot 10^{-3}$ & 0.020 \text{nats} & $1.9\% \; \; (2.3\%)$ \\
$10^6$ & $+1$ & 5 & $0.287$ & 0.672 \text{nats} & $4.9\% \; \; (6.0\%)$ \\
$10^8$ & $+1$ & 5 & $28.7$ & 3.094 \text{nats} & $4.9\% \; \; (6.0\%)$ \\
$10^{10}$ & $+1$ & 5 & $2.87 \cdot 10^3$ & 5.399 \text{nats} & $3.3\% \; \; (4.1\%)$\\
$10^{12}$ & $+1$ & 5 & $2.87 \cdot 10^5$ & 7.702 \text{nats} & $2.7\% \; \; (3.3\%)$\\ \hline
$10^4$ & $-1$ & 5 & $2.87 \cdot 10^{-3}$ & 0.020 \text{nats}
& $3.8 \cdot 10^{-6}$ \\
$10^6$ & $-1$ & 5 & $0.287$ & 0.672 \text{nats} & $9.6 \cdot 10^{-6}$ \\
$10^8$ & $-1$ & 5 & $28.7$ & 3.094 \text{nats} & $9.3 \cdot 10^{-6}$ \\
$10^{10}$ & $-1$ & 5 & $2.87 \cdot 10^3$ & 5.399 \text{nats}
& $6.1 \cdot 10^{-6}$ \\
$10^{12}$ & $-1$ & 5 & $2.87 \cdot 10^5$ & 7.702 \text{nats} & $4.8 \cdot 10^{-6}$\\ \hline
$10^4$ & $+1$ & 6 & $9.87 \cdot 10^{-6}$ & $1.24 \cdot 10^{-4}$ \text{nats}
& $0.2\% \; \; (0.2\%)$ \\
$10^6$ & $+1$ & 6 & $9.87 \cdot 10^{-4}$ & 0.008 \text{nats} & $0.3\% \; \; (0.4\%)$ \\
$10^8$ & $+1$ & 6 & $9.87 \cdot 10^{-2}$ & 0.327 \text{nats} & $0.7\% \; \; (0.8\%)$ \\
$10^{10}$ & $+1$ & 6 & $9.87$ & 2.555 \text{nats} & $1.0\% \; \; (1.2\%)$ \\
$10^{12}$ & $+1$ & 6 & $9.87 \cdot 10^2$ & 4.866 \text{nats} & $0.6\% \; \; (0.7\%)$\\ \hline
\end{tabular}
\end{table*}
Table~\ref{table: the Gaussian problem in this paper} supports the following observations,
which are first listed and then explained:
\begin{itemize}
\item For fixed values of $n$ and $\theta$, the Poisson approximation is
improved by increasing the level $t$.
\item For fixed values of $n$ and $t$, the error bounds for
the Poisson approximation of the entropy improve when the value of $\theta$ is modified
in a way that decreases the lag-1 auto-correlation $\rho$ in
\eqref{eq: the correlation rho between the considered jointly Gaussian RVs}.
\item For fixed values of $n$ and $t$, the effect of the tightened upper bound
of $b_2$ (see \eqref{eq: upper bound on b2 for the problem of the sequence of Gaussian RVs})
on the error bound of the entropy $H(W)$ is more enhanced when $\rho$ is increased (via a
change in the value of $\theta$).
\item For fixed values of $\theta$ and $t$, the error bounds for the Poisson
approximation are weakly dependent on $n$.
\end{itemize}
The explanation of these observations is, respectively, as follows:
\begin{itemize}
\item For fixed values of $n$ and $\theta$, by increasing the value of the positive level $t$,
the probability that a standard Gaussian random variable $Y_i$ (for $i \in \{1, \ldots, n\}$)
exceeds the value $t$ is decreased. The law of small numbers indicates on the enhancement of
the accuracy of the Poisson approximation for $W$ in this case.
\item For fixed values of $n$ and $t$, the expected value of $W$ (i.e., $\lambda_n(t)$
in \eqref{eq: the expected value of W for the considered sequence of Gaussian RVs}) is kept
fixed, and so is the upper bound on $b_1$ in
\eqref{eq: upper bound on b1 for the problem of the sequence of Gaussian RVs}. However,
if the correlation $\rho$ in \eqref{eq: the correlation rho between the considered jointly Gaussian RVs}
is decreased (by a proper change in the value of $\theta$) then the
value of $u$ in \eqref{eq: u} is increased, and the upper bound on $b_2$
(see \eqref{eq: upper bound on b2 for the problem of the sequence of Gaussian RVs}) is
decreased. Since the upper bounds on $b_1$ and $b_3$ are not affected by a change in the
value of $\theta$ and the upper bound on $b_2$ is decreased, then the upper bound
on the total variation distance in Theorem~\ref{theorem: upper bound on the total variation
distance by Arratia et al.} is decreased as well. This also decreases the error
bound that refers to the Poisson approximation of the entropy in
Theorem~\ref{theorem: upper bound on the Poisson approximation of the entropy}. Note that
Table~\ref{table: the Gaussian problem in this paper} compares the situation for
$\theta = \pm 1$, which corresponds respectively to $\rho = \pm \frac{1}{2}$ (these
are the two extreme values of $\rho$).
\item When $n$ and $t$ are fixed, the balance between the upper bounds on $b_1$ and $b_2$
changes significantly while changing the value of $\theta$. To exemplify this numerically, let
$n = 10^8$ and $t=5$ be the length of the sequence of Gaussian random variables and the
considered level, respectively. If $\theta=1$, the upper bounds on $b_1$ and $b_2$ in
\eqref{eq: upper bound on b1 for the problem of the sequence of Gaussian RVs} and
\eqref{eq: upper bound on b2 for the problem of the sequence of Gaussian RVs} are,
respectively, equal to $2.47 \cdot 10^{-5}$  and $0.176$ (the loosened bound on $b_2$
is equal to $0.218$). In this case, $b_2$ dominates $b_1$ and therefore an improvement
in the value of $b_2$ (or its upper bound) also improves the error bound for the Poisson
approximation of the entropy $H(W)$ in
Theorem~\ref{theorem: upper bound on the Poisson approximation of the entropy}.
Consider now the case where $\theta = -1$ (while $n, t$
are kept fixed); this changes the lag-1 autocorrelation $\rho$ in
\eqref{eq: the correlation rho between the considered jointly Gaussian RVs} from its
maximal value $(+\frac{1}{2})$ to its minimal value $(-\frac{1}{2})$. In this case,
the upper bound on $b_1$ does not change, but the new bound on $b_2$ is decreased from
$0.218$ to $6.88 \cdot 10^{-17}$ (and the loosened bound on $b_2$, for $\theta=-1$,
is equal to $9.90 \cdot 10^{-16}$). In the latter case, the situation w.r.t.
the balance between the coefficients $b_1$ and $b_2$ is reversed, i.e.,
the bound on $b_1$ dominates the bound on $b_2$. Hence, the upper bound on the total
variation distance and the error bound that follows from the Poisson approximation of
the entropy $H(W)$ are reduced considerably when $\theta$ changes from $+1$ to $-1$. This
is because, from
Theorem~\ref{theorem: upper bound on the total variation distance by Arratia et al.},
the upper bound on the total variation distance depends linearly on the sum $b_1+b_2$
when $b_3=0$). A similar conclusion also holds w.r.t. the error bound on the entropy
(see Theorem~\ref{theorem: upper bound on the Poisson approximation of the entropy}).
In light of this comparison, the tightened bound on $b_2$ affects the error bound for
the Poisson approximation of $H(W)$ when $\theta=1$, in contrast to the
case when $\theta = -1$.
\item The numerical results in Table~\ref{table: the Gaussian problem in this paper}
show that the accuracy of the Poisson approximation is weakly dependent on the length
$n$ of the sequence $\{Y_i\}_{i=1}^n$. This is attributed to the fact that the
probabilities $p_i$, for $i \in \{1, \ldots, n\}$, are not affected by $n$ but they
are only affected by choice of the level $t$. Hence,
the law of small numbers does not necessarily indicate on an enhanced accuracy of the
Poisson approximation for $H(W)$ when the length of the sequence $n$ is increased.
\end{itemize}
\label{example: the entropy of the number of times that a sequence of Gaussian random variables
exceeds a given level}
\end{example}

\subsection{Proofs of the New Bounds in Section~\ref{subsection: New error bounds on the entropy}}
\label{subsection: Proofs of the new bounds in the first part of this paper}

\subsubsection{Proof of Theorem~\ref{theorem: upper bound on the Poisson approximation of the entropy}}
\label{subsubsection: Proof of the Theorem with the upper bound on the Poisson approximation
of the entropy}
The random variable $W = \sum_{\alpha \in I} X_{\alpha}$
is a sum of Bernoulli random variables where $|I|=n < \infty$,
then $W$ gets values from the set $\{0, 1, \ldots, n\}$,
and $Z$ gets non-negative integer values. Theorem~\ref{theorem: L1 bound on the entropy}
therefore implies that
\begin{equation}
\bigl|H(W)-H(Z)\bigr| \leq \eta \log(M-1) + h(\eta) + \mu
\end{equation}
and we need in the following to calculate proper constants
$\eta, \mu$ and $M$ for the Poisson approximation.
The cardinality of the set of possible values of $W$ is $m = n+1$, so
it follows from \eqref{eq: definition of the parameter M} that $M$ is
given by \eqref{eq: function M in the upper bound on the Poisson approximation of the entropy}.
The parameter $\eta$, which serves as an upper bound on the total variation distance
$d_{\text{TV}}(W,Z)$, is given in
\eqref{eq: function eta in the upper bound on the Poisson approximation of the entropy}
due to the result in
Theorem~\ref{theorem: upper bound on the total variation distance by Arratia et al.}.
The last thing that is now required is the calculation of $\mu$.
Let $$\Pi_{\lambda}(k) \triangleq \frac{e^{-\lambda} \, \lambda^k}{k!},
\quad \forall \, k \in \{0, 1, \ldots\}$$
designate the probability distribution of $Z \sim \text{Po}(\lambda)$,
so $\mu$ is an upper bound on
$\sum_{k=M}^{\infty} \bigl\{-\Pi_{\lambda}(k) \, \log \Pi_{\lambda}(k) \bigr\}$,
which is an infinite sum that only depends on the Poisson distribution.
Straightforward calculation gives that
\begin{eqnarray}
&& \sum_{k=M}^{\infty} \bigl\{-\Pi_{\lambda}(k) \, \log \Pi_{\lambda}(k) \bigr\} \nonumber \\
&& = -\lambda \, \log \lambda \sum_{k=M-1}^{\infty} \Pi_{\lambda}(k)
+ \lambda \sum_{k=M}^{\infty} \Pi_{\lambda}(k)
+ \sum_{k=M}^{\infty} \Pi_{\lambda}(k) \, \log(k!) \, .
\label{eq: bound on the second sum for the derivation of the error bound on the entropy - 1st step}
\end{eqnarray}
From Stirling's formula, for every $k \in \naturals$, the
equality $k ! = \sqrt{2\pi k} \left(\frac{k}{e}\right)^k \, e^{\eta_k}$
holds for some
$\eta_k \in \bigl(\frac{1}{12k+1}, \frac{1}{12k}\bigr)$.
This therefore implies that the third infinite sum on the right-hand side of
\eqref{eq: bound on the second sum for the derivation of the error bound on the entropy - 1st step}
satisfies
\begin{eqnarray}
&& \sum_{k=M}^{\infty} \Pi_{\lambda}(k) \, \log(k!) \nonumber \\
&& \leq \sum_{k=M}^{\infty} \Pi_{\lambda}(k) \, \log\left(\sqrt{2\pi k}
\left(\frac{k}{e}\right)^k \, e^{\frac{1}{12k}} \right) \nonumber \\
&& = \frac{\log(2\pi)}{2} \sum_{k=M}^{\infty} \Pi_{\lambda}(k) +
\sum_{k=M}^{\infty} \Pi_{\lambda}(k) \left[\bigl(k+\frac{1}{2}\bigr) \log(k)-k \right]
+ \frac{1}{12} \sum_{k=M}^{\infty} \frac{\Pi_{\lambda}(k)}{k} \nonumber \\
&& \leq \frac{\log(2\pi)}{2} \sum_{k=M}^{\infty} \Pi_{\lambda}(k) +
\sum_{k=M}^{\infty} \bigl\{k(k-1) \, \Pi_{\lambda}(k) \bigr\}
+ \frac{1}{12} \sum_{k=M}^{\infty} \Pi_{\lambda}(k) \nonumber \\
&& \stackrel{\text{(a)}}{=} \frac{\log(2\pi)}{2} \sum_{k=M}^{\infty} \Pi_{\lambda}(k) +
\lambda^2 \sum_{k=M-2}^{\infty} \Pi_{\lambda}(k)
+ \frac{1}{12} \sum_{k=M}^{\infty} \Pi_{\lambda}(k) \nonumber \\
&& \leq \left( \frac{6\log(2\pi)+1}{12} + \lambda^2 \right) \sum_{k=M-2}^{\infty} \Pi_{\lambda}(k)
\label{eq: bound on the second sum for the derivation of the error bound on the entropy - 2nd step}
\end{eqnarray}
where the equality in~(a) follows from the identity
$k(k-1) \, \Pi_{\lambda}(k) = \lambda^2 \, \Pi_{\lambda}(k-2)$ for every $k \geq 2$.
By combining \eqref{eq: bound on the second sum for the derivation of the error bound
on the entropy - 1st step} and
\eqref{eq: bound on the second sum for the derivation of the error bound on the
entropy - 2nd step}, it follows that
\begin{eqnarray}
&& \sum_{k=M}^{\infty} -\Pi_{\lambda}(k) \, \log \Pi_{\lambda}(k)
\nonumber \\
&& \leq \left(\lambda \, \log\Bigl(\frac{e}{\lambda}\Bigr)\right)_{+} \,
\sum_{k=M-1}^{\infty} \Pi_{\lambda}(k) +
\left( \frac{6\log(2\pi)+1}{12} + \lambda^2 \right) \sum_{k=M-2}^{\infty} \Pi_{\lambda}(k)
\nonumber \\
&& \leq \left[\left(\lambda \, \log\Bigl(\frac{e}{\lambda}\Bigr)\right)_{+} +
\lambda^2 + \frac{6\log(2\pi)+1}{12} \right] \sum_{k=M-2}^{\infty} \Pi_{\lambda}(k).
\label{eq: bound on the second sum for the derivation of the error bound on the entropy - 3rd step}
\end{eqnarray}
Based on Chernoff's bound, since $Z \sim \text{Po}(\lambda)$,
\begin{eqnarray}
&& \sum_{k=M-2}^{\infty} \Pi_{\lambda}(k) \nonumber \\[0.1cm]
&& = \pr(Z \geq M-2) \nonumber \\[0.1cm]
&& \leq \inf_{\theta \geq 0} \left\{e^{-\theta (M-2)} \,
\expectation\bigl[e^{\theta Z}\bigr] \right\} \nonumber \\
&& = \inf_{\theta \geq 0} \left\{e^{-\theta (M-2)} \,
e^{\lambda (e^{\theta}-1)} \right\} \nonumber \\
&& = \exp\left\{ -\left[ \lambda + (M-2) \log \Bigl(\frac{M-2}{\lambda e}\Bigr)
\right] \right\}
\label{eq: bound on the second sum for the derivation of the error bound on
the entropy - 4th step}
\end{eqnarray}
where the last equality follows by substituting the optimized value
$\theta = \log\bigl(\frac{M-2}{\lambda}\bigr)$ in the exponent
(note that $\lambda \leq n = m-1 \leq M-2$, so optimized value of
$\theta$ is indeed non-negative). Hence, by combining
\eqref{eq: bound on the second sum for the derivation of the error
bound on the entropy - 3rd step} and \eqref{eq: bound on the second
sum for the derivation of the error bound on the entropy - 4th step},
it follows that
\begin{equation}
\sum_{k=M}^{\infty} \bigl\{ -\Pi_{\lambda}(k) \, \log \Pi_{\lambda}(k) \bigr\}
\leq \mu
\label{eq: bound on the second sum for the derivation of the error bound on
the entropy - 5th step}
\end{equation}
where the parameter $\mu$ is introduced in
\eqref{eq: function mu in the upper bound on the Poisson approximation of the entropy}.
This completes the proof of
Theorem~\ref{theorem: upper bound on the Poisson approximation of the entropy}.

\vspace*{0.1cm}
\subsubsection{Proof of Corollary~\ref{corollary: upper bound on the Poisson
approximation of the entropy for independent RVs}}
\label{subsubsection: Proof of the corollary with the upper bound on the
entropy - sum of independent Bernoulli RVs}
For proving the right-hand side of
\eqref{eq: an error bound on the Poisson approximation of the entropy
for independent Bernoulli RVs}, which holds under the assumption that
the Bernoulli random variables $\{X_{\alpha}\}_{\alpha \in I}$
are independent, one chooses (similarly to
Remark~\ref{remark: Generalization of Theorem 1 of Barbour and Hall})
the set $B_{\alpha} \triangleq \{\alpha\}$ as
the neighborhood of dependence for every $\alpha \in I$. Note that
this choice of $B_{\alpha}$ is taken because
$\sigma\bigl(X_{\beta})_{\beta \in I \setminus \{\alpha\}}\bigr)$ is
independent of $X_{\alpha}$. From \eqref{eq: b1}--\eqref{eq: b3}, this
choice gives that $b_1 = \sum_{\alpha \in I} p_{\alpha}^2$ and $b_2=b_3=0$ which
therefore implies the right-hand side of \eqref{eq: an error bound on
the Poisson approximation of the entropy for independent Bernoulli RVs}
as a special case of
Theorem~\ref{theorem: upper bound on the Poisson approximation of the entropy}.
Furthermore, due to the maximum entropy result of the Poisson distribution (see
Theorem~\ref{theorem: maximum entropy result for the Poisson distribution}),
then $H(Z) - H(W) \geq 0$. This completes the proof of
Corollary~\ref{corollary: upper bound on the Poisson approximation of the entropy
for independent RVs}.

\vspace*{0.1cm}
\subsubsection{Proof of Proposition~\ref{proposition: a possibly improved error
bound on the Poisson approximation of the entropy for independent Bernoulli RVs}}
\label{subsubsection: Proof of the proposition with the upper bound on the
entropy - sum of independent Bernoulli RVs}
Under the assumption that the Bernoulli random variables $\{X_{\alpha}\}_{\alpha \in I}$
are independent, we rely here on two possible upper bounds on the total variation
distance between the distributions of $W$ and $Z \sim \text{Po}(\lambda)$. The first
bound is the one in \cite[Theorem~1]{BarbourH_1984}, used earlier in
Corollary~\ref{corollary: upper bound on the Poisson approximation of the entropy
for independent RVs}. This bound gets the form
\begin{equation}
d_{\text{TV}}(P_W, \text{Po}(\lambda)) \leq \left( \frac{1-e^{-\lambda}}{\lambda} \right)
\sum_{\alpha \in I} p_{\alpha}^2 = \bigl(1-e^{-\lambda}\bigr) \theta
\label{eq: first upper bound on the total variation distance for proving Proposition 1}
\end{equation}
where $\theta$ is introduced in \eqref{eq: theta}.
The second bound appears in \cite[Eq.~(30)]{CekanaviciusR_2006}, and it improves the
bound in \cite[Eq.~(10)]{Roos_2001} (see also \cite[Eq.~(4)]{Roos_2003}). This bound
gets the form
\begin{equation}
d_{\text{TV}}(P_W, \text{Po}(\lambda)) \leq \frac{3 \theta}{4e
\bigl(1-\sqrt{\theta}\bigr)^{3/2}} \, .
\label{eq: second upper bound on the total variation distance for proving Proposition 1}
\end{equation}
It therefore follows that
\begin{equation}
d_{\text{TV}}(P_W, \text{Po}(\lambda)) \leq \eta
\label{eq: combined bound on the total variation distance for proving Proposition 1}
\end{equation}
where $\eta$ is defined in \eqref{eq: improved eta} to be the minimum of the upper bounds
on the total variation distance
in \eqref{eq: first upper bound on the total variation distance for proving Proposition 1}
and \eqref{eq: second upper bound on the total variation distance for proving Proposition 1}.
The continuation of the proof of this proposition is similar to the proof of
Corollary~\ref{corollary: upper bound on the Poisson approximation of the entropy for
independent RVs}.

\subsection{Generalization: Bounds on the Entropy for a Sum of Non-Negative, Integer-Valued and Bounded Random Variables}
\label{subsection: generalization of the bounds on the entropy for the sum of integer-valued random variables}
We introduce in the following a generalization of the bounds in
Section~\ref{subsection: New error bounds on the entropy} to consider the
accuracy of the Poisson approximation for the entropy of a sum of non-negative,
integer-valued and bounded random variables. The generalized version of
Theorem~\ref{theorem: upper bound on the Poisson approximation of the entropy}
is first introduced, and it is then justified by relying on
the proof of this theorem for sums of Bernoulli random variables with the
approach of Serfling in \cite[Section~7]{Serfling_1978}. This approach enables
to derive an explicit upper bound on the total variation distance between a sum
of non-negative and integer-valued random variables and a Poisson distribution
with the same mean. The requirement that the summands are bounded random variables
is used to obtain an upper bound on the accuracy of the Poisson approximation for
the entropy of a sum of non-negative, integer-valued and bounded random variables.
The following proposition forms a generalized version of
Theorem~\ref{theorem: upper bound on the Poisson approximation of the entropy}.

\begin{proposition}
Let $I$ be an arbitrary finite index set, and let $|I| \triangleq n$. Let
$\{X_{\alpha}\}_{\alpha \in I}$ be non-negative, integer-valued random variables,
and assume that there exists some $A \in \naturals$ such that
$X_\alpha \in \{0, 1, \ldots, A\}$ a.s. for every $\alpha \in I$. Let
\begin{eqnarray}
W \triangleq \sum_{\alpha \in I} X_{\alpha} \, ,
\quad p_{\alpha} \triangleq \pr(X_\alpha = 1) \, ,
\quad q_{\alpha} \triangleq \pr(X_{\alpha} \geq 2) \, ,
\quad \lambda \triangleq \sum_{\alpha \in I} p_{\alpha} \, ,
\quad q \triangleq \sum_{\alpha \in I} q_{\alpha}
\label{eq: notation1 for the generalized version of the entropy bound}
\end{eqnarray}
where $\lambda > 0$ and $q \geq 0$. Furthermore, for every $\alpha \in I$,
let $X'_{\alpha}$ be a Bernoulli random variable that is equal to~1
if $X_{\alpha}=1$, and let it be equal otherwise to zero.
Referring to these Bernoulli random variables, let
\begin{eqnarray}
&& b'_1 \triangleq \sum_{\alpha \in I} \sum_{\beta \in B_{\alpha}} p_{\alpha} p_{\beta}
\label{eq: b1_prime} \\[0.1cm]
&& b'_2 \triangleq \sum_{\alpha \in I} \sum_{\alpha \neq \beta \in B_{\alpha}} p'_{\alpha, \beta},
\quad p'_{\alpha, \beta} \triangleq \expectation(X'_{\alpha} X'_{\beta})
\label{eq: b2_prime} \\[0.1cm]
&& b'_3 \triangleq \sum_{\alpha \in I} s'_{\alpha}, \quad \quad
s'_{\alpha} \triangleq \expectation \bigl| \expectation(X'_{\alpha} - p_{\alpha} \, | \,
\sigma(\{X_{\beta}\})_{\beta \in I \setminus B_{\alpha}}) \bigr|
\label{eq: b3_prime}
\end{eqnarray}
where, for every $\alpha \in I$, the subset $B_{\alpha} \subseteq I$ is determined
arbitrarily such that it includes the element $\alpha$. Furthermore, let
\begin{eqnarray}
&& \eta_A \triangleq 2 (b'_1 + b'_2) \left(\frac{1-e^{-\lambda}}{\lambda}\right)
+ b'_3 \Bigl(1 \wedge \frac{1.4}{\sqrt{\lambda}}\Bigr) + q
\label{eq: function eta_prime in the upper bound on the Poisson approximation of the entropy}
\\[0.3cm]
&& M_A \triangleq \max\left\{nA+2, \frac{1}{1-\eta_A} \right\} \\[0.2cm]
&& \mu_A \triangleq \left[ \Bigl(\lambda \log \bigl(\frac{e}{\lambda}\bigr)\Bigr)_+ \,
+ \lambda^2 + \frac{6 \log(2\pi) + 1}{12} \right]
\, \exp \left\{-\left[\lambda + (M_A-2) \log\left(\frac{M_A-2}{\lambda e} \right)
\right] \right\}
\label{eq: function mu_prime in the upper bound on the Poisson approximation of the entropy}
\end{eqnarray}
provided that $\eta_A < 1$. Then, the difference between the
entropies (to base~$e$) of $W$ and $Z \sim \text{Po}(\lambda)$ satisfies
\begin{equation}
|H(Z) - H(W)| \leq \eta_A \, \log(M_A-1) + h(\eta_A) + \mu_A.
\label{eq: generalized upper bound on the Poisson approximation of the entropy}
\end{equation}
\label{proposition: generalized upper bound on the Poisson approximation of the entropy}
\end{proposition}

\vspace*{0.1cm}
\begin{proof}
Following the approach in \cite[Section~7]{Serfling_1978}, let
$X'_{\alpha} \triangleq 1_{\{X_\alpha = 1\}}$ be a Bernoulli random variable
that is equal to the indicator function of the event $X_{\alpha} = 1$ and
$\pr(X'_{\alpha} = 1) = p_{\alpha}$ for every $\alpha \in I$. Let
$W' \triangleq \sum_{\alpha \in I} X'_{\alpha}$ be the sum of the induced
Bernoulli random variables. From the Chen-Stein method (see
Theorem~\ref{theorem: upper bound on the total variation distance by Arratia et al.})
\begin{equation}
d_{\text{TV}}\bigl(P_{W'}, \text{Po}(\lambda)\bigr) \leq (b'_1 + b'_2) \left(\frac{1-e^{-\lambda}}{\lambda}\right) +
b'_3  \Bigl(1 \wedge \frac{1.4}{\sqrt{\lambda}}\Bigr)
\label{eq: upper bound on the total variation distance between W prime and Z}
\end{equation}
with the constants $b'_1, b'_2$ and $b'_3$ as defined in
\eqref{eq: b1_prime}--\eqref{eq: b3_prime}. Furthermore,
from \cite[Eq.~(7.2)]{Serfling_1978}, it follows that
\begin{eqnarray}
&& d_{\text{TV}}\bigl(P_W, \, P_{W'}\bigr) \nonumber \\[0.1cm]
&& \leq \pr(W' \neq W) \nonumber \\[0.1cm]
&& \leq \sum_{\alpha \in I} \pr\bigl(X'_{\alpha} \neq X_{\alpha}\bigr) \nonumber \\[0.1cm]
&& = \sum_{\alpha \in I} \pr\bigl(X_{\alpha} \geq 2) \nonumber \\
&& = \sum_{\alpha \in I} q_{\alpha} = q.
\label{eq: total variation between W and W prime is less than or equal to q}
\end{eqnarray}
It therefore follows from
\eqref{eq: function eta_prime in the upper bound on the Poisson approximation of the entropy},
\eqref{eq: upper bound on the total variation distance between W prime and Z} and
\eqref{eq: total variation between W and W prime is less than or equal to q} that
$$d_{\text{TV}}\bigl(P_W, \text{Po}(\lambda)\bigr) \leq
d_{\text{TV}}\bigl(P_W, \, P_{W'}\bigr) + d_{\text{TV}}\bigl(P_{W'}, \text{Po}(\lambda)\bigr)
\leq \eta_A.$$
The rest of this proof follows closely the proof of
Theorem~\ref{theorem: upper bound on the Poisson approximation of the entropy}
(note that $P_W(k)=0$ for $k > nA$, so $W$ gets $m \triangleq nA+1$ possible
values). This completes the proof of
Proposition~\ref{proposition: generalized upper bound on the Poisson approximation of
the entropy}.
\end{proof}

\section{Improved Lower Bounds on the Total Variation Distance,
Relative Entropy and Some Related Quantities for Sums of Independent
Bernoulli Random Variables}
\label{section: improved lower bounds on the total variation distance etc.}

This section forms the second part of this work. As in the
previous section, the presentation starts in Section~\ref{subsection: Second
part of the revision of some known results} with a brief review
of some reported results that are relevant to the analysis in
this section. Improved lower bounds on the total variation
distance between the distribution of the sum of independent
Bernoulli random variables and the Poisson distribution with
the same mean are introduced in
Section~\ref{subsection: Improved lower bounds on the total
variation distance}. These improvements are obtained via the
Chen-Stein method, by a non-trivial refinement of the analysis
that was used for the derivation of the original lower bound by Barbour
and Hall (see \cite[Theorem~2]{BarbourH_1984}). Furthermore, the
improved tightness of the new lower bounds
and their connection to the original lower bound are further considered.
Section~\ref{subsection: Improved lower bounds on the relative entropy}
introduces an improved lower bound on the relative entropy between
the above two distributions. The analysis that is used for the derivation
of the lower bound on the relative entropy is based on the lower bounds on
the total variation distance in
Section~\ref{subsection: Improved lower bounds on the total variation distance},
combined with the use of the distribution-dependent
refinement of Pinsker's inequality by Ordentlich and Weinberger
\cite{OrdentlichW_IT2005} (where the latter is specialized to the Poisson
distribution). The lower bound on the relative entropy
is compared to some previously reported upper bounds on the relative entropy
by Kontoyiannis et al. \cite{KontoyiannisHJ_2005} in the context of the
Poisson approximation.
Upper and lower bounds on the Bhattacharyya parameter, Chernoff information
and Hellinger distance between the distribution of the sum of independent Bernoulli
random variables and the Poisson distribution are next derived
in Section~\ref{subsection: Bounds on related quantities}.
The discussion proceeds in
Section~\ref{subsection: Second part of applications of the new bounds}
by exemplifying the use of some of the new bounds that are derived in this section
in the context of the classical binary hypothesis testing.
Finally, Section~\ref{subsection: Proofs of the results in the second part of this paper}
proves the new results that are introduced in
Sections~\ref{subsection: Improved lower bounds on the relative entropy}
and \ref{subsection: Bounds on related quantities}.
It is emphasized that, in contrast to the setting in
Section~\ref{section: Error bounds on the entropy of the sum of Bernoulli random variables}
where the Bernoulli random variables may be dependent summands, the analysis
in this section depends on the assumption that the Bernoulli random
variables are independent. This difference stems from
the derivation of the improved lower bound on the total variation distance in
Section~\ref{subsection: Improved lower bounds on the total variation distance}, which
forms the starting point for the derivation of all the subsequent results that are introduced
in this section, assuming an independence of the summands.

\subsection{Review of Some Essential Results for the Analysis in
Section~\ref{section: improved lower bounds on the total variation distance etc.}}
\label{subsection: Second part of the revision of some known results}

The following definitions of probability metrics are particularized and simplified
to the case of our interest where the probability mass functions are defined on
$\naturals_0$.
\begin{definition}
Let  $P$ and $Q$ be two probability mass functions that are defined on a same countable
set $\mathcal{X}$. The Hellinger distance and
the Bhattacharyya parameter between $P$ and $Q$ are, respectively, given by
\begin{eqnarray}
&& d_{\text{H}}(P, Q) \triangleq \left( \frac{1}{2} \, \sum_{x \in \mathcal{X}}
\Bigl(\sqrt{P(x)} - \sqrt{Q(x)}\Bigr)^2 \, \right)^{\frac{1}{2}}
\label{eq: Hellinger distance} \\[0.1cm]
&& \text{BC}(P, Q) \triangleq \sum_{x \in \mathcal{X}} \sqrt{P(x) \, Q(x)}
\label{eq: Bhattacharyya parameter}
\end{eqnarray}
so, these two probability metrics (including the total variation distance in
Definition~\ref{definition: total variation distance}) are bounded
between~0 and~1.
\label{definition: probability metrics}
\end{definition}

\begin{remark}
In general, these probability metrics are defined in the setting
where $(\mathcal{X}, d)$ is a separable metric space. The interest in this work
is in the specific case where $\mathcal{X} = \naturals_0$ and $d = | \cdot |$.
In this case, the expressions of these probability metrics are simplified as above.
For further study of probability metrics and their properties, the
interested reader is referred to, e.g., \cite[Appendix~A.1]{BarbourHJ_book_1992},
\cite[Chapter~2]{DasGupta_2008} and \cite[Section~3.3]{Reiss_book1989}.
\end{remark}

\begin{remark}
The Hellinger distance is related to the Bhattacharyya parameter
via the equality
\begin{equation}
d_{\text{H}}(P, Q) = \sqrt{1-\text{BC}(P, Q)}.
\label{eq: relation between the Hellinger distance and Bhattacharyya parameter}
\end{equation}
\label{remark: relation between the Hellinger distance and Bhattacharyya parameter}
\end{remark}

\vspace*{-0.5cm}
\begin{definition}
The Chernoff information and relative entropy (a.k.a. divergence or Kullback-Leibler
distance) between two probability mass functions $P$ and $Q$ that are defined on a
countable set $\mathcal{X}$ are, respectively, given by
\vspace*{-0.1cm}
\begin{eqnarray}
&& C(P,Q) \triangleq -\min_{\theta \in [0,1]} \log \left( \sum_{x \in \mathcal{X}}
P^{\theta}(x) Q^{1-\theta}(x) \right) \label{eq: Chernoff information} \\[0.1cm]
&& D(P||Q) \triangleq \sum_{x \in \mathcal{X}} P(x)
\log\left(\frac{P(x)}{Q(x)}\right) \label{eq: relative entropy}
\end{eqnarray}
so $C(P, Q), D(P||Q) \in [0, \infty]$. Throughout this paper, the logarithms are on
base~$e$.
\label{definition: Chernoff information and relative entropy}
\end{definition}

\begin{proposition}
For two probability mass functions $P$ and $Q$
that are defined on the same set $\mathcal{X}$
\begin{equation}
d_{\text{TV}}(P, Q) \leq \sqrt{2} \, d_{\text{H}}(P, Q) \leq \sqrt{D(P||Q)}.
\label{eq: known inequality that relates between the total variation,
Hellinger distance and relative entropy}
\end{equation}
\label{proposition: known inequality that relates between the total
variation, Hellinger distance and relative entropy}
\end{proposition}
The left-hand side of \eqref{eq: known inequality that relates between
the total variation, Hellinger distance and relative entropy}
is proved in \cite[p.~99]{Reiss_book1989}, and the right-hand side is
proved in \cite[p.~328]{Reiss_book1989}.
\begin{remark}
It is noted that the Hellinger
distance in the middle of \eqref{eq: known inequality that relates
between the total variation, Hellinger distance and relative entropy}
is not multiplied by the square-root of~2 in \cite{Reiss_book1989}, due
to a small difference in the definition of this distance where the factor of
one-half on the right-hand side of \eqref{eq: Hellinger distance} does
not appear in the definition of the Hellinger distance in
\cite[p.~98]{Reiss_book1989}. However, this is just a matter of
normalization of this distance (as otherwise,
according to \cite{Reiss_book1989}, the Hellinger distance varies
between 0 and $\sqrt{2}$ instead of the interval $[0,1]$). The definition
of this distance in \eqref{eq: Hellinger distance} is consistent, e.g.,
with \cite{BarbourHJ_book_1992}.
It makes the range of this distance to be between 0 and~1, similarly to the total
variation, local and Kolmogorov-Smirnov distances and also the Bhattacharyya parameter
that are considered in this paper.
\end{remark}

The Chernoff information, $C(P,Q)$, is the best achievable exponent
in the Bayesian probability of error for binary hypothesis testing
(see, e.g., \cite[Theorem~11.9.1]{Cover_Thomas}). Furthermore, if
$X_1, X_2, \ldots, X_N$ are i.i.d. random variables, having
distribution $P$ with prior probability $\pi_1$ and distribution
$Q$ with prior probability $\pi_2$, the following upper bound holds
for the best achievable overall probability of error:
\begin{equation}
P_{\text{e}}^{(N)} \leq \exp\bigl(-N \, C(P,Q) \bigr).
\label{eq: Chernoff bound for the best achievable overall probability
of error in binary hypothesis testing}
\end{equation}

\subsubsection*{A distribution-dependent refinement of Pinsker's inequality
\cite{OrdentlichW_IT2005}}:
Pinsker's inequality provides a lower bound on the relative entropy
in terms of the total variation distance between two probability measures
that are defined on the same set. It states that
\begin{equation}
D(P||Q) \geq 2 \Bigl(d_{\text{TV}}(P,Q)\Bigr)^2.
\label{eq: Pinsker's inequality}
\end{equation}
In \cite{OrdentlichW_IT2005}, a distribution-dependent refinement of Pinsker's inequality
was introduced for an arbitrary pair of probability distributions $P$ and $Q$ that are
defined on $\naturals_0$. It is of the form
\begin{equation}
D(P || Q) \geq \varphi(\pi_Q) \; \Bigl(d_{\text{TV}}(P, Q)\Bigr)^2
\label{eq: a distribution-dependent refinement of Pinsker's inequality}
\end{equation}
where
\begin{equation}
\pi_Q \triangleq \sup_{A \subseteq \naturals_0} \min \bigl\{Q(A), 1-Q(A)\bigr\}
\label{eq: pi_Q for the refinement of Pinsker's inequality}
\end{equation}
and
\begin{equation}
\varphi(p) \triangleq \left\{ \begin{array}{cl}
\frac{1}{1-2p} \; \log\left(\frac{1-p}{p}\right) & \quad \mbox{if $0<p<\frac{1}{2}$} \\
2   & \quad \mbox{if $p = \frac{1}{2}$}
\end{array}
\right.
\label{eq: phi function for the refinement of Pinsker's inequality}
\end{equation}
so $\varphi$ is monotonic decreasing in the interval $(0, \frac{1}{2}]$,
$$\lim_{p \rightarrow 0^+} \varphi(p) = +\infty, \quad \lim_{p \rightarrow \frac{1}{2}^-} \varphi(p) = 2$$
where the latter limit implies that $\varphi$ is left-continuous at one-half.
Note that it follows from \eqref{eq: pi_Q for the refinement of Pinsker's inequality} that
$\pi_Q \in [0, \frac{1}{2}]$.

In Section~\ref{subsection: Improved lower bounds on the relative entropy},
we rely on this refinement of Pinsker's inequality and combine
it with the new lower bound on the total variation distance between
the distribution of a sum of independent Bernoulli random variables
and the Poisson distribution with the same mean that is introduced in
Section~\ref{subsection: Improved lower bounds on the total variation distance}.
The combination of these two bounds provides a new lower bound on the
relative entropy between these two distributions.

\subsection{Improved Lower Bounds on the Total Variation Distance}
\label{subsection: Improved lower bounds on the total variation distance}
In Theorem~\ref{theorem: bounds on the total variation distance - Barbour and Hall 1984},
we introduced the upper and lower bounds on the total variation distance in
\cite[Theorem~1 and~2]{BarbourC_book_2005} (see
also \cite[Theorem~2.M and Corollary~3.D.1]{BarbourHJ_book_1992}). This
shows that these upper and lower bounds are essentially tight, where the
lower bound is about $\frac{1}{32}$ of the upper bound. Furthermore, it was claimed
in \cite[Remark~3.2.2]{BarbourHJ_book_1992} (with no explicit proof) that the constant
$\frac{1}{32}$ in the lower bound on the left-hand side of
\eqref{eq: bounds on the total variation distance - Barbour and Hall 1984} can be
improved to $\frac{1}{14}$. In this section, we obtain further improvements of this
lower bound where, e.g., the ratio of the upper and new lower bounds on the total variation
distance tends to 1.69 in the limit where $\lambda \rightarrow 0$, and this ratio tends to
10.54 in the limit where $\lambda \rightarrow \infty$. As will be demonstrated in the
continuation of Section~\ref{section: improved lower bounds on the total variation distance etc.},
the effect of these improvements is enhanced considerably
when considering improved lower bounds on the relative entropy and some other related
information-theoretic measures. We further study later in this section the implications
of the improvement in lower bounding the total variation distance, originating in this
sub-section, and exemplify these improvements in the context of information theory and
statistics.

Similarly to the proof of \cite[Theorem~2]{BarbourH_1984},
the derivation of the improved lower bound is also based on the Chen-Stein method,
but it follows from a significant modification of the analysis that served to derive
the original lower bound in \cite[Theorem~2]{BarbourH_1984}.
The following upper bound on the total variation distance is taken (as is) from
\cite[Theorem~1]{BarbourH_1984} (this bound also appears in
Theorem~\ref{theorem: bounds on the total variation distance - Barbour and Hall 1984}
here). The motivation for improving the lower bound on the total variation distance
is to take advantage of it to improve the lower bound on the relative entropy (via
Pinsker's inequality or a refinement of it) and some other related quantities,
and then to examine the benefit of this improvement in an information-theoretic
context.

\vspace*{0.1cm}
\begin{theorem}
Let $W = \sum_{i=1}^n X_i$ be a sum of $n$ independent Bernoulli random
variables with $\expectation(X_i) = p_i$ for $i \in \{1, \ldots, n\}$,
and $\expectation(W) = \lambda$. Then, the total variation distance
between the probability distribution of $W$ and the Poisson
distribution with mean $\lambda$ satisfies
\begin{equation}
K_1(\lambda) \, \sum_{i=1}^n p_i^2
\leq d_{\text{TV}}(P_W, \text{Po}(\lambda)) \leq
\left(\frac{1-e^{-\lambda}}{\lambda}\right) \sum_{i=1}^n p_i^2
\label{eq: improved lower bound on the total variation distance}
\end{equation}
where $K_1$ is given by
\begin{equation}
K_1(\lambda) \triangleq \sup_{\small \begin{array}{ll}
& \alpha_1, \alpha_2 \in \reals, \\
& \alpha_2 \leq \lambda + \frac{3}{2}, \\
& \hspace*{0.4cm} \theta > 0 \\
\end{array}} \left( \frac{1-h_{\lambda}(\alpha_1, \alpha_2, \theta)}{2
\, g_{\lambda}(\alpha_1, \alpha_2, \theta)} \right)
\label{eq: K1 in the lower bound on the total variation distance}
\end{equation}
and
\begin{eqnarray}
&& \hspace*{-0.5cm} h_{\lambda}(\alpha_1, \alpha_2, \theta) \triangleq
\frac{3 \lambda + (2-\alpha_2+\lambda)^3 - (1-\alpha_2+\lambda)^3 +
|\alpha_1 - \alpha_2| \, \bigl(2 \lambda + |3-2\alpha_2| \bigr)
\, \exp\left(-\frac{(1-\alpha_2)_+^2}{\theta \lambda}\right)}{\theta \lambda}
\label{eq: h in the lower bound on the total variation distance} \\[0.2cm]
&& \hspace*{-0.5cm} x_+ \triangleq \max\{x, 0\}, \quad
x_+^2 \triangleq \bigl(x_+)^2, \quad
\forall \, x \in \reals \\[0.3cm]
&& \hspace*{-0.5cm} g_{\lambda}(\alpha_1, \alpha_2, \theta) \triangleq
\max \left\{ \, \left| \left(1 + \sqrt{\frac{2}{\theta \lambda e}} \cdot
|\alpha_1 - \alpha_2|\right) \lambda + \max \bigl\{ x(u_i) \bigr\} \right|, \right. \nonumber \\[0.3cm]
&& \hspace*{3.2cm} \left. \left| \left(2 e^{-\frac{3}{2}} + \sqrt{\frac{2}{\theta \lambda e}}  \cdot
|\alpha_1 - \alpha_2|\right) \lambda - \min \bigl\{x(u_i)\bigr\} \right| \,
\right\} \label{eq: g in the lower bound on the total variation distance} \\[0.3cm]
&& \hspace*{-0.5cm} x(u) \triangleq (c_0 + c_1 u + c_2 u^2) \, \exp(-u^2),
\quad \forall \, u \in \reals \label{eq: function x} \\[0.2cm]
&& \hspace*{-0.5cm} \{u_i\} \triangleq \Bigl\{ u \in \reals:
\, 2 c_2 u^3 + 2 c_1 u^2 - 2(c_2 - c_0) u - c_1 = 0\Bigr\}
\label{eq: zeros of a cubic polynomial equation} \\[0.1cm]
&& \hspace*{-0.5cm} c_0 \triangleq (\alpha_2 - \alpha_1) (\lambda - \alpha_2)
\label{eq: c0} \\[0.1cm]
&& \hspace*{-0.5cm} c_1 \triangleq \sqrt{\theta \lambda}
\, (\lambda + \alpha_1 - 2 \alpha_2) \label{eq: c1} \\[0.1cm]
&& \hspace*{-0.5cm} c_2 \triangleq -\theta \lambda.  \label{eq: c2}
\end{eqnarray}
\label{theorem: improved lower bound on the total variation distance}
\end{theorem}

\begin{remark}
The upper and lower bounds on the total variation distance in
\eqref{eq: improved lower bound on the total variation distance}
scale like $\sum_{i=1}^n p_i^2$, similarly to the known bounds in
Theorem~\ref{theorem: bounds on the total variation distance -
Barbour and Hall 1984}. The ratio of the upper and lower bounds in
Theorem~\ref{theorem: bounds on the total variation distance -
Barbour and Hall 1984} tends to~32.00 when either $\lambda$ tends
to zero or infinity. It was obtained numerically that the ratio of
the upper and lower bounds in
Theorem~\ref{theorem: improved lower bound on the total variation distance}
improves by a factor of 18.96 when $\lambda \rightarrow 0$, a factor
of 3.04 when $\lambda \rightarrow \infty$, and at least by a factor
of 2.48 for all $\lambda \in (0, \infty)$. Alternatively, since the
upper bound on the total variation distance in
Theorems~\ref{theorem: bounds on the total variation distance - Barbour and
Hall 1984} and~\ref{theorem: improved lower bound on the total variation distance}
is common, it follows that the ratio of the upper bound and new lower bound
on the total variation distance is reduced to 1.69 when $\lambda \rightarrow 0$,
it is 10.54 when $\lambda \rightarrow \infty$, and it is at most 12.91 for
all $\lambda \in (0, \infty)$.
\label{remark: improvement in the tightness of the new lower bound on
the total variation distance}
\end{remark}

\vspace*{0.1cm}
\begin{remark}
\cite[Theorem~1.2]{DeheuvelsP_AOP86} provides an asymptotic result for the
total variation distance between the distribution of the sum $W$ of $n$ independent
Bernoulli random variables with $\expectation(X_i) = p_i$ and the Poisson
distribution with mean $\lambda = \sum_{i=1}^n p_i$. It shows that when
$\sum_{i=1}^n p_i \rightarrow \infty$ and
$\max_{1 \leq i \leq n} p_i \rightarrow 0$ as $n \rightarrow \infty$ then
\begin{equation}
d_{\text{TV}}(P_W, \text{Po}(\lambda)) \sim \frac{1}{\sqrt{2 \pi e} \;
\lambda} \; \sum_{i=1}^n p_i^2 \, .
\label{eq: asymptotic expression for the total variation distance when lambda
tends to infinity and p_max to zero}
\end{equation}
This implies that the ratio of the upper bound on the total variation distance
in \cite[Theorem~1]{BarbourH_1984} (see Theorems~\ref{theorem: bounds on the
total variation distance - Barbour and Hall 1984} here) and this asymptotic
expression is equal to $\sqrt{2 \pi e} \approx 4.133$.
Therefore, in light of the previous remark (see
Remark~\ref{remark: improvement in the tightness of the new lower bound on
the total variation distance}), it follows that the ratio between the exact
asymptotic value in \eqref{eq: asymptotic expression for the total variation
distance when lambda tends to infinity and p_max to zero} and the new lower
bound in \eqref{eq: improved lower bound on the total variation distance}
is equal to $\frac{10.54}{\sqrt{2 \pi e}} \approx 2.55$. It therefore follows
from Remark~\ref{remark: improvement in the tightness of the new lower bound
on the total variation distance} that in the limit where $\lambda \rightarrow 0$,
the new lower bound on the total variation in
\eqref{eq: improved lower bound on the total variation distance} is smaller
than the exact value by no more than 1.69, and for $\lambda \gg 1$, it is
smaller than the exact asymptotic result by a factor of 2.55.
\label{remark: more on the tightness of the new improved lower bound on the
total variation distance}
\end{remark}

\vspace*{0.1cm}
\begin{remark}
Since $\{u_i\}$ in \eqref{eq: zeros of a cubic polynomial equation} are zeros
of a cubic polynomial equation with real coefficients, then the size of the set
$\{u_i\}$ is either 1 or 3. But since one of the values of $u_i$ is a point
where the global maximum of $x$ is attained, and another value of $u_i$ is the
point where its global minimum is attained (note that
$\lim_{u \rightarrow \pm \infty} x(u) = 0$ and $x$ is differentiable, so
the global maxima and minima of $x$ are attained at finite values where the
derivative of $x$ is equal to zero), then the size of the set $\{u_i\}$ cannot
be~1, which implies that it should be equal to~3.
\label{remark: the size of the set of real zeros is equal to 3}
\end{remark}

\vspace*{0.1cm}
\begin{remark}
The optimization that is required for the computation of $K_1$ in
\eqref{eq: K1 in the lower bound on the total variation distance}
w.r.t. the three parameters $\alpha_1, \alpha_2 \in \reals$ and $\theta \in \reals^+$
is performed numerically. The numerical procedure for the computation of $K_1$ will
be discussed later (after introducing the following corollary).
\end{remark}

\vspace*{0.1cm}
In the following, we introduce a closed-form lower bound on the total variation
distance that is looser than the lower bound in
Theorem~\ref{theorem: improved lower bound on the total variation distance},
but which already improves the lower bound in \cite[Theorem~2]{BarbourH_1984}.
This lower bound follows from Theorem~\ref{theorem: improved lower bound on the total variation distance}
by the special choice of $\alpha_1 = \alpha_2 = \lambda$ that is included in the
optimization set for $K_1$ on the right-hand side of
\eqref{eq: K1 in the lower bound on the total variation distance}.
Following this sub-optimal choice, the lower bound in the next corollary
follows by a derivation of a closed-form expression for the third free
parameter $\theta \in \reals^+$. In fact, this was our first step towards
the derivation of an improved lower bound on the total variation distance.
After introducing the following corollary, we discuss it shortly, and
suggest an optimization procedure for the computing $K_1$ on the
left-hand side of \eqref{eq: improved lower bound on the total variation distance}.

\begin{figure}[here!]  
\begin{center}
\epsfig{file=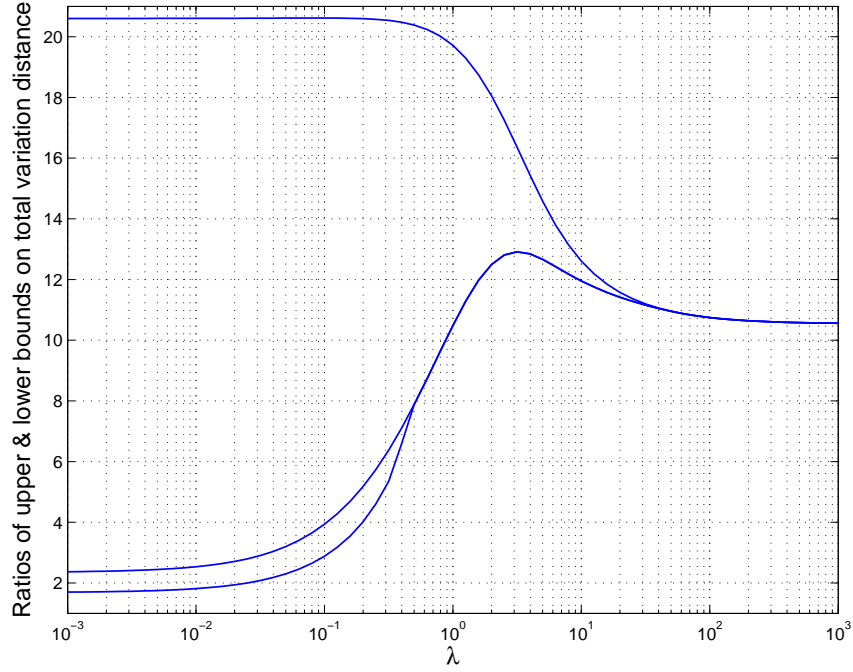,scale=0.65}
\end{center}
\caption{\label{Figure: ratio ot upper and lower bounds on the total variation distance} The
figure presents curves that correspond to ratios of upper and lower bounds on the total
variation distance between the sum of independent Bernoulli random variables and the Poisson
distribution with the same mean $\lambda$. The upper bound on the total variation distance
for all these three curves is the bound by Barbour and Hall (see \cite[Theorem~1]{BarbourH_1984}
or Theorem~\ref{theorem: bounds on the total variation distance - Barbour and Hall 1984}
here). The lower bounds that the three curves refer to them are the following:
the curve at the bottom (i.e., the one which provides the lowest ratio for a fixed
$\lambda$) is the improved lower bound on the total variation distance that is
introduced in Theorem~\ref{theorem: improved lower bound on the total variation distance}.
The curve slightly above it for small values of $\lambda$ corresponds to looser lower bound
when $\alpha_1$ and $\alpha_2$ in
\eqref{eq: K1 in the lower bound on the total variation distance} are set to be equal
(i.e., $\alpha_1 = \alpha_2 \triangleq \alpha$ is their common value), so that the
optimization of $K_1$ for this curve is reduced to be a two-parameter maximization of $K_1$
over the two free parameters $\alpha \in \reals$ and $\theta \in \reals^+$. Finally,
the curve at the top of this figure corresponds to the further loosening of this lower bound
where $\alpha$ is set to be equal to $\lambda$; this leads to a single-parameter maximization
of $K_1$ (over the parameter $\theta \in \reals^+$) whose optimization leads to the closed-form
expression of the lower bound in Corollary~\ref{corollary: improved lower bound on the Chernoff
information between Bernoulli sums of independent RVs and Poisson distribution}. For comparison,
in order to assess the enhanced tightness of the new lower bounds, note that the ratio of the
upper and lower bounds on the total variation distance from
\cite[Theorems~1 and 2]{BarbourH_1984} (or
Theorem~\ref{theorem: bounds on the total variation distance - Barbour and Hall 1984} here)
is roughly equal to 32 for all values of $\lambda$.}
\end{figure}

\vspace*{0.1cm}
\begin{corollary}
Under the assumptions in Theorem~\ref{theorem: improved lower bound on the total variation distance},
then
\begin{equation}
\widetilde{K}_1(\lambda) \, \sum_{i=1}^n p_i^2
\leq d_{\text{TV}}(P_W, \text{Po}(\lambda)) \leq
\left(\frac{1-e^{-\lambda}}{\lambda}\right) \sum_{i=1}^n p_i^2
\label{eq: first improvement of the lower bound on the total variation distance}
\end{equation}
where

\vspace*{-0.8cm}
\begin{eqnarray}
&& \widetilde{K}_1(\lambda) \triangleq \frac{e}{2 \lambda} \; \frac{1 - \frac{1}{\theta}
\, \left(3+\frac{7}{\lambda}\right)}{\theta + 2 e^{-1/2}}
\label{eq: tilted K_1 in the lower bound on the total variation distance} \\[0.3cm]
&& \theta \triangleq 3 + \frac{7}{\lambda} + \frac{1}{\lambda} \cdot
\sqrt{(3\lambda+7)\bigl[(3+2e^{-1/2}) \lambda + 7\bigr]}.
\label{eq: optimal theta for alpha1 and alpha2 equal to lambda}
\end{eqnarray}
Furthermore, the ratio of the upper and lower bounds on the total variation distance
in \eqref{eq: first improvement of the lower bound on the total variation distance}
tends to $\frac{56}{e} \approx 20.601$ as $\lambda \rightarrow 0$, it tends to 10.539
as $\lambda \rightarrow \infty$, and this ratio is monotonic decreasing as a function
of $\lambda \in (0, \infty)$ (see the upper plot in
Figure~\ref{Figure: ratio ot upper and lower bounds on the total variation distance},
and the calculation of the two limits in Section~\ref{subsubsection: Connection of the corollary
with the improved lower bound on the total variation distance to the original lower
bound of Barbour and Hall}).
\label{corollary: lower bound on the total variation distance}
\end{corollary}

\vspace*{0.1cm}
\begin{remark}
The lower bound on the total variation distance on the left-hand side of
\eqref{eq: first improvement of the lower bound on the total variation distance}
improves uniformly the lower bound in \cite[Theorem~2]{BarbourH_1984} (i.e.,
the left-hand side of
Eq.~\eqref{eq: bounds on the total variation distance - Barbour and Hall 1984} here).
The improvement is by factors of 1.55 and 3.03 for $\lambda \rightarrow 0$
and $\lambda \rightarrow \infty$, respectively. Note that this improvement is already
remarkable since the ratio of the upper and lower bounds in \cite[Theorems~1 and 2]{BarbourH_1984}
(Theorem~\ref{theorem: bounds on the total variation distance - Barbour and Hall 1984}
here) is equal to 32 in these two extreme cases, and it is also uniformly upper bounded
by~32 for all values of $\lambda \in (0, \infty)$. Furthermore, in light of
Remark~\ref{remark: improvement in the tightness of the new lower bound on the total variation distance},
the improvement of the lower bound on the total variation distance in
Theorem~\ref{theorem: improved lower bound on the total variation distance} over its
loosened version in Corollary~\ref{corollary: lower bound on the total variation distance}
is especially significant for small values of $\lambda$, but it is marginal
for large values of $\lambda$; this improvement is by a factor of 11.88 in the limit where
$\lambda \rightarrow 0$, but asymptotically there is no improvement
if $\lambda \rightarrow \infty$ where it even holds for $\lambda \geq 20$
(see Figure~\ref{Figure: ratio ot upper and lower bounds on the total variation distance}
where all the curves in this plot merge approximately for $\lambda \geq 20$).
Note, however, that even if $\lambda \rightarrow \infty$, the lower bounds in
Theorem~\ref{theorem: improved lower bound on the total variation distance} and
Corollary~\ref{corollary: lower bound on the total variation distance} improve
the original lower bound in
Theorem~\ref{theorem: bounds on the total variation distance - Barbour and Hall 1984}
by a factor that is slightly above~3.
\label{remark: a comparison between the improved, simplified and original lower bounds
on the total variation distance}
\end{remark}

\vspace*{0.1cm}
\begin{remark}
In light of Corollary~\ref{corollary: lower bound on the total variation distance},
a simplified algorithm is suggested in the following for the computation of $K_1$
in \eqref{eq: K1 in the lower bound on the total variation distance}. In general,
what we compute numerically is a lower bound on $K_1$;
but this is fine since $K_1$ is the coefficient of the lower bound on the left-hand
side of \eqref{eq: K1 in the lower bound on the total variation distance}, so its
replacement by a lower bound still gives a valid lower bound on the total variation
distance. The advantage of the suggested algorithm is its reduced complexity, as
compared to a brute force search over the infinite three-dimensional region for
$(\alpha_1, \alpha_2, \theta)$; the numerical computation that is involved with this
algorithm takes less than a second on a standard PC. The algorithm proceeds as follows:
\begin{itemize}
\item It chooses the initial values $\alpha_1 = \alpha_2 = \lambda$, and $\theta$
as is determined on the right-hand side of
\eqref{eq: optimal theta for alpha1 and alpha2 equal to lambda}. The corresponding
lower bound on the total variation distance from
Theorem~\ref{theorem: improved lower bound on the total variation distance},
for this sub-optimal selection of the three free parameters
$\alpha_1, \alpha_2, \theta$, is equal to the closed-form lower bound in
Corollary~\ref{corollary: lower bound on the total variation distance}.
\item At this point, the algorithm performs several iterations where at each
iteration, it defines a certain three-dimensional grid around the optimized
point from the previous iteration (the zeroth iteration refers to the initial
choice of parameters from the previous item, and to the closed-form lower
bound in Corollary~\ref{corollary: lower bound on the total variation distance}).
At each iteration, the algorithm searches for the optimized point on the new grid
(i.e., it computes the maximum of the expression inside the supremum on the
right-hand side of
\eqref{eq: K1 in the lower bound on the total variation distance} among
all the points of the grid, and it also updates the new location of this point
$(\alpha_1, \alpha_2, \theta)$ for the search that is made in the next iteration.
Note that, from \eqref{eq: K1 in the lower bound on the total variation distance},
the grid should exclude points $(\alpha_1, \alpha_2, \theta)$ when either
$\theta < 0$ or $\alpha_2 > \lambda + \frac{3}{2}$.
\item At the beginning of this recursive procedure, the algorithm take a
very large neighborhood around the point that was selected at the previous
iteration (or the initial selection of the point from the first item). The
size of this neighborhood at each subsequent iteration shrinks, but the
grid also becomes more dense around the new selected point from the previous iteration.
\end{itemize}
It is noted that numerically, the resulting lower bound on $K_1$ seems to
be the exact value in \eqref{eq: K1 in the lower bound on the total variation distance}
and not just a lower bound. However, the reduction in the computational complexity
of (a lower bound on) $K_1$ provides a very fast algorithm. The conclusions of the
last two remarks (i.e.,
Remarks~\ref{remark: a comparison between the improved, simplified and original lower bounds on the total variation distance}
and~\ref{remark: more on the improved lower bounds on the total variation distance}
are supported by Figure~\ref{Figure: ratio ot upper and lower bounds on the total variation distance}.
\label{remark: more on the improved lower bounds on the total variation distance}
\end{remark}

\subsection{Improved Lower Bounds on the Relative Entropy}
\label{subsection: Improved lower bounds on the relative entropy}

The following theorem relies on the new lower bound on the
total variation distance in
Theorem~\ref{theorem: improved lower bound on the total variation distance},
and the distribution-dependent refinement of Pinsker's
inequality in \cite{OrdentlichW_IT2005}. Their combination serves
to derive a new lower bound on the relative entropy between the
distribution of a sum of independent Bernoulli random
variables and a Poisson distribution with the same mean.
The following upper bound on the relative entropy was introduced
in \cite[Theorem~1]{KontoyiannisHJ_2005}.
Together with the new lower bound on the relative entropy, it leads to
the following statement:

\begin{theorem}
In the setting of Theorem~\ref{theorem: improved lower bound on the total variation distance},
the relative entropy between the probability distribution of $W$ and the
Poisson distribution with mean $\lambda = \expectation(W)$ satisfies the following
inequality:
\begin{equation}
K_2(\lambda) \, \left( \sum_{i=1}^n p_i^2 \right)^2 \leq D\bigl(P_W || \text{Po}(\lambda)\bigr)
\leq \frac{1}{\lambda} \sum_{i=1}^n \frac{p_i^3}{1-p_i}
\label{eq: improved lower bound on the relative entropy}
\end{equation}
where
\begin{equation}
K_2(\lambda) \triangleq m(\lambda) \, \bigl(K_1(\lambda)\bigr)^2
\label{eq: coefficient of improved lower bound on the relative entropy}
\end{equation}
with $K_1$ from \eqref{eq: K1 in the lower bound on the total variation distance},
and
\begin{equation}
m(\lambda) \triangleq \left\{ \begin{array}{cl}
\left( \frac{1}{2 e^{-\lambda}-1} \right) \;
\log \left(\frac{1}{e^{\lambda}-1} \right) \quad
& \mbox{if $\lambda \in (0, \log 2)$} \\[0.3cm]
2 \quad & \mbox{if $\lambda \geq \log 2$.}
\end{array}
\right.
\label{eq: the refinement of Pinsker's inequality for the Poisson distribution}
\end{equation}
\label{theorem: improved lower bound on the relative entropy}
\end{theorem}

\begin{remark}
For the sake of simplicity, in order to have a bound in closed-form
(that is not subject to numerical optimization), the lower bound on
the relative entropy on the left-hand side of
\eqref{eq: improved lower bound on the relative entropy} can be
loosened by replacing $K_1(\lambda)$ on the right-hand side of
\eqref{eq: coefficient of improved lower bound on the relative entropy}
with $\widetilde{K}_1(\lambda)$ in
\eqref{eq: tilted K_1 in the lower bound on the total variation distance}
and \eqref{eq: optimal theta for alpha1 and alpha2 equal to lambda}.
In light of
Remark~\ref{remark: a comparison between the improved, simplified and
original lower bounds on the total variation distance},
this possible loosening of the lower bound on the relative entropy
has no effect if $\lambda \geq 30$.
\label{remark: a possible loosening of the lower bound on the relative entropy}
\end{remark}

\begin{remark}
The distribution-dependent refinement of Pinsker's inequality
from \cite{OrdentlichW_IT2005} yields that, when applied to a
Poisson distribution with mean $\lambda$, the coefficient
$m(\lambda)$ in
\eqref{eq: coefficient of improved lower bound on the relative entropy}
is larger than $2$ for $\lambda \in (0, \log 2)$, and it is approximately
equal to $\log\bigl(\frac{1}{\lambda}\bigr)$ for $\lambda \approx 0$.
Hence, for $\lambda \approx 0$, the refinement of
Pinsker's inequality in \cite{OrdentlichW_IT2005} leads to a remarkable
improvement in the lower bound that appears in
\eqref{eq: improved lower bound on the relative entropy}--\eqref{eq:
the refinement of Pinsker's inequality for the Poisson distribution},
which is by approximately a factor of $\frac{1}{2} \, \log\bigl(\frac{1}{\lambda}\bigr)$.
If, however, $\lambda \geq \log 2$ then there is no refinement of Pinsker's inequality
(since $m(\lambda)=2$ in \eqref{eq: the refinement of Pinsker's inequality for the Poisson distribution}).
\label{remark: the effect of the refinement of Pinsker's inequality}
\end{remark}

\begin{remark}
The combination of the original lower bound on the total variation
distance from \cite[Theorem~2]{BarbourH_1984}
(see \eqref{eq: bounds on the total variation distance - Barbour and Hall 1984})
with Pinsker's inequality (see \eqref{eq: Pinsker's inequality}) gives
the following lower bound on the relative entropy:
\begin{equation}
D\bigl(P_W || \, \text{Po}(\lambda)\bigr) \geq \frac{1}{512} \,
\Bigl(1 \wedge \frac{1}{\lambda^2} \Bigr) \,
\left(\sum_{i=1}^n p_i^2 \right)^2.
\label{eq: original lower bound on the relative entropy}
\end{equation}
In light of Remarks~\ref{remark: improvement in the tightness of the new lower bound on the total variation distance}
and~\ref{remark: the effect of the refinement of Pinsker's inequality}, it is possible
to quantify the improvement that is obtained by the new lower bound of
Theorem~\ref{theorem: improved lower bound on the relative entropy}
in comparison to the looser lower bound in
\eqref{eq: original lower bound on the relative entropy}.
The improvement of the new lower bound on the relative entropy is by a factor of
$179.7 \, \log\bigl(\frac{1}{\lambda}\bigr)$ for $\lambda \approx 0$,
a factor of 9.22 for $\lambda \rightarrow \infty$, and at least by a factor of 6.14 for
all $\lambda \in (0, \infty)$. The conclusions in the last two remarks (i.e.,
Remark~\ref{remark: the effect of the refinement of Pinsker's inequality}
and~\ref{remark: comparison to the lower bound on the relative entropy that follows from the
original lower bound on the total variation distance and Pinsker's inequality}) are supported
by Figure~\ref{Figure:
exact_and_bounds_for_the_relative_entropy_between_binomial_and_poisson_distributions}
that refers to the special case of the relative entropy between the binomial and Poisson distributions.
\label{remark: comparison to the lower bound on the relative entropy that follows from the
original lower bound on the total variation distance and Pinsker's inequality}
\end{remark}

\begin{figure}[here!]  
\begin{center}
\epsfig{file=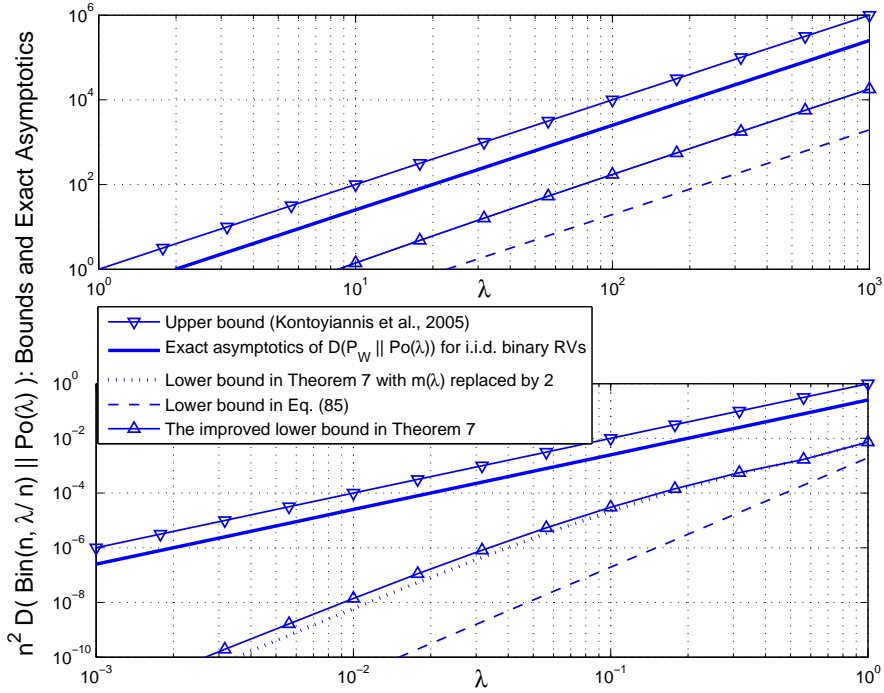,scale=0.65}
\end{center}
\caption{\label{Figure: exact_and_bounds_for_the_relative_entropy_between_binomial_and_poisson_distributions}
This figure refers to the relative entropy between the binomial and Poisson distributions with
the same mean $\lambda$. The horizontal axis refers to $\lambda$, and the vertical axis
refers to a scaled relative entropy $n^2 D(\text{Bin}(n, \frac{\lambda}{n}) || \text{Po}(\lambda))$
($\sum_{i=1}^n X_i \sim \text{Bin}(n, \frac{\lambda}{n})$ when $X_i \sim \text{Bern}(p_i)$
with $p_i \triangleq \frac{\lambda}{n}$ is fixed for all $i \in \{1, \ldots, n\}$).
This scaling of the relative entropy is supported by the upper bound on
the relative entropy by Kontoyiannis et al. (see
\cite[Theorem~1]{KontoyiannisHJ_2005}) that is equal to
$\frac{1}{\lambda} \sum_{i=1}^n \frac{p_i^3}{1-p_i} = \frac{\lambda^2}{n^2} +
O(\frac{1}{n^3})$. It is also supported by the new lower bounds in
Theorems~\ref{theorem: improved lower bound on the relative entropy}
and Eq.~\eqref{eq: original lower bound on the relative entropy}
since the common term in these lower bounds is equal to
$\left(\sum_{i=1}^n p_i^2\right)^2 = \frac{\lambda^4}{n^2}$,
so a multiplication of these lower bounds on the relative entropy by $n^2$
gives an expression that only depends on $\lambda$. It follows from
\cite[Theorem~1]{HarremoesR_2004} (see also \cite[p.~2302]{AdellLY_IT2010})
that $D(\text{Bin}(n, \frac{\lambda}{n}) || \text{Po}(\lambda)) = \frac{\lambda^2}{4n^2} +
O(\frac{1}{n^3})$ (so, the exact value is asymptotically equal to
one-quarter of the upper bound). This figure shows the upper and lower bounds,
as well as the exact asymptotic result, in order to study the tightness of
the existing upper bound and the new lower bounds.
By comparing the dotted and dashed lines, this figure also shows the significant
impact of the refinement of the lower bound on the total variation distance by
Barbour and Hall (see \cite[Theorem~2]{BarbourH_1984}) on the improved
lower bound on the relative entropy (the former improvement is squared via
Pinsker's inequality or its refinement). Furthermore, by comparing the dotted
and solid lines of this figure, it shows that the probability-dependent refinement
of Pinsker's inequality, applied to the Poisson distribution, affects the lower
bound for $\lambda < \log(2)$.}
\end{figure}

\begin{remark}
In \cite[Example~6]{HarremoesV_2011}, it is shown that if
$\expectation(X) \leq \lambda$
then $D\bigl(P_X \, || \, \text{Po}(\lambda)\bigr) \geq
\frac{1}{2 \lambda} \, \bigl(\expectation(X)-\lambda\bigr)^2 \, .$
Since $\expectation(S_n) = \lambda$ then this lower bound on the
relative entropy is not informative for the relative entropy
$D\bigl(P_{S_n} \, || \, \text{Po}(\lambda)\bigr)$.
Theorem~\ref{theorem: improved lower bound on the relative entropy}
and the loosened bound in \eqref{eq: original lower bound on the relative entropy}
are, however, informative in the studied case.

The author was notified in \cite{personal_communications} about
the existence of another recently derived lower bound on the relative
entropy $D\bigl(P_X \, || \, \text{Po}(\lambda)\bigr)$ in terms of the
variance of a random variable $X$ with values in $\naturals_0$
(this lower bound appears in a currently un-published work). The two
bounds were derived independently, based on different approaches.
In the setting where $X = \sum_{i=1}^n X_i$ is a sum of independent Bernoulli
random variables $\{X_i\}_{i=1}^n$ with $\expectation(X_i) = p_i$
and $\lambda = \expectation(X) = \sum_{i=1}^n p_i$, the
two lower bounds on the relative entropy scale like
$\bigl(\sum_{i=1}^n p_i^2 \bigr)^2$ but with a different scaling factor.
\end{remark}

\subsection{Bounds on Related Quantities}
\label{subsection: Bounds on related quantities}

\subsubsection{Bounds on the Hellinger Distance and Bhattacharyya Parameter}
\label{subsubsection: Bounds on the Hellinger distance and Bhattacharyya parameter}

The following proposition introduces a sharpened version of
Proposition~\ref{proposition: known inequality that relates between the total variation,
Hellinger distance and relative entropy}.
\begin{proposition}
Let $P$ and $Q$ be two probability mass functions
that are defined on a same set $\mathcal{X}$. Then,
the following inequality suggests a sharpened version
of the inequality in
\eqref{eq: known inequality that relates between the total variation, Hellinger distance
and relative entropy}
\begin{equation}
\sqrt{1 - \sqrt{1 - \bigl( d_{\text{TV}}(P, Q) \bigr)^2}}
\leq d_{\text{H}}(P, Q)
\leq \sqrt{1-\exp\biggl(-\frac{D(P||Q)}{2} \biggr)}
\label{eq: sharpened inequality that relates between the total variation, Hellinger
distance and relative entropy}
\end{equation}
and
\begin{equation}
\exp\biggl(-\frac{D(P||Q)}{2} \biggr) \leq \text{BC}(P, Q)
\leq \sqrt{1 - \bigl( d_{\text{TV}}(P, Q) \bigr)^2}.
\label{eq: sharpened inequality that relates between the total variation, Bhattacharyya
parameter and relative entropy}
\end{equation}
\label{proposition: sharpened inequality that relates between the total variation,
Hellinger distance, Bhattacharyya parameter and relative entropy}
\end{proposition}

\begin{remark}
A comparison of the upper and lower bounds on the Hellinger distance in
\eqref{eq: sharpened inequality that relates between the total variation, Hellinger
distance and relative entropy} or the Bhattacharyya parameter in
\eqref{eq: sharpened inequality that relates between the total variation, Bhattacharyya
parameter and relative entropy}
gives the following lower bound on the relative entropy in terms of the total
variation distance:
\begin{equation}
D(P||Q) \geq \log \left(\frac{1}{1-\bigl(d_{\text{TV}}(P,Q)\bigr)^2} \right).
\label{eq: a lower bound  on the relative entropy in terms of the total variation distance}
\end{equation}
It is noted that \eqref{eq: a lower bound  on the relative entropy in terms of the total
variation distance}
also follows from the combination of the last two inequalities in \cite[p.~741]{HoeffdingW_1958}.
It is tighter than Pinsker's inequality (see \eqref{eq: Pinsker's inequality} if
$d_{\text{TV}}(P,Q) \geq 0.893$, having also the advantage of giving the right bound
for the relative entropy $(\infty)$ when the total variation distance approaches to~1. However,
\eqref{eq: a lower bound  on the relative entropy in terms of the total variation distance}
is a slightly looser bound on the relative entropy in comparison to Vajda's lower
bound \cite{Vajda_IT1970} that reads:
\begin{equation}
D(P||Q) \geq \log \left(\frac{1+d_{\text{TV}}(P,Q)}{1-d_{\text{TV}}(P,Q)} \right) -
\frac{2d_{\text{TV}}(P,Q)}{1+d_{\text{TV}}(P,Q)} \, .
\label{eq: Vajda's lower bound on the relative entropy}
\end{equation}
\end{remark}

\vspace*{0.1cm}
\begin{corollary}
Under the assumptions in Theorem~\ref{theorem: improved lower bound on the total variation distance},
the Hellinger distance and Bhattacharyya parameter satisfy the following upper
and lower bounds:
\begin{equation}
\sqrt{1 - \sqrt{1-\bigl(K_1(\lambda)\bigr)^2 \left(\sum_{i=1}^n p_i^2 \right)^2}}
\leq d_{\text{H}}(P_W, \text{Po}(\lambda)) \leq \sqrt{1-\exp\left(-\frac{1}{2\lambda}
\sum_{i=1}^n \frac{p_i^3}{1-p_i} \right)}
\label{eq: upper and lower bounds on the Hellinger distance in the context of Poisson approximation}
\end{equation}
and
\begin{equation}
\exp\left(-\frac{1}{2\lambda} \sum_{i=1}^n \frac{p_i^3}{1-p_i} \right)
\leq \text{BC}(P_W, \text{Po}(\lambda))
\leq \sqrt{1-\bigl(K_1(\lambda)\bigr)^2 \left(\sum_{i=1}^n p_i^2 \right)^2}
\label{eq: upper and lower bounds on the Bhattacharyya parameter in the context of Poisson
approximation}
\end{equation}
where $K_1$ on the left-hand side of
\eqref{eq: upper and lower bounds on the Hellinger distance in the context of Poisson approximation}
and the right-hand side of
\eqref{eq: upper and lower bounds on the Bhattacharyya parameter in the context of
Poisson approximation}
is introduced in \eqref{eq: K1 in the lower bound on the total variation distance}.
\label{corollary: upper and lower bounds on the Hellinger distance and Bhattacharyya
parameter in the context of Poisson approximation}
\end{corollary}

\vspace*{0.1cm}
\begin{corollary}
Let $\{S_n\}_{n=1}^{\infty}$ be a sequence of random variables where
$S_n \triangleq \sum_{i=1}^n X_i^{(n)}$ is a sum of $n$ independent Bernoulli
random variables $\{X_i^{(n)}\}_{i=1}^n$ with $\pr(X_i^{(n)}=1) = p_i^{(n)}$
(note that, for $n \neq m$, the binary random variables $X_i^{(n)}$ and $X_j^{(m)}$
may be dependent).
Assume that $\expectation(S_n) = \sum_{i=1}^n p_i^{(n)} = \lambda$
for some $\lambda \in (0, \infty)$ and every $n \in \naturals$, and that
there exist some fixed constants $c_1, c_2 > 0$ such that
$$ \frac{c_1 \lambda}{n} \leq p_i^{(n)} \leq \frac{c_2 \lambda}{n}, \quad
\forall \, i \in \{1, \ldots, n\} $$
(which implies that $c_1 \leq 1$ and $c_2 \geq 1$, and $c_1 = c_2 = 1$
if and only if the binary random variables $\{X_i^{(n)}\}_{i=1}^n$ are i.i.d.).
Then, the following asymptotic results hold:
\begin{eqnarray}
&& D\bigl(P_{S_n} || \text{Po}(\lambda) \bigr) = O\Bigl(\frac{1}{n^2}\Bigr)
\label{eq: asymptotic scaling of the relative entropy} \\[0.1cm]
&& d_{\text{TV}}\bigl(P_{S_n}, \text{Po}(\lambda) \bigr) = O\Bigl(\frac{1}{n}\Bigr)
\label{eq: asymptotic scaling of the total variation distance} \\[0.1cm]
&& d_{\text{H}}\bigl(P_{S_n}, \text{Po}(\lambda) \bigr) = O\Bigl(\frac{1}{n}\Bigr)
\label{eq: asymptotic scaling of the Hellinger distance} \\[0.1cm]
&& \text{BC}\bigl(P_{S_n}, \text{Po}(\lambda) \bigr) = 1 - O\Bigl(\frac{1}{n^2}\Bigr)
\label{eq: asymptotic scaling of the Bhattacharyya parameter}
\end{eqnarray}
so, the relative entropy between the distribution of $S_n$ and the Poisson distribution
with mean $\lambda$ scales like $\frac{1}{n^2}$, the total variation and Hellinger
distances scale like $\frac{1}{n}$, and the gap of the Bhattacharyya parameter
to~1 scales like~$\frac{1}{n^2}$.
\label{corollary: asymptotic results of the relative entropy and related quantities for
independent Bernoulli sums}
\end{corollary}

\vspace*{0.1cm}
\subsubsection{Bounds on the Chernoff Information}
\label{subsubsection: Bounds on the Chernoff information}

\begin{proposition}
Let $P$ and $Q$ be two probability mass functions
that are defined on a same set $\mathcal{X}$. Then, the
Chernoff information between $P$ and $Q$ is lower bounded
in terms of the total variation distance as follows:
\begin{equation}
C(P, Q) \geq -\frac{1}{2} \, \log\Bigl(1-\bigl(d_{\text{TV}}(P,Q) \bigr)^2 \Bigr).
\label{eq: lower bound on the Chernoff information in terms of the total variation distance}
\end{equation}
\label{proposition: lower bound on the Chernoff information in terms of the total variation distance}
\end{proposition}

\vspace*{0.1cm}
\begin{corollary}
Under the assumptions in Theorem~\ref{theorem: improved lower bound on the total variation distance},
the Chernoff information satisfies the following lower bound:
\begin{equation}
C(P_W, \text{Po}(\lambda)) \geq -\frac{1}{2} \; \log\left(1 -
\bigl(K_1(\lambda)\bigr)^2 \left( \sum_{i=1}^n p_i^2 \right)^2  \right)
\label{eq: improved lower bound on the Chernoff information between Bernoulli sums of
independent RVs and Poisson distribution}
\end{equation}
where $K_1$ is introduced in \eqref{eq: K1 in the lower bound on the total variation distance}.
\label{corollary: improved lower bound on the Chernoff information between Bernoulli sums
of independent RVs and Poisson distribution}
\end{corollary}

\begin{remark}
Remark~\ref{remark: a possible loosening of the lower bound on the relative entropy} also
applies to
Corollaries~\ref{corollary: upper and lower bounds on the Hellinger distance and Bhattacharyya
parameter in the context of Poisson approximation}
and~\ref{corollary: improved lower bound on the Chernoff information between Bernoulli sums of independent RVs and Poisson distribution}.
\label{remark: a possible loosening of the lower bounds on the related quantities}
\end{remark}

\begin{remark}
The combination of
Proposition~\ref{proposition: lower bound on the Chernoff information in terms of the total variation distance}
with the lower bound on the total variation distance in \cite[Theorem~2]{BarbourH_1984}
(see Theorem~\ref{theorem: bounds on the total variation distance - Barbour and Hall 1984}
here) gives the following looser lower bound on the Chernoff information:
\begin{equation}
C(P_W, \text{Po}(\lambda)) \geq -\frac{1}{2} \; \log\left(1 - \frac{1}{1024} \;
\Bigl(1 \wedge \frac{1}{\lambda^2} \Bigr) \left( \sum_{i=1}^n p_i^2 \right)^2  \right).
\label{eq: looser lower bound on the Chernoff information between Bernoulli sums of
independent RVs and Poisson distribution}
\end{equation}
The impact of the tightened lower bound in
\eqref{eq: improved lower bound on the Chernoff information between Bernoulli sums of
independent RVs and Poisson distribution},
as compared to the bound in
\eqref{eq: looser lower bound on the Chernoff information between Bernoulli sums of
independent RVs and Poisson distribution}
is exemplified in
Section~\ref{subsection: Second part of applications of the new bounds}
in the context of the Bayesian approach for binary hypothesis testing.
\label{remark: looser lower bound on the Chernoff information between Bernoulli sums
of independent RVs and Poisson distribution}
\end{remark}

\subsection{Applications of the New Bounds in
Section~\ref{section: improved lower bounds on the total variation distance etc.}}
\label{subsection: Second part of applications of the new bounds}

In the following, we consider the use of the new bounds in
Section~\ref{section: improved lower bounds on the total variation distance etc.}
for binary hypothesis testing.

\vspace*{0.1cm}
\begin{example}[Application of the Chernoff-Stein lemma
and lower bounds on the relative entropy] The Chernoff-Stein lemma
considers the asymptotic error exponent in binary hypothesis testing when
one of the probabilities of error is held fixed, and the other one has to be
made as small as possible (see, e.g., \cite[Theorem~11.8.3]{Cover_Thomas}).

Let $\{Y_j\}_{j=1}^N$ be a sequence of non-negative, integer-valued i.i.d.
random variables with $\expectation(Y_1) = \lambda$ for some
$\lambda \in (0, \infty)$. Let $Y_1 \sim Q$ where we consider the following
two hypothesis:
\begin{itemize}
\item $H_1$: $Q=P_1$ where $Y_j$, for $j \in \{1, \ldots, N\}$, is a sum of
$n$ binary random variables $\{X_{i,j}\}_{i=1}^n$ with
$\expectation(X_{i,j}) = p_i$ and $\sum_{i=1}^n p_i = \lambda$. It is assumed
that the elements of the sequence $\{X_{i,j}\}$ are independent, and
$n \in \naturals$ is fixed.
\item $H_2$: $Q=P_2$ is the Poisson distribution with mean $\lambda$
(i.e., $Y_1 \sim \text{Po}(\lambda)$).
\end{itemize}
Note that in this case, if one of the $Y_j$ exceeds the value $n$ then $H_1$
is rejected automatically, so one may assume that $n \gg \max\{\lambda, 1\}$.
More explicitly, if $Y_j \sim \text{Po}(\lambda)$ for $j \in \{1, \ldots, N\}$,
the probability of this event to happen is upper bounded (via the union and
Chernoff bounds) by
\begin{equation}
\pr(\exists \, j \in \{1, \ldots, N\}: \, Y_j \geq n+1)
\leq N \, \exp\left\{-\left[\lambda+(n+1)
\log\left(\frac{n+1}{\lambda e}\right)\right]\right\}
\label{eq: the probability for an element of the sequence Y of Poisson random
variables to exceed the value n}
\end{equation}
so, if $n \geq 10 \max\{\lambda, 1\}$, this probability is typically very
small.

For an arbitrary $N \in \naturals$, let $A_N$ be an acceptance region for hypothesis~1.
Using standard notation, let
\begin{equation}
\alpha_N \triangleq P_1^N(A_N^{\text{c}}), \quad \beta_N \triangleq P_2^N(A_N)
\label{eq: two types of error in binary hypothesis testing}
\end{equation}
be the two types of error probabilities. Following \cite[Theorem~11.8.3]{Cover_Thomas},
for an arbitrary $\varepsilon \in \bigl(0, \frac{1}{2}\bigr)$, let
\begin{equation*}
\beta_N^{\varepsilon} \triangleq \min_{A_N \subseteq \mathcal{Y}^N: \,
\alpha_N < \varepsilon} \beta_N
\end{equation*}
where $\mathcal{Y} \triangleq \{0, 1, \ldots, n\}$ is the alphabet that is associated
with hypothesis $H_1$. Then, the best asymptotic exponent of $\beta_N^{\varepsilon}$
in the limit where $\varepsilon \rightarrow 0$ is
\begin{equation*}
\lim_{\varepsilon \rightarrow 0} \lim_{N \rightarrow \infty} \frac{1}{N} \, \log \beta_N^{\varepsilon} = -D(P_1 || P_2).
\end{equation*}
From \cite[Eqs.~(11.206), (11.207) and (11.227)]{Cover_Thomas}, for
the relative entropy typical set that is defined by
\begin{equation}
A_{N}^{(\varepsilon)}(P_1 || P_2) \triangleq \left\{ \underline{y} \in \mathcal{Y}^N:
\left| \frac{1}{N} \, \log \left(\frac{P_1^N(\underline{y})}{P_2^N(\underline{y})}\right)
-D(P_1 || P_2) \right| \leq \varepsilon \right\}
\end{equation}
then, it follows from the AEP for relative entropy that $\alpha_N < \varepsilon$
for $N$ large enough (see, e.g., \cite[Theorem~11.8.1]{Cover_Thomas}). Furthermore,
for every $N$ (see, e.g, \cite[Theorem~11.8.2]{Cover_Thomas}),
\begin{equation}
\beta_N < \exp\Bigl(-N \bigl(D(P_1||P_2) - \varepsilon \bigr)\Bigr).
\label{eq: upper bound on the error probability of the second type}
\end{equation}
The error probability of the second type $\beta_N$ is treated here separately from $\alpha_N$.
In this case, a lower bound on the relative entropy $D(P_1||P_2)$ gives an exponential
upper bound on $\beta_N$. Let $\varepsilon \rightarrow 0$ (more explicitly, let
$\varepsilon$ be chosen to be small enough as compared to a lower bound on $D(P_1||P_2)$).
In the following two simple examples, we calculate the improved
lower bound in Theorem~\ref{theorem: improved lower bound on the relative entropy},
and compare it to the lower bound in \eqref{eq: original lower bound on the relative entropy}.
More importantly, we study the impact of
Theorem~\ref{theorem: improved lower bound on the relative entropy}
on the reduction of the number of samples $N$ that are required for achieving
$\beta_N < \varepsilon$. The following two cases are used to exemplify this issue:
\begin{enumerate}
\item Let the probabilities $\{p_i\}_{i=1}^n$ (that correspond to hypothesis~1) be given by
$$p_i = \frac{i \, p_n}{n}, \quad \forall \, i \in \{1, \ldots, n\}.$$
For $\lambda \in (0, \infty)$, in order to satisfy the equality
$\sum_{i=1}^n p_i = \lambda$ then $p_n = \frac{2 \lambda}{n+1}$, and
$ \sum_{i=1}^n p_i^2 = \frac{2 \lambda^2}{3} \, \frac{2n+1}{n(n+1)} \, .$
From Theorem~\ref{theorem: improved lower bound on the relative entropy}, the improved
lower bound on the relative entropy reads
\begin{equation}
D(P_1 || P_2) \geq K_2(\lambda) \left(\frac{2 \lambda^2}{3} \,
\frac{2n+1}{n(n+1)} \right)^2
\label{eq: improved lower bound on the relative entropy for case 1}
\end{equation}
where $K_2$ is introduced in \eqref{eq: coefficient of improved lower bound on the relative
entropy}, and the weaker lower bound in \eqref{eq: original lower bound on the relative entropy}
gets the form
\begin{equation}
D(P_1 || P_2) \geq \left(\frac{\lambda^4}{1152}\right) \,
\min\left\{1, \frac{1}{\lambda^2}\right\} \left(\frac{2n+1}{n(n+1)}\right)^2 \, .
\label{eq: weak lower bound on the relative entropy for case 1}
\end{equation}
Lets examine the two bounds on the relative entropy for $\lambda = 10$ and
$n=100$ to find accordingly a proper value of $N$ such that $\beta_N < 10^{-10}$,
and choose $\varepsilon = 10^{-10}$. Note that the probability of the event that
one of the $N$ Poisson random variables $\{Y_j\}_{j=1}^N$, under hypothesis $H_2$,
exceeds the value $n$ is upper bounded in \eqref{eq: the probability for an element of the
sequence Y of Poisson random variables to exceed the value n} by $1.22 N \cdot 10^{-62}$,
so it is neglected for all reasonable amounts of samples $N$. In this setting, the two
lower bounds on the relative entropy in
\eqref{eq: improved lower bound on the relative entropy for case 1} and
\eqref{eq: weak lower bound on the relative entropy for case 1}, respectively,
are equal to $2.47 \cdot 10^{-4}$ and $3.44 \cdot 10^{-5}$ nats. For these two lower
bounds, the exponential upper bound in
\eqref{eq: upper bound on the error probability of the second type} ensures that
$\beta_N < 10^{-10}$ for $N \geq 9.32 \cdot 10^4$ and $N \geq 6.70 \cdot 10^5$,
respectively. Hence, the improved lower bound on the relative entropy in
Theorem~\ref{theorem: improved lower bound on the relative entropy} implies here
a reduction in the required number of samples by a factor of 7.17.
\item In the second case, assume that the probabilities $\{p_i\}_{i=1}^n$ scale
exponentially in $i$ (instead of the linear scaling in the previous case). Let
$\alpha \in (0,1)$ and consider the case where
$$p_i = p_1 \alpha^{i-1}, \quad \forall \, i \in \{1, \ldots, n\}.$$
For $\lambda \in (0, \infty)$, in order to hold the equality
$\sum_{i=1}^n p_i = \lambda$ then
$p_1 = \frac{\lambda (1-\alpha)}{1-\alpha^n}$, and
$\sum_{i=1}^n p_i^2 = \frac{\lambda^2 (1-\alpha)}{1+\alpha}
\, \frac{1+\alpha^n}{1-\alpha^n}$. Hence, the improved lower bound in
Theorem~\ref{theorem: improved lower bound on the relative entropy} and
the other bound in \eqref{eq: original lower bound on the relative entropy}
imply respectively that
\begin{eqnarray}
D(P_1 || P_2) \geq \lambda^4 \, K_2(\lambda) \left( \frac{1-\alpha}{1+\alpha} \;
\frac{1+\alpha^n}{1-\alpha^n} \right)^2
\label{eq: improved lower bound on the relative entropy for case 2}
\end{eqnarray}
and
\vspace*{-0.5cm}
\begin{eqnarray}
D(P_1 || P_2) \geq \Bigl(\frac{\lambda^4}{512}\Bigr) \,
\min\left\{1, \frac{1}{\lambda^2}\right\}
\left( \frac{1-\alpha}{1+\alpha} \; \frac{1+\alpha^n}{1-\alpha^n} \right)^2.
\label{eq: weak lower bound on the relative entropy for case 2}
\end{eqnarray}
The choice $\alpha = 0.05$, $\lambda = 0.1$ and $n=100$, implies that
the two lower bounds on the relative entropy in
\eqref{eq: improved lower bound on the relative entropy for case 2} and
\eqref{eq: weak lower bound on the relative entropy for case 2} are respectively
equal to $2.48 \cdot 10^{-5}$ and $1.60 \cdot 10^{-7}$. The exponential upper bound in
\eqref{eq: upper bound on the error probability of the second type} therefore
ensures that $\beta_N < 10^{-10}$ for $N \geq 9.26 \cdot 10^5$ and $N \geq 1.44 \cdot 10^8$,
respectively. Hence, the improvement in
Theorem~\ref{theorem: improved lower bound on the relative entropy}
leads in this case to the conclusion that one can achieve the target error probability
of the second type while reducing the number of samples $\{Y_j\}_{j=1}^N$ by
a factor of 155.
\end{enumerate}
\label{example: Application of the Chernoff-Stein lemma and the new lower
bound on the relative entropy}
\end{example}

\vspace*{0.1cm}
\begin{example}[Application of the lower bounds on the Chernoff
information to binary hypothesis testing]
We turn to consider binary hypothesis testing with
the Bayesian approach (see, e.g., \cite[Section~11.9]{Cover_Thomas}).
In this setting, one wishes to minimize the overall probability of error
while we refer to the two hypotheses in Example~\ref{example: Application of the
Chernoff-Stein lemma and the new lower bound on the relative entropy}.
The best asymptotic exponent in the Bayesian approach is the Chernoff
information (see \eqref{eq: Chernoff information}), and the overall
error probability satisfies the following exponential upper bound:
\begin{equation}
P_{\text{e}}^{(N)} \leq \exp\bigl(-N \, C(P_1, P_2) \bigr)
\label{eq: exponential upper bound on the overall probability in terms
of Chernoff information}
\end{equation}
so, a lower bound on the Chernoff information provides an upper bound
on the overall error probability. In the following, the two lower bounds
on the Chernoff information in \eqref{eq: improved lower bound on the
Chernoff information between Bernoulli sums of independent RVs and
Poisson distribution} and \eqref{eq: looser lower bound on the Chernoff
information between Bernoulli sums of independent RVs and Poisson distribution},
and the advantage of the former lower bound is studied
in the two cases of Example~\ref{example: Application of the Chernoff-Stein
lemma and the new lower bound on the relative entropy} in order to
examine the impact of its improved tightness on the reduction of the
number of samples $N$ that are required to achieve an overall error
probability below $\varepsilon = 10^{-10}$. We refer, respectively, to
cases~1 and~2 of Example~\ref{example: Application of the Chernoff-Stein
lemma and the new lower bound on the relative entropy}.
\begin{enumerate}
\item In case~1 of Example~\ref{example: Application of the Chernoff-Stein
lemma and the new lower bound on the relative entropy}, the two lower
bounds on the Chernoff information in
Corollary~\ref{corollary: improved lower bound on the Chernoff information
between Bernoulli sums of independent RVs and Poisson distribution} and
Remark~\ref{remark: looser lower bound on the Chernoff information between
Bernoulli sums of independent RVs and Poisson distribution} (following the
calculation of $\sum_{i=1}^n p_i^2$ for these two cases) are
\begin{equation*}
C(P_1, P_2) \geq \left\{ \begin{array}{ll}
-\frac{1}{2} \, \log \left(1- \, \bigl(K_1(\lambda)\bigr)^2
\left(\frac{2 \lambda^2}{3} \, \frac{2n+1}{n(n+1)}\right)^2 \right)
& \quad \mbox{From Eq.~\eqref{eq: improved lower bound on the
Chernoff information between Bernoulli sums of independent RVs and
Poisson distribution} (Corollary~\ref{corollary: improved lower bound
on the Chernoff information between Bernoulli sums of independent RVs
and Poisson distribution})} \\[0.4cm]
-\frac{1}{2} \, \log \left(1-\frac{\lambda^4}{2304} \, \min\Bigl\{1,
\frac{1}{\lambda^2}\Bigr\} \left(\frac{2n+1}{n(n+1)}\right)^2 \right)
& \quad \mbox{From Eq.~\eqref{eq: looser lower bound on the
Chernoff information between Bernoulli sums of independent RVs and
Poisson distribution} (Remark~\ref{remark: looser lower bound on the
Chernoff information between Bernoulli sums of independent RVs and
Poisson distribution}).}
\end{array}
\right.
\end{equation*}
As in the first case of Example~\ref{example: Application of the
Chernoff-Stein lemma and the new lower bound on the relative entropy},
let $\lambda = 10$ and $n = 100$. The lower bounds on the Chernoff
information are therefore equal to
\begin{equation}
C(P_1, P_2) \geq \left\{ \begin{array}{ll}
6.16 \cdot 10^{-5}
& \quad \mbox{From Eq.~\eqref{eq: improved lower bound on the
Chernoff information between Bernoulli sums of independent RVs and
Poisson distribution}} \\[0.4cm]
8.59 \cdot 10^{-6}
& \quad \mbox{From Eq.~\eqref{eq: looser lower bound on the
Chernoff information between Bernoulli sums of independent RVs and
Poisson distribution}.}
\end{array}
\right.
\label{eq: lower bounds on the Chernoff information for case 1 of the example}
\end{equation}
Hence, in order to achieve the target $P_{\text{e}}^{(N)} \leq 10^{-10}$
for the overall error probability, the lower bounds on the Chernoff information
in \eqref{eq: lower bounds on the Chernoff information for case 1 of the example}
and the exponential upper bound on the overall error probability in
\eqref{eq: exponential upper bound on the overall probability in terms
of Chernoff information} imply that
\begin{equation}
N \geq \left\{ \begin{array}{ll}
3.74 \cdot 10^5
& \quad \mbox{From Eqs.~\eqref{eq: improved lower bound on the
Chernoff information between Bernoulli sums of independent RVs
and Poisson distribution} and \eqref{eq: exponential upper bound on
the overall probability in terms of Chernoff information}} \\[0.4cm]
2.68 \cdot 10^6
& \quad \mbox{From Eqs.~\eqref{eq: looser lower bound on the
Chernoff information between Bernoulli sums of independent RVs
and Poisson distribution} and \eqref{eq: exponential upper bound
on the overall probability in terms of Chernoff information}}
\end{array}
\right.
\label{eq: lower bounds on N for case 1 of the example}
\end{equation}
so, the number of required samples is approximately reduced by a factor of 7.
\item For the second case in Example~\ref{example: Application of
the Chernoff-Stein lemma and the new lower bound on the relative entropy},
the lower bounds on the Chernoff information in
Eqs.~\eqref{eq: improved lower bound on the Chernoff information between
Bernoulli sums of independent RVs and Poisson distribution} and
\eqref{eq: looser lower bound on the Chernoff information between Bernoulli
sums of independent RVs and Poisson distribution} read
\begin{equation*}
C(P_1, P_2) \geq \left\{ \begin{array}{ll}
-\frac{1}{2} \, \log \left(1- \lambda^4 \, \bigl(K_1(\lambda)\bigr)^2 \,
\left(\frac{1-\alpha}{1+\alpha} \, \frac{1+\alpha^n}{1-\alpha^n}
\right)^2 \right)
& \quad \mbox{From Eq.~\eqref{eq: improved lower bound on the
Chernoff information between Bernoulli sums of independent RVs and
Poisson distribution}} \\[0.4cm]
-\frac{1}{2} \, \log \left(1-\frac{\lambda^4}{1024} \, \min\Bigl\{1,
\frac{1}{\lambda^2}\Bigr\} \left(\frac{1-\alpha}{1+\alpha} \, \frac{1+\alpha^n}{1-\alpha^n}\right)^2 \right)
& \quad \mbox{From Eq.~\eqref{eq: looser lower bound on the
Chernoff information between Bernoulli sums of independent RVs and
Poisson distribution}}
\end{array}
\right.
\end{equation*}
so, the same choice of parameters $\alpha = 0.05$, $\lambda = 0.1$
and $n=100$ as in Example~\ref{example: Application of the Chernoff-Stein
lemma and the new lower bound on the relative entropy} implies that
\begin{equation}
C(P_1, P_2) \geq \left\{ \begin{array}{ll}
4.93 \cdot 10^{-6}
& \quad \mbox{From Eq.~\eqref{eq: improved lower bound on the
Chernoff information between Bernoulli sums of independent RVs and
Poisson distribution}} \\[0.4cm]
4.00 \cdot 10^{-8}
& \quad \mbox{From Eq.~\eqref{eq: looser lower bound on the
Chernoff information between Bernoulli sums of independent RVs and
Poisson distribution}.}
\end{array}
\right.
\label{eq: lower bounds on the Chernoff information for case 2 of the example}
\end{equation}
\end{enumerate}
For obtaining the target $P_{\text{e}}^{(N)} \leq 10^{-10}$
for the overall error probability, the lower bounds on the
Chernoff information in
\eqref{eq: lower bounds on the Chernoff information for case 2 of the example}
and the exponential upper bound on the overall error probability in
\eqref{eq: exponential upper bound on the overall probability in terms
of Chernoff information} imply that
\begin{equation}
N \geq \left\{ \begin{array}{ll}
4.68 \cdot 10^6
& \quad \mbox{From Eqs.~\eqref{eq: improved lower bound on the
Chernoff information between Bernoulli sums of independent RVs
and Poisson distribution} and \eqref{eq: exponential upper bound on
the overall probability in terms of Chernoff information}} \\[0.4cm]
5.76 \cdot 10^8
& \quad \mbox{From Eqs.~\eqref{eq: looser lower bound on the
Chernoff information between Bernoulli sums of independent RVs
and Poisson distribution} and \eqref{eq: exponential upper bound
on the overall probability in terms of Chernoff information}}
\end{array}
\right.
\label{eq: lower bounds on N for case 2 of the example}
\end{equation}
so, the improved lower bound on the Chernoff information implies
in this case a reduction in the required number of samples $N$
by a factor of 123.
\label{example: The application of the new lower bound on the Chernoff
information with the Bayesian approach for binary hypothesis testing}
\end{example}

\subsection{Proofs of the New Results in Section~\ref{section: improved
lower bounds on the total variation distance etc.}}
\label{subsection: Proofs of the results in the second part of this paper}

\subsubsection{Proof of Theorem~\ref{theorem: improved lower bound on the
total variation distance}}
\label{subsubsection: Proof of the theorem with the improved lower bound on the total
variation distance}
The proof of Theorem~\ref{theorem: improved lower bound on the total variation distance}
starts similarly to the proof of \cite[Theorem~2]{BarbourH_1984}. However, it significantly
deviates from the original analysis in order to derive an improved lower bound on the
total variation distance.
In the following, we introduce the proof of Theorem~\ref{theorem: improved lower bound
on the total variation distance}.

Let $\{X_i\}_{i=1}^n$ be independent Bernoulli random variables with $\expectation(X_i) = p_i$.
Let $W \triangleq \sum_{i=1}^n X_i$,
$V_i \triangleq \sum_{j \neq i} X_j$ for every $i \in \{1, \ldots, n\}$, and
$Z \sim \text{Po}(\lambda)$ with mean $\lambda \triangleq \sum_{i=1}^n p_i$.
From the basic equation of the Chen-Stein method, the equality
\begin{equation}
\expectation[\lambda f(Z+1) - Z f(Z)] = 0.
\label{eq: basic equation of the Chen-Stein method for Poisson approximation}
\end{equation}
holds for an arbitrary bounded function $f: \naturals_0 \rightarrow \reals$.
Furthermore
\begin{eqnarray}
&& \expectation\bigl[\lambda f(W+1) - W f(W)\bigr] \nonumber \\
&& = \sum_{j=1}^n p_j \, \expectation\bigl[f(W+1)\bigr] - \sum_{j=1}^n
\expectation\bigl[X_j f(W) \bigr] \nonumber \\
&& = \sum_{j=1}^n p_j \, \expectation\bigl[f(W+1)\bigr] -
\sum_{j=1}^n p_j \, \expectation\bigl[f(V_j+1) \, | \, X_j = 1 \bigr] \nonumber \\
&&  \stackrel{(\text{a})}{=} \sum_{j=1}^n p_j \, \expectation\bigl[f(W+1) - f(V_j+1) \bigr]
\nonumber \\
&& = \sum_{j=1}^n p_j^2 \, \expectation\bigl[f(W+1) - f(V_j+1) \, | \, X_j = 1 \bigr] \nonumber \\
&& = \sum_{j=1}^n p_j^2 \, \expectation\bigl[f(V_j+2) - f(V_j+1) \, | \, X_j = 1 \bigr]
\nonumber \\
&&  \stackrel{(\text{b})}{=} \sum_{j=1}^n p_j^2 \, \expectation\bigl[f(V_j+2) - f(V_j+1) \bigr]
\label{eq: first step in the derivation of the improved lower bound on total variation distance}
\end{eqnarray}
where equalities~(a) and (b) hold since $X_j$ and $V_j$ are independent random variables
for every $j \in \{1, \ldots, n\}$. By subtracting
\eqref{eq: basic equation of the Chen-Stein method for Poisson approximation} from
\eqref{eq: first step in the derivation of the improved lower bound on total variation distance},
it follows that for an arbitrary bounded function $f: \naturals_0 \rightarrow \reals$
\begin{equation}
\expectation\bigl[\lambda f(W+1) - W f(W)\bigr] -
\expectation\bigl[\lambda f(Z+1) - Z f(Z)\bigr]
= \sum_{j=1}^n p_j^2 \, \expectation\bigl[f(V_j+2) - f(V_j+1) \bigr].
\label{eq: second step in the derivation of the improved lower bound on total variation distance}
\end{equation}
In the following, an upper bound on the left-hand side of
\eqref{eq: second step in the derivation of the improved lower bound on total variation distance}
is derived, based on total variation distance between the two distributions of $W$ and $Z$.
\begin{eqnarray}
&& \expectation\bigl[\lambda f(W+1) - W f(W)\bigr] -
\expectation\bigl[\lambda f(Z+1) - Z f(Z)\bigr] \nonumber \\
&& = \sum_{k=0}^{\infty} \, \bigl( \lambda f(k+1) - k f(k) \bigr) \,
\bigl( \pr(W=k) - \pr(Z=k) \bigr) \nonumber \\
&& \leq \sum_{k=0}^{\infty} \, \bigl| \lambda f(k+1) - k f(k) \bigr| \,
\bigl| \pr(W=k) - \pr(Z=k) \bigr|
\label{eq: intermediate step that is used later to derive a lower bound on the local distance} \\
&& \leq \sup_{k \in \naturals_0} \bigl| \lambda f(k+1) - k f(k) \bigr| \,
\sum_{k=0}^{\infty} \bigl| \pr(W=k) - \pr(Z=k) \bigr| \nonumber \\
&& = 2 d_{\text{TV}}(P_W, \, \text{Po}(\lambda)) \,
\sup_{k \in \naturals_0} \bigl| \lambda f(k+1) - k f(k) \bigr|
\label{eq: third step in the derivation of the improved lower bound on total variation distance}
\end{eqnarray}
where the last equality follows from
\eqref{eq: the L1 distance is twice the total variation distance}.
Hence, the combination of
\eqref{eq: second step in the derivation of the improved lower bound on total variation distance}
and
\eqref{eq: third step in the derivation of the improved lower bound on total variation distance}
gives the following lower bound on the total variation distance:
\begin{eqnarray}
d_{\text{TV}}(P_W, \, \text{Po}(\lambda)) \geq
\frac{ \dsum_{j=1}^n \Bigl\{p_j^2 \, \expectation\bigl[f(V_j+2) - f(V_j+1) \bigr] \Bigr\}}{2 \,
\sup_{k \in \naturals_0} \bigl| \lambda f(k+1) - k f(k) \bigr|}
\label{eq: fourth step in the derivation of the improved lower bound on total variation distance}
\end{eqnarray}
which holds, in general, for an arbitrary bounded function $f: \naturals_0 \rightarrow \reals$.

At this point, we deviate from the proof of \cite[Theorem~2]{BarbourH_1984} by generalizing and
refining (in a non-trivial way) the original analysis. The general problem
with the current lower bound in \eqref{eq: fourth step in the derivation of the improved lower
bound on total variation distance} is that it is not calculable in closed form for a given $f$, so one needs
to choose a proper function $f$ and derive a closed-form expression for a lower bound on the right-hand side of
\eqref{eq: fourth step in the derivation of the improved lower bound on total variation distance}.
To this end, let
\begin{equation}
f(k) \triangleq (k-\alpha_1) \, \exp\biggl(-\frac{(k-\alpha_2)^2}{\theta \lambda}\biggr) \, ,
\quad \forall \, k \in \naturals_0
\label{eq: proposed f for the derivation of the improved lower bound on the total variation
distance}
\end{equation}
where $\alpha_1, \alpha_2 \in \reals$ and $\theta \in \reals^+$ are fixed constants (note that
$\theta$ in \eqref{eq: proposed f for the derivation of the improved lower bound on the total variation distance}
needs to be positive for $f$ to be a bounded function). In order
to derive a lower bound on the total variation distance, we calculate a lower bound on the
numerator and an upper bound on the denominator of the right-hand side
of \eqref{eq: fourth step in the derivation of the improved lower bound on total variation distance}
for the function $f$ in \eqref{eq: proposed f for the derivation of the improved lower bound on the
total variation distance}. Referring to the numerator of the right-hand side
of \eqref{eq: fourth step in the derivation of the improved lower bound on total variation
distance} with $f$ in \eqref{eq: proposed f for the derivation of the improved lower bound on the
total variation distance}, for every $j \in \{1, \ldots, n\}$,
\begin{eqnarray}
&& f(V_j+2) - f(V_j+1) \nonumber \\[0.1cm]
&& = \int_{V_j + 1 - \alpha_2}^{V_j + 2 - \alpha_2} \frac{\mathrm{d}}{\mathrm{d}u}
\left( (u+\alpha_2-\alpha_1) \, \exp\Bigl(-\frac{u^2}{\theta \lambda}\Bigr) \right) \,
\mathrm{d}u  \nonumber \\[0.2cm]
&& = \int_{V_j + 1 - \alpha_2}^{V_j + 2 - \alpha_2} \left(1 - \frac{2u
(u+\alpha_2-\alpha_1)}{\theta \lambda} \right) \exp\Bigl(-\frac{u^2}{\theta \lambda}\Bigr) \,
\mathrm{d}u \nonumber \\[0.2cm]
&& = \int_{V_j + 1 - \alpha_2}^{V_j + 2 - \alpha_2} \left(1 - \frac{2 u^2}{\theta \lambda}\right)
\, \exp\Bigl(-\frac{u^2}{\theta \lambda}\Bigr) \, \mathrm{d}u -
\frac{2(\alpha_2 - \alpha_1)}{\theta \lambda} \int_{V_j + 1 - \alpha_2}^{V_j + 2 - \alpha_2}
u \, \exp\Bigl(-\frac{u^2}{\theta \lambda}\Bigr) \, \mathrm{d}u \nonumber \\[0.2cm]
&& = \int_{V_j + 1 - \alpha_2}^{V_j + 2 - \alpha_2} \left(1 - \frac{2 u^2}{\theta \lambda}\right)
\, \exp\Bigl(-\frac{u^2}{\theta \lambda}\Bigr) \, \mathrm{d}u \nonumber \\
&& \hspace*{0.4cm} - (\alpha_2 - \alpha_1) \left[ \exp\biggl(-\frac{(V_j + 2 - \alpha_2)^2}{\theta
\lambda}\biggr) - \exp\biggl(-\frac{(V_j + 1 - \alpha_2)^2}{\theta \lambda}\biggr) \right].
\label{eq: fifth step in the derivation of the improved lower bound on total variation distance}
\end{eqnarray}
We rely in the following on the inequality
$$(1-2x) \, e^{-x} \geq 1-3x, \quad \forall \, x \geq 0.$$
Applying it to the integral on the right-hand side of
\eqref{eq: fifth step in the derivation of the improved lower bound on total variation distance}
gives that
\begin{eqnarray}
&& f(V_j+2) - f(V_j+1) \nonumber \\[0.1cm]
&& \geq \int_{V_j + 1 - \alpha_2}^{V_j + 2 - \alpha_2} \left(1 - \frac{3 u^2}{\theta
\lambda}\right) \, \mathrm{d}u -
(\alpha_2 - \alpha_1) \left[ \exp\biggl(-\frac{(V_j + 2 - \alpha_2)^2}{\theta \lambda}\biggr) -
\exp\biggl(-\frac{(V_j + 1 - \alpha_2)^2}{\theta \lambda}\biggr) \right] \nonumber \\[0.1cm]
&& \geq 1 - \frac{\bigl(V_j + 2 - \alpha_2\bigr)^3 - \bigl(V_j + 1 - \alpha_2\bigr)^3}{\theta
\lambda} \nonumber \\[0.1cm]
&& \hspace*{0.4cm} - \bigl|\alpha_2 - \alpha_1\bigr| \cdot
\left| \exp\biggl(-\frac{(V_j + 2 - \alpha_2)^2}{\theta \lambda}\biggr) -
\exp\biggl(-\frac{(V_j + 1 - \alpha_2)^2}{\theta \lambda}\biggr) \right|.
\label{eq: sixth step in the derivation of the improved lower bound on total variation distance}
\end{eqnarray}
In order to proceed, note that if $x_1, x_2 \geq 0$ then (based on the mean-value theorem
of calculus)
\begin{eqnarray*}
&& | e^{-x_2} - e^{-x_1} | \\
&& = \bigl|e^{-c} \, (x_1 - x_2)\bigr|   \quad \mbox{for some} \; \; c \in [x_1, x_2] \\[0.1cm]
&& \leq e^{-\min\{x_1, x_2\}} \, |x_1 - x_2|
\end{eqnarray*}
which, by applying it to the second term on the right-hand side of
\eqref{eq: sixth step in the derivation of the improved lower bound on total variation distance},
gives that for every $j \in \{1, \ldots, n\}$
\begin{eqnarray}
&&\left| \exp\biggl(-\frac{(V_j + 2 - \alpha_2)^2}{\theta \lambda}\biggr) -
\exp\biggl(-\frac{(V_j + 1 - \alpha_2)^2}{\theta \lambda}\biggr) \right|
\nonumber \\[0.1cm]
&& \leq \exp\left(-\frac{\min\Bigl\{(V_j+2-\alpha_2)^2, \, (V_j+1-\alpha_2)^2
\Bigr\}}{\theta \lambda}\right) \cdot
\left(\frac{(V_j+2-\alpha_2)^2-(V_j+1-\alpha_2)^2}{\theta \lambda}\right) \, .
\label{eq: seventh step in the derivation of the improved lower bound on total variation distance}
\end{eqnarray}
Since $V_j = \sum_{i \neq j} X_i \geq 0$ then
\begin{eqnarray}
&& \min\Bigl\{(V_j+2-\alpha_2)^2, \, (V_j+1-\alpha_2)^2\Bigr\} \nonumber \\[0.1cm]
&& \geq \left\{ \begin{array}{cl}
0 \quad & \mbox{if $\alpha_2 \geq 1$} \\[0.1cm]
(1-\alpha_2)^2 \quad & \mbox{if $\alpha_2 < 1$}
\end{array}
\right. \nonumber \\[0.1cm]
&& = \bigl(1-\alpha_2\bigr)_{+}^2
\label{eq: 8th step in the derivation of the improved lower bound on total variation distance}
\end{eqnarray}
where $$x_+ \triangleq \max\{x,0\}, \quad x_+^2 \triangleq \bigl(x_+ \bigr)^2, \quad
\forall \, x \in \reals.$$ Hence, the combination
of the two inequalities in
\eqref{eq: seventh step in the derivation of the improved lower bound
on total variation distance}--\eqref{eq: 8th step in the derivation
of the improved lower bound on total variation distance} gives that
\begin{eqnarray}
&&\left| \exp\biggl(-\frac{(V_j + 2 - \alpha_2)^2}{\theta \lambda}\biggr) -
\exp\biggl(-\frac{(V_j + 1 - \alpha_2)^2}{\theta \lambda}\biggr) \right|
\nonumber \\[0.1cm]
&& \leq \exp\left(-\frac{(1-\alpha_2)_+^2}{\theta \lambda} \right) \cdot
\left(\frac{\left|(V_j + 2 - \alpha_2)^2 -
(V_j + 1 - \alpha_2)^2 \right|}{\theta \lambda} \right) \nonumber \\[0.1cm]
&& = \exp\left(-\frac{(1-\alpha_2)_+^2}{\theta \lambda} \right) \cdot
\frac{\left|2V_j + 3 - 2 \alpha_2 \right|}{\theta \lambda}
\nonumber \\[0.1cm]
&& \leq \exp\left(-\frac{(1-\alpha_2)_+^2}{\theta \lambda} \right) \cdot
\frac{2V_j + \left|3 - 2 \alpha_2 \right|}{\theta \lambda}
\label{eq: ninth step in the derivation of the improved lower bound on total variation distance}
\end{eqnarray}
and therefore, a combination of the inequalities in
\eqref{eq: sixth step in the derivation of the improved lower bound on total variation distance}
and
\eqref{eq: ninth step in the derivation of the improved lower bound on total variation distance}
gives that
\begin{eqnarray}
&& f(V_j+2) - f(V_j+1) \nonumber \\[0.1cm]
&& \geq 1 - \frac{\bigl(V_j + 2 - \alpha_2\bigr)^3 - \bigl(V_j + 1 - \alpha_2\bigr)^3}{\theta
\lambda} \nonumber \\[0.1cm]
&& \hspace*{0.4cm} - \bigl|\alpha_2 - \alpha_1\bigr| \cdot
\exp\left(-\frac{(1-\alpha_2)_+^2}{\theta \lambda} \right) \cdot
\frac{2V_j + \left|3 - 2 \alpha_2 \right|}{\theta \lambda} \; .
\label{eq: 10th step in the derivation of the improved lower bound on total variation distance}
\end{eqnarray}
Let $U_j \triangleq V_j - \lambda$, then
\begin{eqnarray}
&& f(V_j+2) - f(V_j+1) \nonumber \\[0.1cm]
&& \geq 1 - \frac{\bigl(U_j + \lambda + 2 - \alpha_2\bigr)^3 -
\bigl(U_j + \lambda + 1 - \alpha_2\bigr)^3}{\theta
\lambda} \nonumber \\[0.1cm]
&& \hspace*{0.4cm} - \bigl|\alpha_2 - \alpha_1\bigr| \cdot
\exp\left(-\frac{(1-\alpha_2)_+^2}{\theta \lambda} \right) \cdot
\frac{2U_j + 2\lambda + \left|3 - 2 \alpha_2 \right|}{\theta \lambda} \nonumber \\[0.1cm]
&& = 1 - \frac{3 U_j^2 + 3 \bigl(3-2\alpha_2 + 2 \lambda\bigr) U_j +
(2-\alpha_2+\lambda)^3 - (1-\alpha_2+\lambda)^3}{\theta \lambda} \nonumber \\[0.1cm]
&& \hspace*{0.4cm} - \bigl|\alpha_2 - \alpha_1\bigr| \cdot
\exp\left(-\frac{(1-\alpha_2)_+^2}{\theta \lambda} \right) \cdot
\frac{2U_j + 2\lambda + \left|3 - 2 \alpha_2 \right|}{\theta \lambda} \; .
\label{eq: 11th step in the derivation of the improved lower bound on total variation distance}
\end{eqnarray}
In order to derive a lower bound on the numerator of the right-hand side of
\eqref{eq: fourth step in the derivation of the improved lower bound on total variation distance},
for the function $f$ in \eqref{eq: proposed f for the derivation of the improved lower bound on the
total variation distance}, we need to calculate the expected value of the right-hand side of
\eqref{eq: 11th step in the derivation of the improved lower bound on total variation distance}.
To this end, the first and second moments of $U_j$ are calculated as follows:
\begin{eqnarray}
&& \expectation(U_j) \nonumber \\
&& = \expectation(V_j) - \lambda \nonumber \\
&& = \sum_{i \neq j} p_i - \sum_{i=1}^n p_i \nonumber \\
&& = -p_j \label{eq: first moment of U_j} \\
\text{and} \nonumber \\
&& \expectation(U_j^2) \nonumber \\[0.1cm]
&& = \expectation\bigl( (V_j - \lambda)^2 \bigr) \nonumber \\[0.1cm]
&& = \expectation \left[ \, \left(\sum_{i \neq j} (X_i - p_i) - p_j \right)^2 \, \right]
\nonumber \\
&& \stackrel{(\text{a})}{=} \sum_{i \neq j} \expectation\bigl[(X_i - p_i)^2\bigr] + p_j^2
\nonumber \\
&& \stackrel{(\text{b})}{=} \sum_{i \neq j} p_i(1-p_i) + p_j^2 \nonumber \\
&& = \sum_{i \neq j} p_i - \sum_{i \neq j} p_i^2 + p_j^2 \nonumber \\
&& = \lambda - p_j - \sum_{i \neq j} p_i^2 + p_j^2.
\label{eq: second moment of U_j}
\end{eqnarray}
where equalities~(a) and~(b) hold since, by assumption, the binary random variables $\{X_i\}_{i=1}^n$
are independent and $\expectation(X_i)=p_i$,
$\text{Var}(X_i) = p_i (1-p_i)$. By taking expectations on both sides of
\eqref{eq: 11th step in the derivation of the improved lower bound on total variation distance},
one obtains from \eqref{eq: first moment of U_j} and \eqref{eq: second moment of U_j} that
\begin{eqnarray}
&& \expectation\bigl[f(V_j+2) - f(V_j+1)\bigr] \nonumber \\[0.2cm]
&& \geq 1 - \frac{3 \Bigl(\lambda - p_j - \sum_{i \neq j} p_i^2 + p_j^2\Bigr)
+ 3 \bigl(3-2\alpha_2 + 2 \lambda\bigr) \bigl(-p_j \bigr) +
(2-\alpha_2+\lambda)^3 - (1-\alpha_2+\lambda)^3}{\theta \lambda} \nonumber \\[0.2cm]
&& \hspace*{0.4cm} - \bigl|\alpha_2 - \alpha_1\bigr| \cdot
\exp\left(-\frac{(1-\alpha_2)_+^2}{\theta \lambda} \right) \cdot
\left(\frac{-2p_j + 2\lambda + \left|3 - 2 \alpha_2 \right|}{\theta \lambda} \right)
\nonumber \\[0.2cm]
&& = 1 - \frac{3 \lambda + (2-\alpha_2+\lambda)^3 - (1-\alpha_2+\lambda)^3
- \Bigl[3p_j(1-p_j) + 3 \sum_{i \neq j} p_i^2
+ 3 \bigl(3-2\alpha_2 + 2 \lambda\bigr) p_j \Bigr]}{\theta \lambda} \nonumber \\[0.2cm]
&& \hspace*{0.4cm} - \biggl(\frac{\bigl|\alpha_2 - \alpha_1\bigr| \,
\bigl(2\lambda - 2 p_j+ \left|3 - 2 \alpha_2 \right| \bigr)}{\theta \lambda} \biggr)
\cdot \exp\left(-\frac{(1-\alpha_2)_+^2}{\theta \lambda} \right) \nonumber \\[0.2cm]
&& \geq 1 - \frac{3 \lambda + (2-\alpha_2+\lambda)^3 - (1-\alpha_2+\lambda)^3
- \bigl(9-6\alpha_2 + 6 \lambda\bigr) p_j}{\theta \lambda} \nonumber \\[0.2cm]
&& \hspace*{0.4cm} - \biggl(\frac{\bigl|\alpha_2 - \alpha_1\bigr| \,
\bigl(2\lambda+ \left|3 - 2 \alpha_2 \right| \bigr)}{\theta \lambda} \biggr)
\cdot \exp\left(-\frac{(1-\alpha_2)_+^2}{\theta \lambda} \right) \; .
\label{eq: 12th step in the derivation of the improved lower bound on total variation distance}
\end{eqnarray}
Therefore, from \eqref{eq: 12th step in the derivation of the improved lower bound on total variation distance},
the following lower bound on the right-hand side of
\eqref{eq: fourth step in the derivation of the improved lower bound on total variation distance}
holds
\begin{eqnarray}
&& \hspace*{-1.5cm} \sum_{j=1}^n \Bigl\{ p_j^2 \,
\expectation \bigl[f(V_j+2) - f(V_j+1)\bigr] \Bigr\} \geq
\left( \frac{3\bigl(3-2\alpha_2+2\lambda\bigr)}{\theta \lambda} \right)
\sum_{j=1}^n p_j^3 \nonumber \\
&& \hspace*{-1.2cm} + \left(1-\frac{3 \lambda + (2-\alpha_2+\lambda)^3 - (1-\alpha_2+\lambda)^3
+ |\alpha_1 - \alpha_2| \bigl(2\lambda+|3-2\alpha_2|\bigr)
\exp\left(-\frac{(1-\alpha_2)_+^2}{\theta \lambda} \right)}{\theta \lambda} \right)
\sum_{j=1}^n p_j^2 \, .
\label{eq: 12.5th step in the derivation of the improved lower bound on total variation distance}
\end{eqnarray}
Note that if $\alpha_2 \leq \lambda + \frac{3}{2}$, which is a condition that
is involved in the maximization of
\eqref{eq: K1 in the lower bound on the total variation distance}, then the first term on
the right-hand side of
\eqref{eq: 12.5th step in the derivation of the improved lower bound on total variation distance}
can be removed, and the resulting lower bound on the numerator of the right-hand side of
\eqref{eq: fourth step in the derivation of the improved lower bound on total variation distance}
gets the form
\begin{equation}
\sum_{j=1}^n \Bigl\{ p_j^2 \, \expectation\bigl[f(V_j+2) - f(V_j+1)\bigr] \Bigr\}
\geq \Bigl(1 - h_{\lambda}(\alpha_1, \alpha_2, \theta) \Bigr) \sum_{j=1}^n p_j^2
\label{eq: 13th step in the derivation of the improved lower bound on total variation distance}
\end{equation}
where the function $h_{\lambda}$ is introduced in \eqref{eq: h in the lower bound on the total
variation distance}.

We turn now to derive an upper bound on the denominator of the right-hand side of
\eqref{eq: fourth step in the derivation of the improved lower bound on total variation distance}.
Therefore, we need to derive a closed-form upper bound on
$\sup_{k \in \naturals_0} \bigl| \lambda \, f(k+1) - k \, f(k) \bigr|$
with the function $f$ in \eqref{eq: proposed f for the derivation of the improved lower
bound on the total variation distance}. For every $k \in \naturals_0$
\begin{equation}
\lambda \, f(k+1) - k \, f(k) = \lambda \, \bigl[ f(k+1) - f(k) \bigr] + (\lambda - k) \, f(k).
\label{eq: 14th step in the derivation of the improved lower bound on total variation distance}
\end{equation}
In the following, we derive bounds on each of the two terms on the right-hand side of
\eqref{eq: 14th step in the derivation of the improved lower bound on total variation distance},
and we start with the first term.
Let $$t(u) \triangleq (u+\alpha_2-\alpha_1) \, \exp\left(-\frac{u^2}{\theta \lambda}\right),
\quad \forall \, u \in \reals$$
then $f(k) = t(k-\alpha_2)$ for every $k \in \naturals_0$, and by the mean value of calculus
\begin{eqnarray}
&& f(k+1) - f(k) \nonumber \\
&& = t(k+1-\alpha_2) - t(k-\alpha_2) \nonumber \\
&& = t'(c_k) \quad \mbox{for some} \; c_k \in [k-\alpha_2, k+1-\alpha_2] \nonumber \\
&& = \left(1-\frac{2 c_k^2}{\theta \lambda}\right) \, \exp\left(-\frac{c_k^2}{\theta
\lambda}\right) + \left(\frac{2(\alpha_1-\alpha_2) c_k}{\theta \lambda}\right) \,
\exp\left(-\frac{c_k^2}{\theta \lambda}\right) \, .
\label{eq: 15th step in the derivation of the improved lower bound on total variation distance}
\end{eqnarray}
By referring to the first term on the right-hand side of
\eqref{eq: 15th step in the derivation of the improved lower bound on total variation distance},
let $$p(u) \triangleq (1-2u) e^{-u}, \quad \forall \, u \geq 0$$ then the global maximum and
minimum of $p$ over the non-negative real line are obtained at
$u=0$ and $u = \frac{3}{2}$, respectively, and therefore
$$ -2 e^{-\frac{3}{2}} \leq p(u) \leq 1, \quad \forall \, u \geq 0.$$
Let $u = \frac{c_k^2}{\theta \lambda}$, then it follows that the first term on the right-hand side
of \eqref{eq: 15th step in the derivation of the improved lower bound on total variation distance}
satisfies the inequality
\begin{equation}
-2 e^{-\frac{3}{2}} \leq \Bigl(1-\frac{2 c_k^2}{\theta \lambda}\Bigr) \,
\exp\Bigl(-\frac{c_k^2}{\theta \lambda}\Bigr) \leq 1.
\label{eq: 16th step in the derivation of the improved lower bound on total variation distance}
\end{equation}
Furthermore, by referring to the second term on the right-hand side of
\eqref{eq: 15th step in the derivation of the improved lower bound on total variation distance},
let $$q(u) \triangleq u e^{-u^2}, \quad \forall \, u \in \reals$$
then the global maximum and minimum of $q$ over the real line are obtained at
$u = +\frac{\sqrt{2}}{2}$ and $u = -\frac{\sqrt{2}}{2}$, respectively, and therefore
$$-\frac{1}{2} \sqrt{\frac{2}{e}} \leq q(u) \leq +\frac{1}{2} \sqrt{\frac{2}{e}} \; ,
\quad \forall \, u \in \reals.$$
Let this time $u = \sqrt{\frac{c_k}{\theta \lambda}}$, then it follows that the second term on the
right-hand side
of \eqref{eq: 15th step in the derivation of the improved lower bound on total variation distance}
satisfies
\begin{equation}
\left| \biggl(\frac{2(\alpha_1 - \alpha_2) c_k}{\theta \lambda} \biggr) \cdot
\exp\biggl(-\frac{c_k^2}{\theta \lambda}\biggr) \right| \leq
\sqrt{\frac{2}{\theta \lambda e}} \cdot |\alpha_1 - \alpha_2|.
\label{eq: 17th step in the derivation of the improved lower bound on total variation distance}
\end{equation}
Hence, by combining the equality in
\eqref{eq: 15th step in the derivation of the improved lower bound on total variation distance}
with the two inequalities in
\eqref{eq: 16th step in the derivation of the improved lower bound on total variation distance}
and
\eqref{eq: 17th step in the derivation of the improved lower bound on total variation distance},
it follows that the first term on the right-hand side of
\eqref{eq: 14th step in the derivation of the improved lower bound on total variation distance}
satisfies
\begin{equation}
-\left(2 \lambda e^{-\frac{3}{2}} + \sqrt{\frac{2\lambda}{\theta e}} \cdot |\alpha_1 -
\alpha_2| \right) \leq \lambda \bigl[f(k+1) - f(k) \bigr]
\leq \lambda + \sqrt{\frac{2\lambda}{\theta e}} \cdot |\alpha_1 - \alpha_2| \, ,
\quad \forall \, k \in \naturals_0.
\label{eq: 18th step in the derivation of the improved lower bound on total variation distance}
\end{equation}
We continue the analysis by a derivation of bounds on the second term of the right-hand side
of \eqref{eq: 14th step in the derivation of the improved lower bound on total variation distance}.
For the function $f$ in \eqref{eq: proposed f for the derivation of the improved lower bound on
the total variation distance}, it is equal to
\begin{eqnarray}
&& (\lambda - k) \, f(k) \nonumber \\
&& = (\lambda-k) (k-\alpha_1) \exp\biggl(-\frac{(k-\alpha_2)^2}{\theta \lambda}\biggr) \nonumber \\
&& = \bigl[ (\lambda-\alpha_2)+(\alpha_2-k) \bigr] \, \bigl[ (k-\alpha_2)+(\alpha_2-\alpha_1)
\bigr] \, \exp\biggl(-\frac{(k-\alpha_2)^2}{\theta \lambda}\biggr) \nonumber \\
&& = \Bigl[ (\lambda-\alpha_2) (k-\alpha_2) +
(\alpha_2-\alpha_1) (\lambda-\alpha_2) - (k-\alpha_2)^2
+ (\alpha_1 - \alpha_2) (k-\alpha_2) \Bigr] \, \exp\biggl(-\frac{(k-\alpha_2)^2}{\theta
\lambda}\biggr) \nonumber \\[0.1cm]
&& = \bigl[ \sqrt{\theta \lambda} \, (\lambda-\alpha_2) \, v_k - \theta \lambda \, v_k^2
- \sqrt{\theta \lambda} \, (\alpha_2 - \alpha_1) \, v_k
+ (\alpha_2 - \alpha_1) (\lambda - \alpha_2) \bigr] \, e^{-v_k^2} \, , \quad
v_k \triangleq \frac{k-\alpha_2}{\sqrt{\theta \lambda}} \; \; \forall \, k \in \naturals_0
\nonumber \\
&& = (c_0 + c_1 v_k + c_2 v_k^2) \, e^{-v_k^2}
\label{eq: 19th step in the derivation of the improved lower bound on total variation distance}
\end{eqnarray}
where the coefficients $c_0, c_1$ and $c_2$ are introduced in
Eqs.~\eqref{eq: c0}--\eqref{eq: c2}, respectively.
In order to derive bounds on the left-hand side of
\eqref{eq: 19th step in the derivation of the improved lower bound on total variation distance},
lets find the global maximum and minimum of the function $x$
in \eqref{eq: function x}:
$$x(u) \triangleq (c_0 + c_1 u + c_2 u^2) e^{-u^2} \, \quad \forall \, u \in \reals.$$
Note that $\lim_{u \rightarrow \pm \infty} x(u) = 0$ and $x$ is differentiable
over the real line, so the global maximum and minimum of $x$ are attained at
finite points and their corresponding values are finite. By setting the derivative of $x$ to zero,
the candidates for the global maximum and minimum of $x$ over the real line are the real zeros
$\{u_i\}$ of the cubic polynomial equation in \eqref{eq: zeros of a cubic polynomial equation}.
Note that by their definition in \eqref{eq: zeros of a cubic polynomial equation}, the values
of $\{u_i\}$ are {\em independent} of the value of $k \in \naturals_0$, and also the
size of the set $\{u_i\}$ is equal to~3 (see Remark~\ref{remark: the size of the set of real zeros
is equal to 3}). Hence, it follows from
\eqref{eq: 19th step in the derivation of the improved lower bound on total variation distance}
that
\begin{equation}
\min_{i \in \{1, 2, 3\}} \{x(u_i)\} \leq (\lambda - k) \, f(k)
\leq \max_{i \in \{1, 2, 3\}} \{x(u_i)\} \, , \quad  \forall \, k \in \naturals_0
\label{eq: 20th step in the derivation of the improved lower bound on total variation distance}
\end{equation}
where these bounds on the second term on the right-hand side of
\eqref{eq: 14th step in the derivation of the improved lower bound on total variation distance}
are independent of the value of $k \in \naturals_0$.

In order to get bounds on the left-hand side of
\eqref{eq: 14th step in the derivation of the improved lower bound on total variation distance},
note that from the bounds on the first and second terms on the right-hand side of
\eqref{eq: 14th step in the derivation of the improved lower bound on total variation distance}
(see
\eqref{eq: 18th step in the derivation of the improved lower bound on total variation distance}
and
\eqref{eq: 20th step in the derivation of the improved lower bound on total variation distance},
respectively) then for every $k \in \naturals_0$
\begin{eqnarray}
&& \min_{i \in \{1, 2, 3\}} \{x(u_i)\} - \left(2 \lambda e^{-\frac{3}{2}} +
\sqrt{\frac{2\lambda}{\theta e}} \cdot |\alpha_1 - \alpha_2| \right) \nonumber \\[0.1cm]
&& \leq \lambda \, f(k+1) - k \, f(k) \nonumber \\[0.1cm]
&& \leq \max_{i \in \{1, 2, 3\}} \{x(u_i)\}
+ \lambda + \sqrt{\frac{2\lambda}{\theta e}} \cdot |\alpha_1 - \alpha_2|
\label{eq: 21st step in the derivation of the improved lower bound on total variation distance}
\end{eqnarray}
which yields that the following inequality is satisfied:
\begin{equation}
\sup_{k \in \naturals_0} \left| \lambda \, f(k+1) - k \, f(k) \right|
\leq g_{\lambda}(\alpha_1, \alpha_2, \theta)
\label{eq: 22nd step in the derivation of the improved lower bound on total variation distance}
\end{equation}
where the function $g_{\lambda}$ is introduced in
\eqref{eq: g in the lower bound on the total variation distance}.
Finally, by combining the inequalities in
Eqs.~\eqref{eq: fourth step in the derivation of the improved lower bound on total variation distance},
\eqref{eq: 13th step in the derivation of the improved lower bound on total variation distance}
and
\eqref{eq: 22nd step in the derivation of the improved lower bound on total variation distance},
the lower bound on the total variation distance in
\eqref{eq: improved lower bound on the total variation distance} follows. The existing
upper bound on the total variation distance in
\eqref{eq: improved lower bound on the total variation distance}
was derived in \cite[Theorem~1]{BarbourH_1984} (see
Theorem~\ref{theorem: bounds on the total variation distance - Barbour and Hall 1984} here).
This completes the proof of
Theorem~\ref{theorem: improved lower bound on the total variation distance}.

\vspace*{0.1cm}
\subsubsection{Proof of Corollary~\ref{corollary: lower bound on the total variation distance}}
\label{subsubsection: Proof of the corollary with the improved lower bound on the total
variation distance}
Corollary~\ref{corollary: lower bound on the total variation distance} follows as a special
case of Theorem~\ref{theorem: improved lower bound on the total variation distance} when
the proposed function $f$ in \eqref{eq: proposed f for the derivation of the improved lower
bound on the total variation distance} is chosen such that two of its three free
parameters (i.e., $\alpha_1$ and $\alpha_2$) are determined sub-optimally, and its third
parameter ($\theta$) is determined optimally in terms of the sub-optimal selection of
the two other parameters. More explicitly, let $\alpha_1$ and $\alpha_2$ in
\eqref{eq: proposed f for the derivation of the improved lower bound on the total variation
distance} be set to be equal to $\lambda$ (i.e., $\alpha_1 = \alpha_2 = \lambda$). From
\eqref{eq: c0}--\eqref{eq: c2}, this setting implies that $c_0 = c_1 = 0$ and
$c_2 = -\theta \lambda < 0$ (since $\theta, \lambda > 0$). The cubic polynomial
equation in \eqref{eq: zeros of a cubic polynomial equation}, which corresponds to this
(possibly sub-optimal) setting of $\alpha_1$ and $\alpha_2$, is $$2 c_2 u^3  - 2 c_2 u = 0$$
whose zeros are $u = 0, \pm 1$. The
function $x$ in \eqref{eq: function x} therefore gets the form
$$x(u) = c_2 u^2 e^{-u^2} \, \quad \forall \, u \in \reals$$
so $x(0)=0$ and $x(\pm 1) = \frac{c_2}{e} < 0$. It implies that
$$\min_{i \in \{1, 2, 3\}} x(u_i) = \frac{c_2}{e}, \quad
\max_{i \in \{1, 2, 3\}} x(u_i) = 0,$$ and therefore $h_{\lambda}$ and
$g_{\lambda}$ in \eqref{eq: h in the lower bound on the total variation distance} and
\eqref{eq: g in the lower bound on the total variation distance}, respectively,
are simplified to
\begin{eqnarray}
&& h_{\lambda}(\lambda, \lambda, \theta) = \frac{3 \lambda+7}{\theta \lambda} \, ,
\label{eq: simplified h for the corollary on the total variation distance} \\
&& g_{\lambda}(\lambda, \lambda, \theta)
= \lambda \, \max\bigl\{1, 2 e^{-\frac{3}{2}} + \theta e^{-1} \bigr\}.
\label{eq: simplified g for the corollary on the total variation distance}
\end{eqnarray}
This sub-optimal setting of $\alpha_1$ and $\alpha_2$ in \eqref{eq: proposed f for the
derivation of the improved lower bound on the total variation distance} implies that the
coefficient $K_1$ in \eqref{eq: K1 in the lower bound on the total variation distance}
is replaced with a loosened version
\begin{equation}
K'_1(\lambda) \triangleq \sup_{\theta > 0} \left(\frac{1-h_{\lambda}(\lambda,
\lambda, \theta)}{2 g_{\lambda}(\lambda, \lambda, \theta)}\right).
\label{eq: loosened coefficient for the corollary on the total variation distance}
\end{equation}
Let $\theta \geq e - \frac{2}{\sqrt{e}}$, then
\eqref{eq: simplified g for the corollary on the total variation distance} is simplified
to $g_{\lambda}(\lambda, \lambda, \theta)
= \lambda \, \bigl(2 e^{-\frac{3}{2}} + \theta e^{-1}\bigr)$. It therefore
follows from \eqref{eq: improved lower bound on the total variation distance},
\eqref{eq: K1 in the lower bound on the total variation distance}
and  \eqref{eq: simplified h for the corollary on the total variation distance}--\eqref{eq:
loosened coefficient for the corollary on the total variation distance} that
\begin{equation}
d_{\text{TV}}\bigl(P_W, \text{Po}(\lambda)\bigr) \geq \widetilde{K}_1(\lambda) \,
\sum_{i=1}^n p_i^2
\label{eq: loosened lower bound on the total variation distance - Corollary}
\end{equation}
where
\begin{equation}
\widetilde{K}_1(\lambda)
= \sup_{\theta \geq e - \frac{2}{\sqrt{e}}} \, \left( \frac{1-\frac{3\lambda+7}{\theta
\lambda}}{2\lambda \bigl(2 e^{-\frac{3}{2}} + \theta e^{-1}\bigr)} \right)
\label{eq: a possibly further loosened coefficient for the corollary on the total variation distance}
\end{equation}
and, in general, $K'_1(\lambda) \geq \widetilde{K}_1(\lambda)$ due to the
above restricted constraint on $\theta$ (see \eqref{eq: loosened coefficient
for the corollary on the total variation distance} versus
\eqref{eq: a possibly further loosened coefficient for the corollary on the
total variation distance}). Differentiation of the function inside the
supremum w.r.t. $\theta$ and by setting its derivative
to zero, one gets the following quadratic equation in $\theta$:
$$\lambda \, \theta^2 - 2(3\lambda+7) \, \theta - 2(3\lambda+7) e^{-1} = 0$$
whose positive solution is the optimized value of $\theta$
in \eqref{eq: optimal theta for alpha1 and alpha2 equal to lambda}.
Furthermore, it is clear that this value of $\theta$ in
\eqref{eq: optimal theta for alpha1 and alpha2 equal to lambda}
is larger than, e.g., 3, so it satisfies the constraint
in \eqref{eq: a possibly further loosened coefficient
for the corollary on the total variation distance}. This completes
the proof of Corollary~\ref{corollary: lower bound on the total variation distance}.

\vspace*{0.1cm}
\subsubsection{Discussion on the Connections of
Theorem~\ref{theorem: improved lower bound on the total variation distance} and
Corollary~\ref{corollary: lower bound on the total
variation distance} to \cite[Theorem~2]{BarbourH_1984}}
\label{subsubsection: Connection of the corollary with the improved lower bound on the total
variation distance to the original lower bound of Barbour and Hall}
As was demonstrated in the previous sub-section, Theorem~\ref{theorem: improved lower bound
on the total variation distance} implies the satisfiability of the lower bound on the
total variation distance in Corollary~\ref{corollary: lower bound on the total
variation distance}. In the following, it is proved that Corollary~\ref{corollary: lower bound
on the total variation distance} implies the lower bound on the total variation distance
in \cite[Theorem~2]{BarbourH_1984} (see also
Theorem~\ref{theorem: bounds on the total variation distance - Barbour and Hall 1984} here),
and the improvement in the tightness of the lower bound
in Corollary~\ref{corollary: lower bound on the total variation distance} is explicitly quantified
in the two extreme cases where $\lambda \rightarrow 0$ and $\lambda \rightarrow \infty$. The
observation that Corollary~\ref{corollary: lower bound on the total variation distance} provides a
tightened lower bound, as compared to \cite[Theorem~2]{BarbourH_1984}, is justified by the
fact that the lower bound in \eqref{eq: loosened lower bound on the total variation distance - Corollary}
with the coefficient
$\widetilde{K}_1(\lambda)$ in \eqref{eq: a possibly further loosened coefficient for the corollary
on the total variation distance} was loosened in the proof of \cite[Theorem~2]{BarbourH_1984} by a
sub-optimal selection of the parameter $\theta$ which leads to a lower bound on $\widetilde{K}_1(\lambda)$
(the sub-optimal selection of $\theta$ in the proof of
\cite[Theorem~2]{BarbourH_1984} is $\theta = 21 \max\bigl\{1, \frac{1}{\lambda}\bigr\}$).
On the other hand, the optimized value of $\theta$ that is used
in \eqref{eq: optimal theta for alpha1 and alpha2 equal to lambda} provides an exact
closed-form expression for $\widetilde{K}_1(\lambda)$ in
\eqref{eq: a possibly further loosened coefficient for the corollary on the total variation distance},
and it leads to the derivation of the bound in
Corollary~\ref{corollary: lower bound on the total variation distance}. This therefore
justifies the observation that the lower bound on the total variation distance in
Corollary~\ref{corollary: lower bound on the total variation distance} implies the original lower
bound in \cite[Theorem~2]{BarbourH_1984}.

From \cite[Theorems~1 and 2]{BarbourH_1984}, the ratio between
the upper and lower bounds on the total variation distance (these bounds also appear in
\eqref{eq: bounds on the total variation distance - Barbour and Hall 1984}) is equal to 32
in the two extreme cases where $\lambda \rightarrow 0$ or $\lambda \rightarrow \infty$. In
order to quantify the improvement that is obtained by
Corollary~\ref{corollary: lower bound on the total variation distance} (that follows by
the optimal selection of the parameter $\theta$), we calculate in the following the
ratio of the same upper bound and the new lower bound in this corollary at these two
extreme cases. In the limit where one lets $\lambda$ tend to infinity, this
ratio tends to
\begin{eqnarray}
&& \lim_{\lambda \rightarrow \infty} \frac{\left(\frac{1-e^{-\lambda}}{\lambda}\right)
\, \sum_{i=1}^n p_i^2}{\left(\frac{1-\frac{3\lambda+7}{\lambda \theta}}{2 \lambda
\bigl(2 e^{-{3/2}}+\theta \, e^{-1}\bigr)} \right)\, \sum_{i=1}^n p_i^2} \quad
\quad (\theta = \theta(\lambda) \;
\mbox{is given in Eq.~\eqref{eq: optimal theta for alpha1 and alpha2 equal to lambda}})
\nonumber \\[0.1cm]
&& = 2 \lim_{\lambda \rightarrow \infty} \frac{2 e^{-{3/2}}+\theta \,
e^{-1}}{1-\frac{3\lambda+7}{\lambda \theta}} \nonumber \\[0.1cm]
&& = \frac{2}{e} \, \lim_{\lambda \rightarrow \infty} \frac{\theta \bigl(2 e^{-{1/2}}+\theta
\bigr)}{\theta-\bigl(3+\frac{7}{\lambda}\bigr)} \nonumber \\[0.1cm]
&& \stackrel{\text{(a)}}{=} \frac{2 \left(3+\sqrt{3(3+2e^{-1/2})} \right) \,
\left(3+2e^{-1/2}+\sqrt{3(3+2e^{-1/2})} \right)}{e \, \sqrt{3(3+2e^{-1/2})}} \nonumber \\[0.1cm]
&& = \frac{2}{e} \, \left(3+\sqrt{3(3+2e^{-1/2})} \right) \,
\left(1+\sqrt{1+\frac{2}{3} \cdot e^{-1/2}} \right) \nonumber \\[0.1cm]
&& = \frac{6}{e} \, \left(1+\sqrt{1+\frac{2}{3} \cdot e^{-1/2}} \right)^2 \approx 10.539
\label{eq: limit of the ratio between the upper and lower bounds on the total variation
distance when lambda tends to infinity}
\end{eqnarray}
where equality~(a) holds since, from \eqref{eq: optimal theta for alpha1 and alpha2 equal to lambda},
$\lim_{\lambda \rightarrow \infty} \theta = 3+\sqrt{3(3+2e^{-1/2})}$.
Furthermore, the limit of this ratio when $\lambda$ tends to zero is equal to
\begin{eqnarray}
&& 2 \, \lim_{\lambda \rightarrow 0} \left(\frac{1-e^{-\lambda}}{\lambda} \right) \,
\lim_{\lambda \rightarrow 0} \left(\frac{\lambda \bigl(2 e^{-{3/2}}+\theta \, e^{-1} \bigr)}{1-
\frac{3\lambda+7}{\lambda \theta}}\right) \nonumber \\[0.1cm]
&& = 2 \,  \lim_{\lambda \rightarrow 0} \left(\frac{\lambda \theta \, (2 e^{-{3/2}}+\theta \,
e^{-1})}{\theta-\bigl(3+\frac{7}{\lambda}\bigr)}\right) \nonumber \\[0.1cm]
&& \stackrel{\text{(a)}}{=} \frac{28}{e} \, \lim_{\lambda \rightarrow 0}
\left(\frac{2 e^{-{1/2}}+\theta)}{\theta-\bigl(3+\frac{7}{\lambda}\bigr)}\right) \nonumber \\
&& \stackrel{\text{(b)}}{=} \frac{56}{e} \approx 20.601
\label{eq: limit of the ratio between the upper and lower bounds on the total variation
distance when lambda tends to zero}
\end{eqnarray}
where equalities~(a) and (b) hold since, from
\eqref{eq: optimal theta for alpha1 and alpha2 equal to lambda}, it follows that
$\lim_{\lambda \rightarrow 0} (\lambda \theta) = 14$. Note that the two limits in
\eqref{eq: limit of the ratio between the upper and lower bounds on the total variation distance when lambda tends to infinity}
and \eqref{eq: limit of the ratio between the upper and lower bounds on the total variation distance when lambda tends to zero}
are indeed consistent with the limits of the upper curve in
Figure.~\ref{Figure: ratio ot upper and lower bounds on the total variation distance} (see
p.~\pageref{Figure: ratio ot upper and lower bounds on the total variation distance}). This
implies that Corollary~\ref{corollary: lower bound on the total variation distance} improves
the original lower bound on the total variation distance in \cite[Theorem~2]{BarbourH_1984}
by a factor of $\frac{32}{10.539} \approx 3.037$ in the limit where $\lambda \rightarrow \infty$,
and it improves it by a factor of $\frac{32}{20.601} \approx 1.553$ in the other extreme case
where $\lambda \rightarrow 0$ while still having a closed-form expression lower bound in
Corollary~\ref{corollary: lower bound on the total variation distance} where the only reason
for this improvement that is related to the optimal choice of the free parameter $\theta$,
versus its sub-optimal choice in the proof of \cite[Theorem~2]{BarbourH_1984}, shows a sensitivity
of the resulting lower bound to the selection of $\theta$. This observation in fact motivated
us to further improve the lower bound on the total variation distance in
Theorem~\ref{theorem: improved lower bound on the total variation distance} by introducing the
two additional parameters $\alpha_1, \alpha_2 \in \reals$ of the proposed function $f$
in \eqref{eq: proposed f for the derivation of the improved lower bound on the total variation distance}
(which, according to the proof in the previous sub-section, are set to be both
equal to $\lambda$). The further improvement in the lower bound at the expense of a feasible
increase in computational complexity is shown in the plot of
Figure.~\ref{Figure: ratio ot upper and lower bounds on the total variation distance}
(by comparing the upper and lower curves of this plot which correspond to
the ratio of the upper bound in \cite[Theorem~1]{BarbourH_1984} and new lower bounds in
Corollary~\ref{corollary: lower bound on the total variation distance} and
Theorem~\ref{theorem: improved lower bound on the total variation distance}, respectively.
It is interesting to note that no improvement is obtained however in
Theorem~\ref{theorem: improved lower bound on the total variation distance}, as
compared to Corollary~\ref{corollary: lower bound on the total variation distance},
for $\lambda \geq 20$, as is shown in
Figure~\ref{Figure: ratio ot upper and lower bounds on the total variation distance}
(since the the upper and lower curves in this plot merge for $\lambda \geq 20$, and their
common limit in the extreme case where $\lambda \rightarrow \infty$ is given in \eqref{eq: limit
of the ratio between the upper and lower bounds on the total variation distance when lambda
tends to infinity}; this therefore implies that the two new lower bounds in Theorem~\ref{theorem:
improved lower bound on the total variation distance} and Corollary~\ref{corollary: lower bound
on the total variation distance} coincide for these values of $\lambda$; however, for this
range of values of $\lambda$, the lower bound on the total variation distance
in Corollary~\ref{corollary: lower bound on the total variation distance} has the advantage of
being expressed in closed form (i.e., there is no need for a numerical optimization of this
bound). Due to the above discussion, another important reasoning for our motivation to improve
the lower bound on the total variation distance in
Theorem~\ref{theorem: improved lower bound on the total variation distance} and
Corollary~\ref{corollary: lower bound on the total variation distance} is that the
factors of improvements that are obtained by these lower bounds (as compared to the
original bound) are squared, according to Pinsker's inequality, when one wishes to
derive lower bounds on the relative entropy, and this improvement becomes significant
in many inequalities in information theory and statistics where the relative entropy
appears in the error exponent (as is exemplified in
Section~\ref{subsection: Second part of applications of the new bounds}).
Finally, it is noted that the reason for introducing
this type of discussion, which partially motivates our paper, in a sub-section that refers
to proofs (of the second half of this work) is because this kind of discussion follows
directly from the proofs of Theorem~\ref{theorem: improved lower bound on the total variation distance}
and Corollary~\ref{corollary: lower bound on the total variation distance}, and
therefore it was introduced here.

\vspace*{0.1cm}
\subsubsection{Proof of Theorem~\ref{theorem: improved lower bound on the relative entropy}}
\label{subsubsection: Proof of the theorem with the improved lower bound on the relative entropy}
In the following we prove Theorem~\ref{theorem: improved lower bound on the relative entropy}
by obtaining a lower bound on the relative entropy between the distribution $P_W$ of a sum of
independent Bernoulli random variables $\{X_i\}_{i=1}^n$ with $X_i \sim \text{Bern}(p_i)$ and
the Poisson distribution $\text{Po}(\lambda)$ with mean $\lambda \triangleq \sum_{i=1}^n p_i$.
A first lower bound on the relative entropy follows from a combination of Pinsker's inequality
(see Eq.~\eqref{eq: Pinsker's inequality}) with the lower bound on the total variation distance
between these distributions (see
Theorem~\ref{theorem: improved lower bound on the total variation distance}). The combination
of the two gives that
\begin{equation}
D\bigl(P_W \, || \, \text{Po}(\lambda)\bigr) \geq 2 \bigl(K_1(\lambda)\bigr)^2 \,
\left(\sum_{i=1}^n p_i^2 \right)^2 \, .
\label{eq: first lower bound on the relative entropy based on Pinsker's inequality and the
lower bound on the total variation distance}
\end{equation}
This lower bound can be tightened via the distribution-dependent refinement of Pinsker's
inequality in \cite{OrdentlichW_IT2005}, which is introduced
shortly in Section~\ref{subsection: Second part of the revision of some known results}.
Following the technique of this refinement, let $Q \triangleq \Pi_{\lambda}$ be the
probability mass function that corresponds to the Poisson distribution $\text{Po}(\lambda)$,
i.e., $$Q(k) = \frac{e^{-\lambda} \, \lambda^k}{k !} \, \quad \forall \, k \in \naturals_0.$$
If $\lambda \leq \log 2$ then $Q(0) = e^{-\lambda} \geq \frac{1}{2}$. Hence, from
\eqref{eq: pi_Q for the refinement of Pinsker's inequality}, the maximization of
$\min \bigl\{Q(A), \, 1-Q(A)\bigr\}$ over all the subsets $A \subseteq \naturals_0$
is obtained for $A = \{0\}$ (or, symmetrically, for $A = \naturals_0 \setminus \{0\} = \naturals$)
which implies that, if $\lambda \leq \log 2$, one gets from
Eqs.~\eqref{eq: a distribution-dependent refinement of Pinsker's inequality},
\eqref{eq: pi_Q for the refinement of Pinsker's inequality} and
\eqref{eq: phi function for the refinement of Pinsker's inequality} that
\begin{equation}
D\bigl(P_W \, || \, \text{Po}(\lambda)\bigr) \geq \varphi(\Pi_Q) \bigl(K_1(\lambda)\bigr)^2 \,
\left(\sum_{i=1}^n p_i^2 \right)^2 \, .
\label{eq: refined lower bound on the relative entropy based on Pinsker's inequality and the
lower bound on the total variation distance}
\end{equation}
where
\begin{equation}
\Pi_{Q} = \min \bigl\{e^{-\lambda}, 1-e^{-\lambda} \bigr\} = 1-e^{-\lambda}
\end{equation}
and, since $\Pi_{Q} < \frac{1}{2}$ then
\begin{eqnarray}
&& \varphi(\Pi_Q) = \frac{1}{1-2 \Pi_Q} \cdot \log\left(\frac{1-\Pi_Q}{\Pi_Q}\right)
\nonumber \\[0.2cm]
&& \hspace*{1.2cm} = \left( \frac{1}{2 e^{-\lambda}-1} \right) \;
\log \left(\frac{1}{e^{\lambda}-1} \right) \, .
\label{eq: Pi_Q for the Poisson distribution with mean lambda}
\end{eqnarray}
Hence, the combination of
\eqref{eq: first lower bound on the relative entropy based on Pinsker's inequality and the lower bound on the total variation distance},
\eqref{eq: refined lower bound on the relative entropy based on Pinsker's inequality and the
lower bound on the total variation distance}, and
\eqref{eq: Pi_Q for the Poisson distribution with mean lambda}
gives the lower bound on the relative entropy in
Theorem~\ref{theorem: improved lower bound on the relative entropy} (see
Eqs.~\eqref{eq: improved lower bound on the relative entropy},
\eqref{eq: coefficient of improved lower bound on the relative entropy} and
\eqref{eq: the refinement of Pinsker's inequality for the Poisson distribution}
in this theorem).
The upper bound on the considered relative entropy is a known result (see
\cite[Theorem~1]{KontoyiannisHJ_2005}), which is cited here in order to have
both upper and lower bounds in the same inequality (see
Eq.~\eqref{eq: improved lower bound on the relative entropy}). This completes
the proof of Theorem~\ref{theorem: improved lower bound on the relative entropy}.

\vspace*{0.1cm}
\subsubsection{Proof of Proposition~\ref{proposition: sharpened inequality that relates
between the total variation, Hellinger distance, Bhattacharyya parameter and relative entropy}}
\label{subsubsection: Proof of the sharpened inequality that relates between the total
variation, Hellinger distance, Bhattacharyya parameter and relative entropy}
We start by proving the tightened upper and lower bounds on the Hellinger
distance in terms of the total variation distance and relative entropy between the two
considered distributions. These refined bounds in
\eqref{eq: sharpened inequality that relates between the total variation, Hellinger
distance and relative entropy} improve the original bounds in
\eqref{eq: known inequality that relates between the total variation,
Hellinger distance and relative entropy}. It is noted that the left-hand side of
\eqref{eq: known inequality that relates between the total variation, Hellinger distance
and relative entropy} is proved in \cite[p.~99]{Reiss_book1989}, and the right-hand side
is proved in \cite[p.~328]{Reiss_book1989}. The following is the proof of the refined bounds
on the Hellinger distance in \eqref{eq: sharpened inequality that relates between the total variation, Hellinger distance and relative entropy}.

Lets start with the proof of the left-hand side of
\eqref{eq: sharpened inequality that relates between the total variation, Hellinger
distance and relative entropy}. To this end, let $P$ and $Q$ be two probability mass
functions that are defined on a same set $\mathcal{X}$.
From \eqref{eq: the L1 distance is twice the total variation distance},
\eqref{eq: Hellinger distance} and the Cauchy-Schwartz inequality
\begin{eqnarray}
&& d_{\text{TV}}(P,Q) \nonumber \\[0.1cm]
&& = \frac{1}{2} \, \sum_{x \in \mathcal{X}} \left| P(x) - Q(x) \right| \nonumber \\[0.1cm]
&& = \frac{1}{2} \, \sum_{x \in \mathcal{X}} \left| \sqrt{P(x)} - \sqrt{Q(x)} \, \right| \,
\left( \sqrt{P(x)} + \sqrt{Q(x)} \, \right) \nonumber \\[0.1cm]
&& \leq \frac{1}{2} \, \left( \sum_{x \in \mathcal{X}} \left( \sqrt{P(x)} - \sqrt{Q(x)} \,
\right)^2 \right)^{\frac{1}{2}} \, \left( \sum_{x \in \mathcal{X}} \left( \sqrt{P(x)} +
\sqrt{Q(x)} \, \right)^2 \right)^{\frac{1}{2}} \nonumber \\[0.1cm]
&& = d_{\text{H}}(P,Q) \cdot \left( 1 + \sum_{x \in \mathcal{X}} \sqrt{P(x) \, Q(x)}
\right)^{\frac{1}{2}} \nonumber \\[0.1cm]
&& = d_{\text{H}}(P,Q) \, \Bigl(2-\bigl(d_{\text{H}}(P,Q)\bigr)^2 \Bigr)^{\frac{1}{2}} \, .
\label{eq: 1st step in the derivation of the refined bounds for Hellinger distance}
\end{eqnarray}
Let $c \triangleq \bigl(d_{\text{TV}}(P,Q)\bigr)^2$ and
$x \triangleq \bigl(d_{\text{H}}(P,Q)\bigr)^2$,
then it follows by squaring both sides of
\eqref{eq: 1st step in the derivation of the refined bounds for Hellinger distance}
that $x (2-x) \geq c$, which therefore implies that
\begin{equation}
1 - \sqrt{1-c} \leq x \leq 1 + \sqrt{1-c} \, .
\label{eq: 2nd step in the derivation of the refined bounds for Hellinger distance}
\end{equation}
The right-hand side of \eqref{eq: 2nd step in the derivation of the refined bounds
for Hellinger distance} is satisfied automatically since $0 \leq d_{\text{H}}(P,Q) \leq 1$
implies that $x \leq 1$. The left-hand side of
\eqref{eq: 2nd step in the derivation of the refined bounds for Hellinger distance}
gives the lower bound on the left-hand side of
\eqref{eq: sharpened inequality that relates between the total variation, Hellinger
distance and relative entropy}. Next, we prove the upper bound on the right-hand side of
\eqref{eq: sharpened inequality that relates between the total variation, Hellinger
distance and relative entropy}. By Jensen's inequality

\vspace*{-0.7cm}
\begin{eqnarray}
&& \bigl(d_{\text{H}}(P,Q)\bigr)^2 \nonumber \\[0.1cm]
&& = \frac{1}{2} \, \sum_{x \in \mathcal{X}} \left\{ \left( \sqrt{P(x)} - \sqrt{Q(x)} \,
\right)^2 \right\} \nonumber \\[0.1cm]
&& = 1 - \sum_{x \in \mathcal{X}} \sqrt{P(x) \, Q(x)} \nonumber \\[0.1cm]
&& = 1 - \sum_{x \in \mathcal{X}} P(x) \, \sqrt{\frac{Q(x)}{P(x)}} \nonumber \\[0.1cm]
&& = 1 - \sum_{x \in \mathcal{X}} P(x) \, e^{\frac{1}{2} \,
\log \left(\frac{Q(x)}{P(x)}\right)} \nonumber \\[0.1cm]
&& \leq 1 - e^{\frac{1}{2} \sum_{x \in \mathcal{X}} P(x) \,
\log \left(\frac{Q(x)}{P(x)}\right)} \nonumber \\[0.1cm]
&& = 1 - e^{-\frac{1}{2} \, D(P||Q)}
\end{eqnarray}
which completes the proof of
\eqref{eq: sharpened inequality that relates between the total variation, Hellinger
distance and relative entropy}. The other bound on the Bhattacharyya
parameter in \eqref{eq: sharpened inequality that relates between the total variation,
Bhattacharyya parameter and relative entropy} follows from
\eqref{eq: sharpened inequality that relates between the total variation, Hellinger distance
and relative entropy} and the simple relation in
\eqref{eq: relation between the Hellinger distance and Bhattacharyya parameter}
between the Bhattacharyya parameter and Hellinger distance. This completes the proof of
Proposition~\ref{proposition: sharpened inequality that relates between the total variation,
Hellinger distance, Bhattacharyya parameter and relative entropy}.

\begin{remark}
The weaker bounds in
\eqref{eq: known inequality that relates between the total variation, Hellinger distance
and relative entropy}, proved in \cite{Reiss_book1989}, follow from their refined version in
\eqref{eq: sharpened inequality that relates between the total variation, Hellinger distance
and relative entropy} by using the simple inequalities
\begin{eqnarray*}
&& \sqrt{1-x} \leq 1-\frac{x}{2} \, , \quad \forall \, x \in [0,1] \\
\mbox{and} \\
&& e^{-x} \geq 1-x, \quad \forall \, x \geq 0.
\end{eqnarray*}
\end{remark}

\vspace*{0.1cm}
\subsubsection{Proof of Corollary~\ref{corollary: upper and lower bounds on the Hellinger
distance and Bhattacharyya parameter in the context of Poisson approximation}}
\label{subsubsection: Proof of the bounds on the Hellinger distance and Bhattacharyya
parameter in the context of Poisson approximation} This corollary is a direct consequence
of Theorems~\ref{theorem: improved lower bound on the total variation distance}
and~\ref{theorem: improved lower bound on the relative entropy}, and
Proposition~\ref{proposition: sharpened inequality that relates
between the total variation, Hellinger distance, Bhattacharyya parameter and relative entropy}.

\vspace*{0.1cm}
\subsubsection{Proof of Corollary~\ref{corollary: asymptotic results of the relative entropy
and related quantities for independent Bernoulli sums}}
\label{subsubsection: Proof of some asymptotic results of the relative entropy and related quantities for independent Bernoulli sums}
Under the conditions in Corollary~\ref{corollary: asymptotic results of the relative entropy and
related quantities for independent Bernoulli sums}, the asymptotic scaling of the
total variation distance, relative entropy, Hellinger distance and Bhattacharyya parameter
follow from their (upper and lower) bounds in Theorems~\ref{theorem: improved lower bound
on the total variation distance} and \ref{theorem: improved lower bound on the relative entropy}
and Eqs.~\eqref{eq: upper and lower bounds on the Hellinger distance in the context of Poisson approximation}
and \eqref{eq: upper and lower bounds on the Bhattacharyya parameter in the context of Poisson approximation}, respectively.
This completes the proof of
Corollary~\ref{corollary: asymptotic results of the relative entropy
and related quantities for independent Bernoulli sums}.

\vspace*{0.1cm}
\subsubsection{Proof of Proposition~\ref{proposition: lower bound on the Chernoff information
in terms of the total variation distance}}
\label{subsubsection: Proof of a lower bound on the Chernoff information in terms of the total variation distance}
Let $P$ and $Q$ be two arbitrary probability mass functions that are defined
on a same set $\mathcal{X}$. We derive in the following the lower bound on the Chernoff information
in terms of the total variation distance between $P$ and $Q$, as is stated in
\eqref{eq: lower bound on the Chernoff information in terms of the total variation distance}.
\begin{eqnarray*}
&& C(P,Q) \stackrel{(\text{a})}{\geq} -\log \left( \sum_{x \in \mathcal{X}}
\sqrt{P(x) \, Q(x)} \right) \\
&& \hspace*{1.5cm} \stackrel{(\text{b})}{=} -\log \, \text{BC}(P,Q) \\
&& \hspace*{1.5cm} \stackrel{(\text{c})}{=} -\log \, \left(1-\bigl(d_{\text{H}}(P,Q)\bigr)^2
\right) \\
&& \hspace*{1.5cm} \stackrel{(\text{d})}{\geq} -\frac{1}{2} \log
\Bigl( 1- \bigl(d_{\text{TV}}(P,Q) \bigr)^2 \Bigr)
\end{eqnarray*}
where inequality~(a) follows by selecting the possibly sub-optimal choice
$\theta = \frac{1}{2}$ in \eqref{eq: Chernoff information}, equality~(b)
holds by definition of the Bhattacharyya parameter (see \eqref{eq: Bhattacharyya parameter}),
equality~(c) follows from the equality in \eqref{eq: relation between the Hellinger
distance and Bhattacharyya parameter} that relates the Hellinger distance and
Bhattacharyya parameter, and inequality~(d) follows from the lower bound on the
Hellinger distance in terms of the total variation distance (see
\eqref{eq: sharpened inequality that relates between the total variation, Hellinger
distance and relative entropy}).
This completes the proof of Proposition~\ref{proposition: lower bound on the Chernoff information
in terms of the total variation distance}.

\vspace*{0.1cm}
\subsubsection{Proof of Corollary~\ref{corollary: improved lower bound on the Chernoff information
between Bernoulli sums of independent RVs and Poisson distribution}}
\label{Proof of the improved lower bound on the Chernoff information between Bernoulli sums of independent RVs and Poisson distribution}
This corollary is a direct consequence of the lower bound on the total variation distance in
Theorem~\ref{theorem: improved lower bound on the total variation distance}, and the lower
bound on the Chernoff information in terms of the total variation distance in
Proposition~\ref{proposition: lower bound on the Chernoff information
in terms of the total variation distance}.

\section*{Acknowledgment} I thank Ioannis Kontoyiannis for inviting
me to present part of this work at the 2012 Information Theory Workshop (ITW 2012) in
Lausanne, Switzerland, September 2012. I also thank Louis H. Y. Chen for
expressing his interest in this work during the 2012 International Workshop on Applied
Probability that took place in June 2012 in Jerusalem, Israel. These two occasions
were stimulating for the writing of this paper. I am thankful to Peter Harremo{\"{e}}s
for personal communications during the 2012 International Symposium
on Information Theory (ISIT 2012) at MIT, and to Abraham J. Wyner for notifying me
(during ISIT 2012 as well) about his related work to the Chen-Stein method and Poisson
approximation in the context of pattern recognition and the Lempel-Ziv algorithm
\cite{Wyner_IT97}.

\end{document}